\documentclass[fleqn,twocolumns,usenatbib]{mnras}

\usepackage[T1]{fontenc}
\usepackage{ae,aecompl}

\usepackage{graphicx}	
\usepackage{amsmath}	

\usepackage{amssymb}	

\title[Nitrogen X-ray ISM absorption]{Nitrogen X-ray absorption in the local ISM}

\author[Gatuzz et al.]{
Efrain Gatuzz$^{1}$\thanks{E-mail: egatuzz@mpe.mpg.de}, 
Javier~A.~Garc\'ia$^{2,3}$
and Timothy~R.~Kallman$^{4}$ 
\\
$^{1}$Max-Planck-Institut f\"ur extraterrestrische Physik, Gie{\ss}enbachstra{\ss}e 1, 85748 Garching, Germany\\
$^{2}$Cahill Center for Astronomy and Astrophysics, California Institute of Technology, Pasadena, CA 91125, USA\\
$^{3}$Dr. Karl Remeis-Observatory and Erlangen Centre for Astroparticle Physics, Sternwartstr. 7, 96049 Bamberg, Germany\\
$^{4}$NASA Goddard Space Flight Center, Greenbelt, MD 20771, USA\\ 
}

\date{Accepted XXX. Received YYY; in original form ZZZ}
\pubyear{2020}
\begin{document}

 \defcitealias{gud12}{GUD+12}
 \defcitealias{nas10}{NAS+10}
 \defcitealias{moo02}{MOO+02} 
 \defcitealias{kna06}{KNA+06} 
 \defcitealias{mey97}{MEY+97}

 \label{firstpage}
\pagerange{\pageref{firstpage}--\pageref{lastpage}}
\maketitle
\begin{abstract}
Nitrogen is one of the most abundant metals in the interstellar medium (ISM), and thus it constitutes an excellent test to study a variety of astrophysical environments, ranging from nova to active galactic nuclei. We present a detailed analysis of the gaseous component of the N K~edge using high-resolution {\it XMM-Newton} spectra of 12 Galactic and 40 extragalactic sources. For each source, we have estimated column densities for {\rm N}~{\sc i}, {\rm N}~{\sc ii}, {\rm N}~{\sc iii}, {\rm N}~{\sc v}, {\rm N}~{\sc vi} and {\rm N}~{\sc vii} ionic species, which trace the cold, warm and hot phases of the local Galactic interstellar medium. We have found that the cold-warm component column densities decrease with the Galactic latitude while the hot component does not. Moreover, the cold column density distribution is in good agreement with UV measurements. This is the first detailed analysis of the nitrogen K-edge absorption due to ISM using high-resolution X-ray spectra.

\end{abstract}


\section{Introduction}\label{sec_in}

The interstellar medium (ISM), defined as gas and dust between stars, is a key ingredient in the Galactic dynamics. The ISM shows multiple phases, characterized by different gas temperatures which vary from 10 to 10$^{6}$ K \citep[e.g][]{mck77,fal05,dra11,jen11,sta18}. In this sense, high-resolution X-ray spectroscopy is a powerful technique to study such environment because, due to their high energy, X-ray photons interact with the cold (including molecules and dust), warm, and hot components \citep{jue04,jue06,pin10,lia13,pin13,luo14,nic16a,gup17,gat18a,gat18c}.    

Among the most abundant metals in the ISM, nitrogen constitutes an excellent diagnostic tool to study a variety of astrophysical environments. For example, \citet{nes03} identified H- and He-like N absorption lines by a white dwarf outflow following the outburst of nova V4743~Sagittarii. \citet{ste05a} identified an {\rm N}~{\sc vi} Ly$\alpha$ absorption line associated to a warm absorber in the {\it XMM-Newton} X-ray spectra of the Seyfert 1 galaxy IC 4329A. \citet{ste05b} identified a {\rm N}~{\sc v} K$\alpha$ absorption line as $29.42$ \AA\ in the outflow of the Seyfert 1 galaxy NGC~5548. \citet{smi07} performed a detailed analysis of the soft X-ray spectrum of the Seyfert 1 galaxy Mrk~509 using {\it XMM-Newton} observations. They found three warm absorber phases traced, among others, by {\rm N}~{\sc vi} and {\rm N}~{\sc vii} absorption lines. \citet{ram08} identified {\rm N}~{\sc vi} K$\alpha$ and {\rm N}~{\sc vi} K$\beta$ absorption lines in the warm absorber of the MR~2251-178 quasar, tracing an outflow of ionized material. \citet{nes11} analyzed {\it XMM-Newton} observations of the fast classical nova V2491~Cyg. They found absorption lines due to {\rm N}~{\sc i} associated to the ISM and  photospheric lines ionic species such as {\rm N}~{\sc vi} and {\rm N}~{\sc vii}, tracing the dynamics of the ejecta.

In order to model the complexity of the photoabsorption K-edge, located at the 24--32 \AA wavelength band, the accuracy of atomic data is crucial, to avoid misidentification and misinterpretation of the observed absorption features. In last few years we have performed benchmarking of the atomic data by comparing theoretical calculations, astronomical observations and laboratory measurements for species found in the ISM such as carbon \citep{gat18b}, oxygen \citep{gar05,gat13a,gat13b,gor13}, neon \citep{gat15}, magnesium \citep{has14} and silicon \citep{gat20}. Following such studies, we present an analysis of the N K-edge absorption region using {\it XMM-Newton} observations of low-mass X-ray binaries (LMXBs) and extragalactic sources.  We describe the data sample and the spectral fitting procedure in Section~\ref{sec_xray_data}. Section~\ref{sec_dis} shows a discussion of the results obtained from the fits. Finally, Section~\ref{sec_con} summarizes the main results of our analysis.

\section{X-ray observations and spectral fitting}\label{sec_xray_data} 
We have compiled a data sample of 12 LMXBs and 40 extragalactic sources from the {\it XMM-Newton} Science Archive (XSA).  We selected observations with more than 1000 counts in the 24--32 \AA\ band (i.e., the N photoionization K-edge absorption region). In order to get an unbiased sample, we did not impose any constraints in the significance of the detection for a particular line (e.g. {\rm N}~{\sc vi} K$\alpha$ detection). Tables~\ref{tab_lmxbs} and ~\ref{tab_ex} list the analyzed Galactic and extragalactic sources, respectively, including the Galactic coordinates, ${\rm HI}$ column densities, which are taken from \citet{wil13}, total exposure times and total number of counts of all observations in the N K-edge. The distances to the sources are included for all LMXBs.  Observations were reduced using the Science Analysis System (SAS\footnote{https://www.cosmos.esa.int/web/xmm-newton/sas}, version 18.0.0) including background subtraction and following the standard procedure to obtain high-resolution spectra from the Reflection Grating Spectrometers \citep[RGS,][]{denh01}. In particular, we follow the SAS thread to reduce RGS data and extract spectra of point-like sources. Using the {\tt rgsproc} task, the procedure included the filtering of events and exposure, to exclude flaring background, and the creation of response matrices. Each spectrum was rebinned to have at least 1 count per channel.
 
For each source, all observations were fitted simultaneously using the {\sc xspec} package (version 12.10.1\footnote{\url{https://heasarc.gsfc.nasa.gov/xanadu/xspec/}}) in the 24--32 \AA\ wavelength region. The continuum was modeled with a {\tt powerlaw*constant} model, where the Photon-Index was set free to vary but tied among all the observations of the same source, while the {\tt constant} accounts for differences in the normalization among them.  Moreover, we used {\tt cash} statistics \citep{cas79} in the spectral fitting analysis.

\subsection{Nitrogen photoabsorption cross-sections}\label{sec_n_cross} 
\begin{figure}
\centering
\includegraphics[scale=0.35]{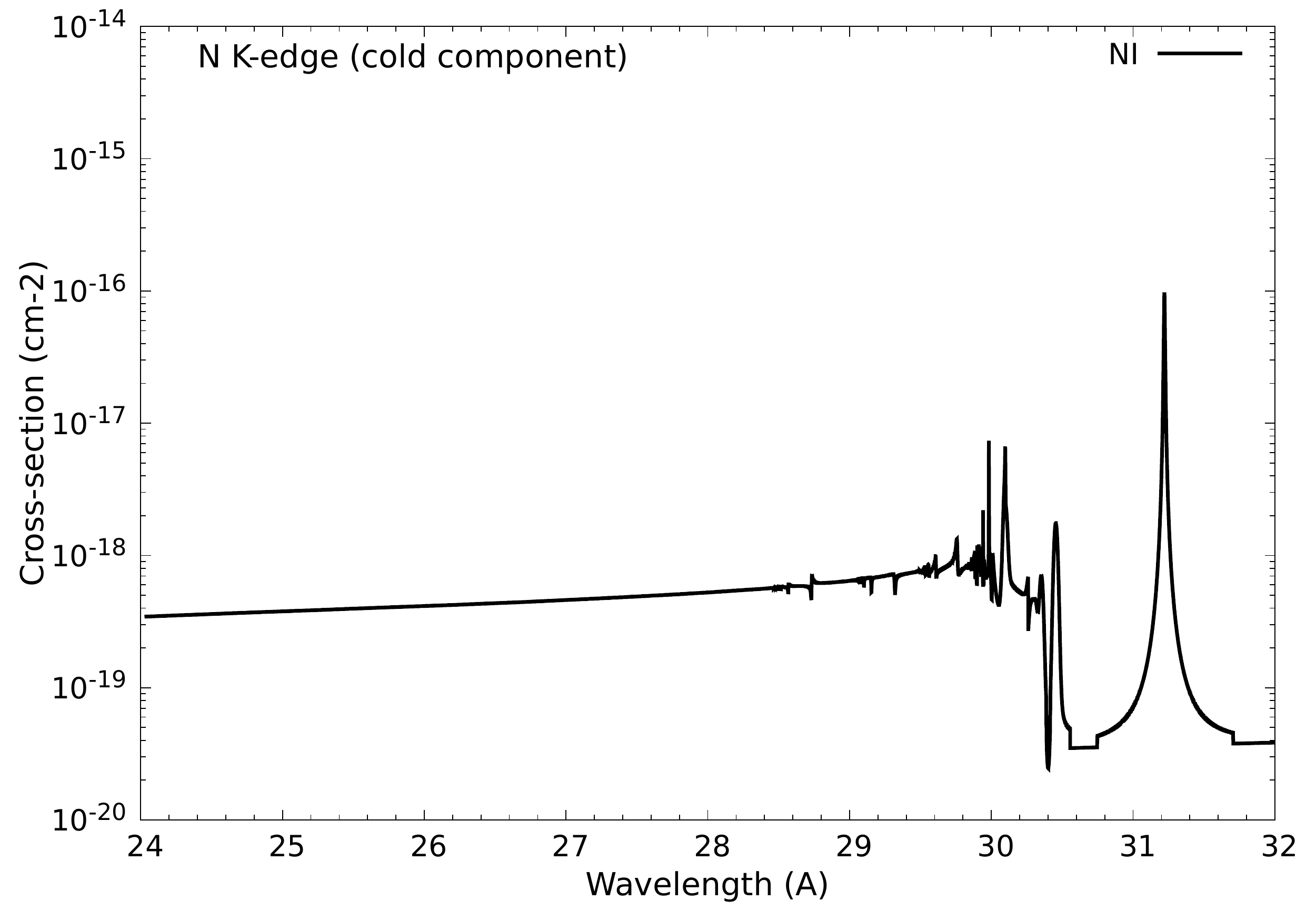}\\
\includegraphics[scale=0.35]{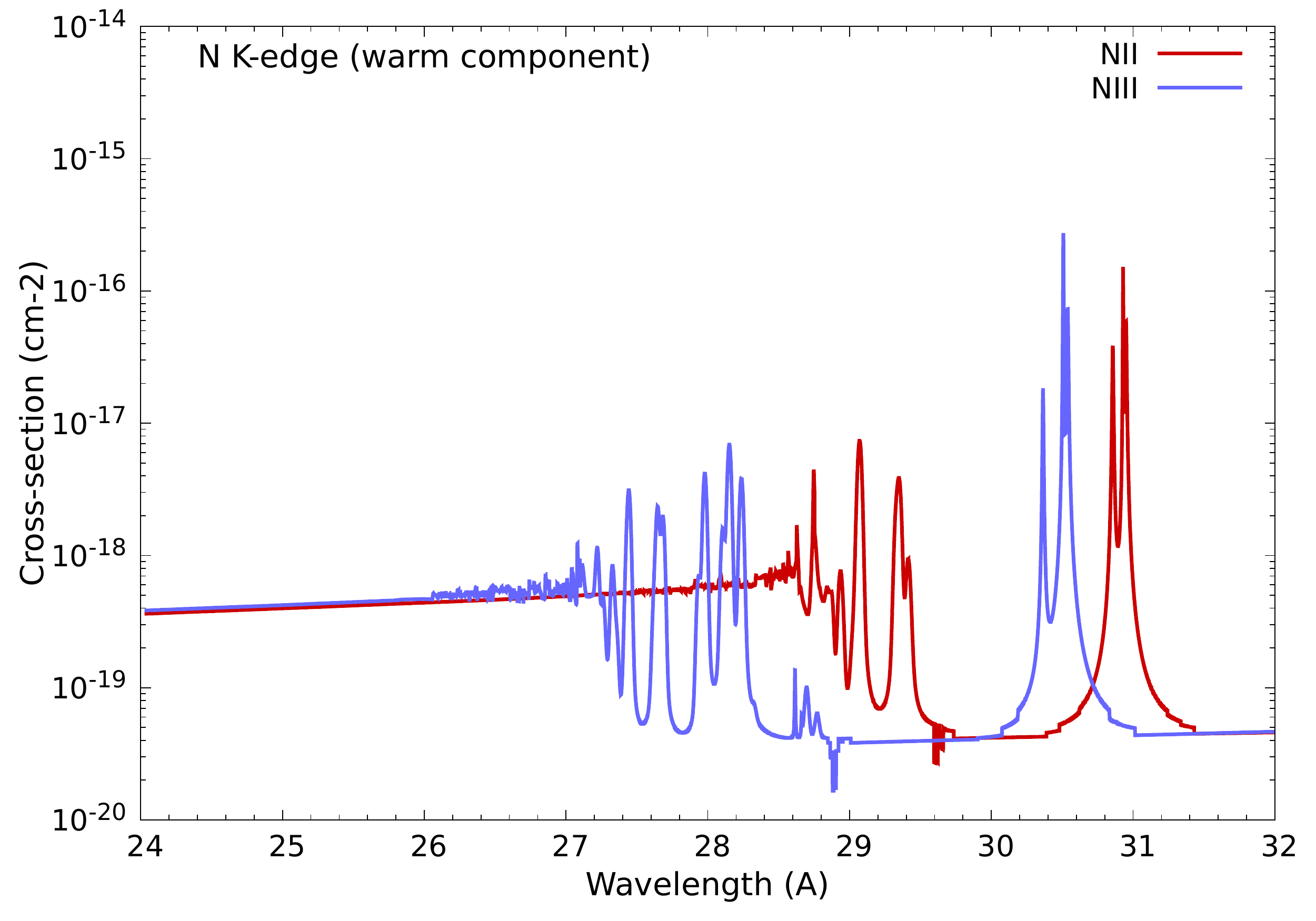}\\
\includegraphics[scale=0.35]{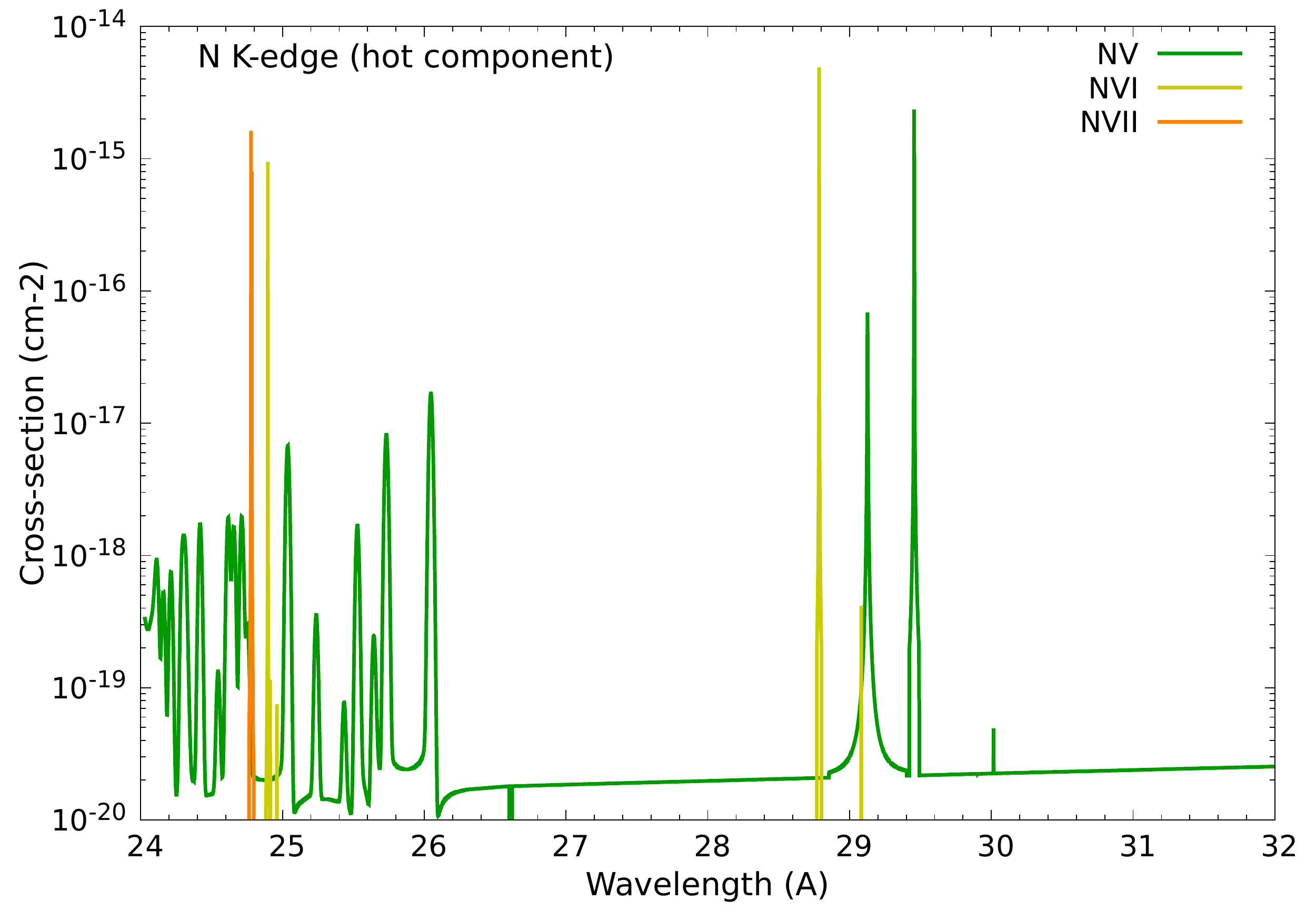}
\caption{ {\rm N}~{\sc i} (top panel), {\rm N}~{\sc ii}, {\rm N}~{\sc iii} (middle panel), {\rm N}~{\sc v}, {\rm N}~{\sc vi} and {\rm N}~{\sc vii} (bottom panel) photoabsorption cross sections computed by \citet{gar09a} and included in the model. }\label{fig_n_cross}
\end{figure}

We use the K-edge photoionization cross-sections for {\rm N}~{\sc ii}-{\rm N}~{\sc v} computed by \citet{gar09a}, which include detailed calculations of atomic properties of K-vacancy states for all ions of the nitrogen isonuclear sequence. Importantly, the smearing of the K-edge due to both Auger and radiation damping is taken into account. In the cases of {\rm N}~{\sc vi} and {\rm N}~{\sc vii}, we used the K-edge photoionization cross-sections from the Opacity Project \citep{bad05}. Figure~\ref{fig_n_cross} shows the photoabsorption cross-sections included in the model, which includes the cold, warm and hot phases of the ISM.

 We included these N photoabsorption cross-sections in a modified version of the {\tt ISMabs} model \citep{gat15}, thus allowing the column densities for the ionic species of interest to be free parameters in the data fitting. We fixed the ${\rm HI}$ {\tt ISMabs} column densities to the values provided by \citet{wil13}. For each source, the column densities were linked between the different observations. We note that a detailed benchmarking of the main doublet/triplet resonance line positions (see Figure~\ref{fig_n_cross}) cannot be performed given that the differences in their relative positions are smaller than the instrumental spectral resolution. For example, for {\rm N}~{\sc ii} K$\alpha$ and {\rm N}~{\sc iii} K$\alpha$ we have a minimum separation for contiguous peaks of $\Delta\lambda\sim 22.1$ m\AA\ and $\Delta\lambda\sim 31.2$ m\AA, respectively, while the RGS resolution in the N K-edge region is $\Delta\lambda\sim 60$ m\AA. Therefore, we used the atomic data as computed by \citet{gar09a}, which estimated the accuracy of the K-threshold to be within 1~eV.

\begin{table*}
\caption{\label{tab_lmxbs}List of Galactic observations analyzed.}
\centering
\begin{tabular}{lcccccccc}
\hline
Source   & Galactic &Distance     &$N({\rm HI})$& Exposure& Number counts\\
 &Coordinates &(kpc)   &   &  Time (ks)& (24-32 \AA)\\
\hline 
4U~1254--69 		& $(303.4,-6.4)$   &  $13.0\pm 3.0^{a}$ &$	3.46	$ & 233& 66936  \\	 	
4U~1543--62  		& $(321.76,-6.34)$ &  $7.0^{b}$         &$	3.79	$ & 50& 35103  \\	  
4U~1636--53 		& $(332.9,-4.8)$   & $6.0\pm 0.5^{c}$   &$	4.04	$ & 379& 137151  \\   
4U~1735--44 		& $(346.0,-6.9)$   & $9.4\pm 1.4^{d}$   &$	3.96	$ & 107& 85494 \\  
4U~1820-30          & $(2.7,-7.9)$     & $7.6\pm 0.4^{e}$   &$  1.33    $ & 81& 381038 \\
4U~1957+11 		    & $(51.31,-9.33)$  & $20< d <40 ^{f}$   &$1.98      $ & 85& 65478  \\  
Aql~X--1    		&$(37.7,-4.1)$     &$5.2\pm 0.8^{d}$    &$	4.30	$ & 165& 43229  \\  
Cygnus~X-2          &$(87.3,-11.3)$    &$13.4 \pm 2^{d}$    &$ 1.90     $ & 342& 633707\\  
GS~1826--238	 	&$(9.3,-6.1)$      &$6.7^{c}$			&$	3.00	$ & 347&  87798 \\
GX~9+9  	  		&$(8.5,9.0)$       &$4.4^{g}$			& $	3.31	$ & 80& 108956  \\ 
HETEJ1900.1-2455    &$(11.30,-12.87)$  &$5^{h}$			    & $	1.76	$ & 71& 57691  \\ 
SAXJ1808.4-3658     &$(355.39,-8.15)$  &$2.8^{c}$			& $	1.76	$ & 173& 568390  \\ 
Serpens~X--1        &$(36.12,4.84)$    &$11.1\pm 1.6^{d}$	& $	5.42	$ & 66& 20858   \\   
Swift~J1753.5--0127	&$(24.89,12.18)$   & $ 8.42^{+4.32}_{-2.85}$ $^{i}$	& $	2.98$& 249& 428804 \\  
\hline
\multicolumn{6}{l}{ $N({\rm HI})$ in units of $10^{21}$cm$^{-2}$ }\\
\multicolumn{6}{l}{Distances obtained from $^a$\citet{int03};$^b$\citet{wang04} }\\
\multicolumn{6}{l}{$^c$\citet{gall08};$^d$\citet{jon04};$^e$\citet{kuu03} }\\ 
\multicolumn{6}{l}{$^f$\citet{gom15};$^g$\citet{gri02};$^h$\citet{hyn04}; $^i$\citet{gan19}.}
\end{tabular} 
\end{table*}

\begin{table}
\caption{\label{tab_ex}List of extragalactic observations analyzed.}
\scriptsize
\centering
\begin{tabular}{lcccccccc}
\hline
Source   & Galactic      &$N({\rm HI})$& Exposure& Number of counts\\
 &Coordinates    &    &Time (ks)& (24-32 \AA) \\
\hline 
1ES~1028+511	&$(	161.44	;	54.44	)$&$	1.26	$&300& 61972 \\
1ES~1553+113	&$(	21.91	;	43.96	)$&$	4.35	$& 2065& 396199 \\ 
1H~0414+009	&$(	191.81	;	-33.16	)$&$	13.70	$& 91& 12194 \\
1H~0707--495	&$(	260.17	;	-17.67	)$&$	6.55 $&1395& 127334 \\
1H~1219+301	&$(	186.36	;	82.73	)$&$0.20$& 29& 29318\\ 	
1H~1426+428	&$(	77.49	;	64.90	)$&$	1.14 $& 425& 145627 \\
3C~120	&$(	190.37	;	-27.40	)$&$	19.30	$& 305& 19907 \\
3C~273	&$(	289.95	;	64.36	)$&$	1.78	$& 1426& 912126 \\
3C~279	&$(	305.10	;	57.06	)$&$	2.22	$& 153&10181 \\ 
3C~390.3	&$(	111.44	;	27.07	)$&$	4.51 $& 123& 29626 \\ 
Ark~120	&$(	201.69	;	-21.13	)$&$	1.40	$& 773& 203191 \\ 
Ark~564	&$(	92.14	;	-25.34	)$&$	6.74	$& 890& 556839 \\ 
ESO~141--G055	&$(	338.18	;	-26.71	)$&$	6.41 $& 268& 69813 \\
ESO~198--G24	&$(	271.64	;	-57.95	)$&$	3.27 $& 178& 20460 \\
Fairall~9	&$(	295.07	;	-57.83	)$&$	3.43 $&328& 42360 \\
H2356--309	&$(	12.84	;	-78.04	)$&$	1.48 $& 702& 91678 \\
H1821+643	&$(	94.00	;	27.42	)$&$0.39	$& 121& 22990 \\
HE1143-1810	&$(	281.85	;	41.71	)$&$0.34	$& 184& 51334 \\ 	
IRAS13349+2438	&$(	20.60	;	79.32	)$&$	1.07	$& 199& 24152 \\
IC4329A	&$(	317.50	;	30.92	)$&$0.56	$& 165& 19327 \\ 	 
IZw1	&$(	123.75	;	-50.17	)$&$	6.01	$& 382& 50616\\
MCG-6-30-15	&$(	313.29	;	27.68	)$&$0.47	$& 794& 341504 \\ 
MR2251--178	&$(	46.20	;	-61.33	)$&$	2.67	$& 592& 131180 \\ 
Mrk~279	&$(	115.04	;	46.86	)$&$	1.72	$& 188& 74129 \\ 
Mrk~421	&$(	179.83	;	65.03	)$&$	2.01	$&3004&7.37$\times 10^{6}$ \\
Mrk~501	&$(	63.60	;	38.86	)$&$	1.66	$& 200& 109344 \\
Mrk~509	&$(	35.97	;	-29.86	)$&$	5.04	$&835& 498007 \\
Mrk~766	&$(	190.68	;	82.27	)$&$ 0.19$& 739& 223087 \\
Mrk~841	&$(	11.21	;	54.63	)$&$	2.43 $& 206& 35132 \\ 
NGC~3783	&$(	287.46	;	22.95	)$&$	13.80	$& 486& 46906 \\
NGC~4593	&$(	297.48	;	57.40	)$&$	2.04	$& 386& 97579 \\
NGC~5548	&$(	31.96	;	70.50	)$&$	1.69	$& 1025& 108103 \\
NGC~7213	&$(	349.59	;	-52.58	)$&$	1.12	$& 181& 22796 \\
NGC~7469	&$(	83.10	;	-45.47	)$&$	5.24	$& 857& 229214 \\
PG1116+215	&$(	223.36	;	68.21	)$&$	1.43	$& 393& 49771 \\ 
PKS~0548--32	&$(	237.57	;	-26.14	)$&$	2.87	$& 254& 58464 \\
PKS~0558--504	&$(	257.96	;	-28.57	)$&$	4.18	$& 933& 420921 \\
PKS~2005--489	&$(	350.37	;	-32.60	)$&$	4.66	$& 253& 22609 \\
PKS~2155--304	&$(	17.73	;	-52.25	)$&$	1.63	$& 2000& 2.53$\times 10^{6}$ \\
Tons~180	&$(	138.99	;	-85.07	)$&$	1.54	$& 222& 42199 \\ 
\hline
\multicolumn{5}{l}{ $N({\rm HI})$ in units of $10^{20}$cm$^{-2}$ } 
\end{tabular} 
\end{table}

\section{Results from spectral fits}\label{sec_dis}
  
\begin{figure*}
          \centering
\includegraphics[scale=0.45]{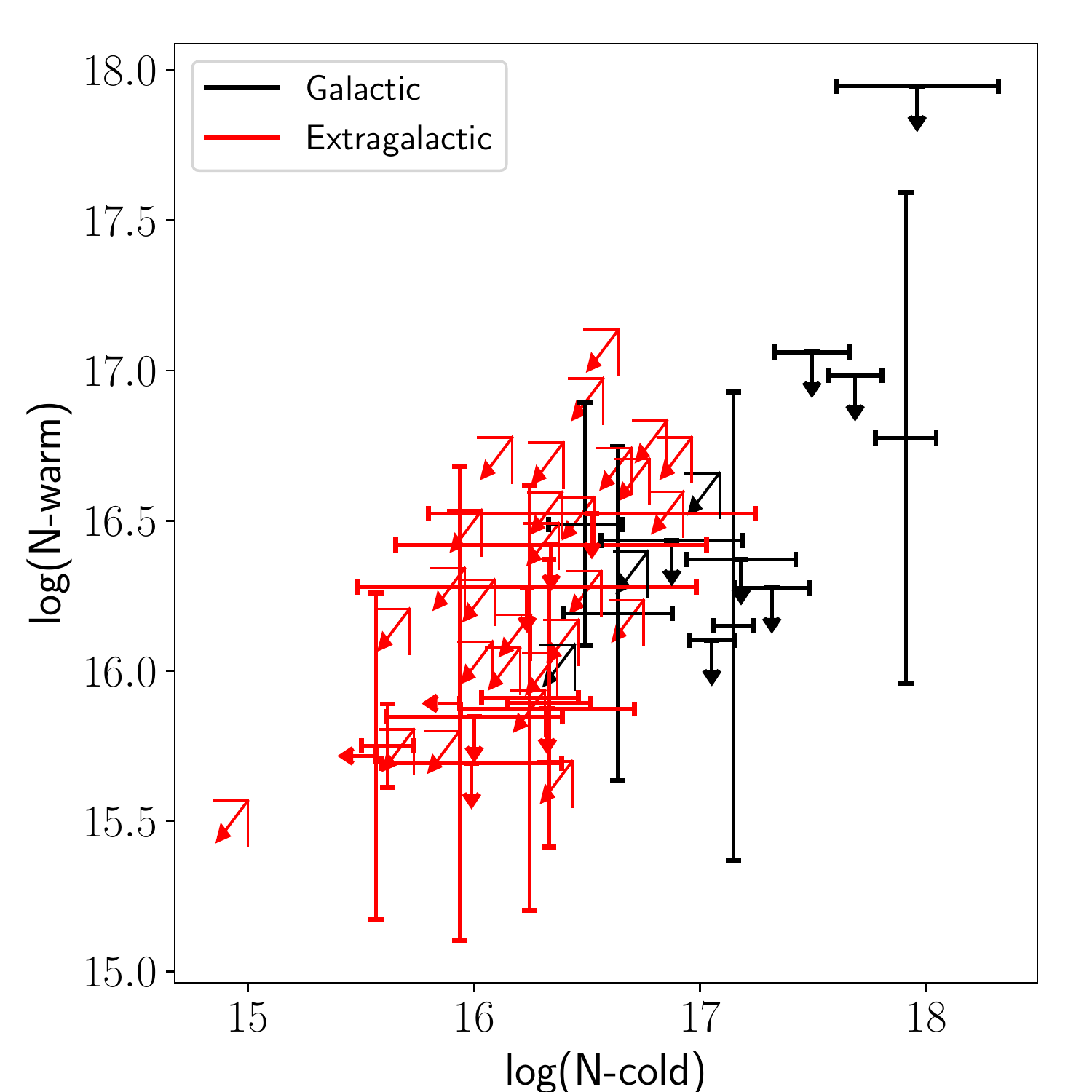}
\includegraphics[scale=0.45]{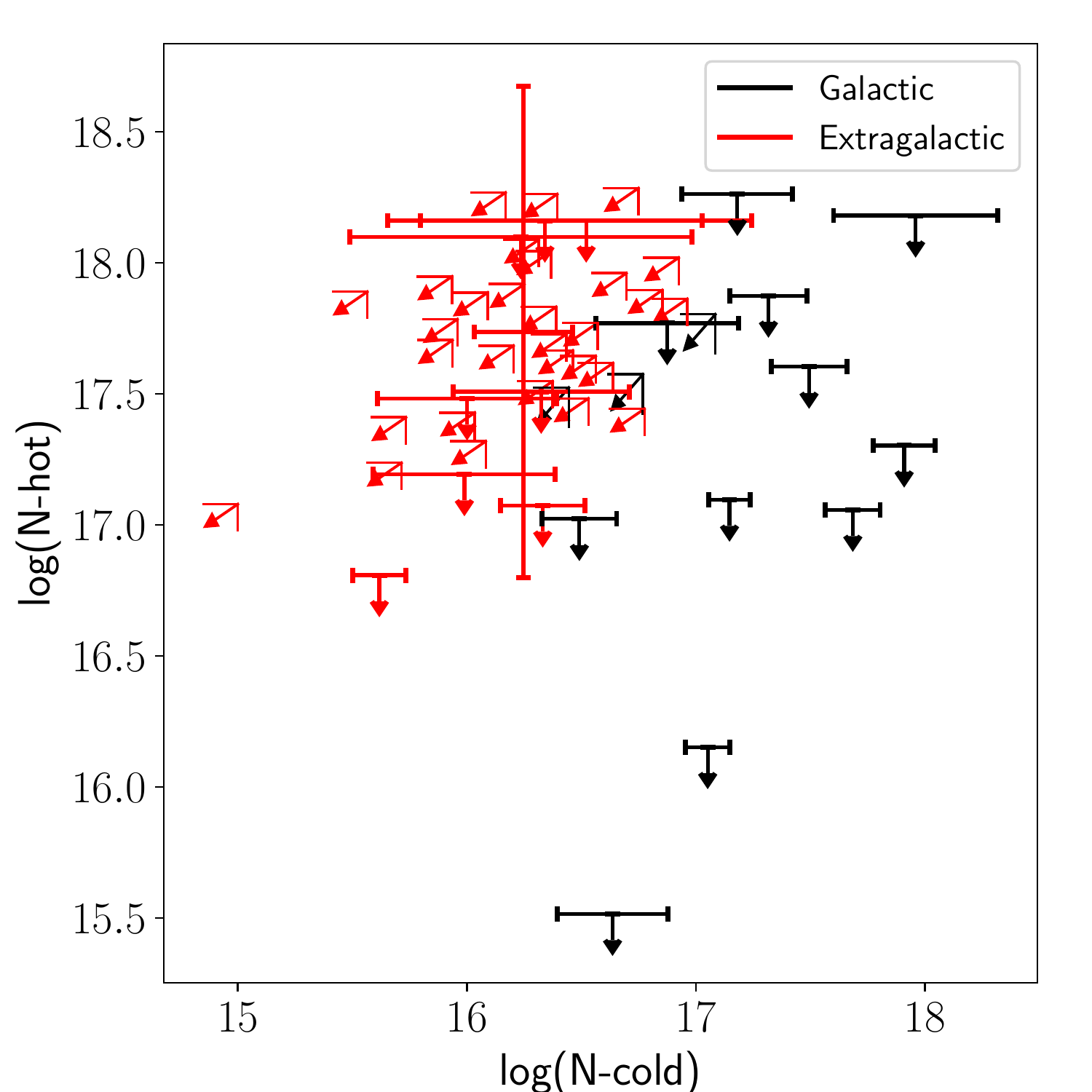} 
      \caption{Best fit column densities for the cold ({\rm N}~{\sc i}), warm ({\rm N}~{\sc ii}+{\rm N}~{\sc iii}) and hot ({\rm N}~{\sc v}+{\rm N}~{\sc vi}+{\rm N}~{\sc vii}) ISM phases. Black data points correspond to LMXB while red points correspond to extragalactic sources.}\label{fig_data_density_fractions}
   \end{figure*}

\begin{figure}
          \centering
\includegraphics[scale=0.43]{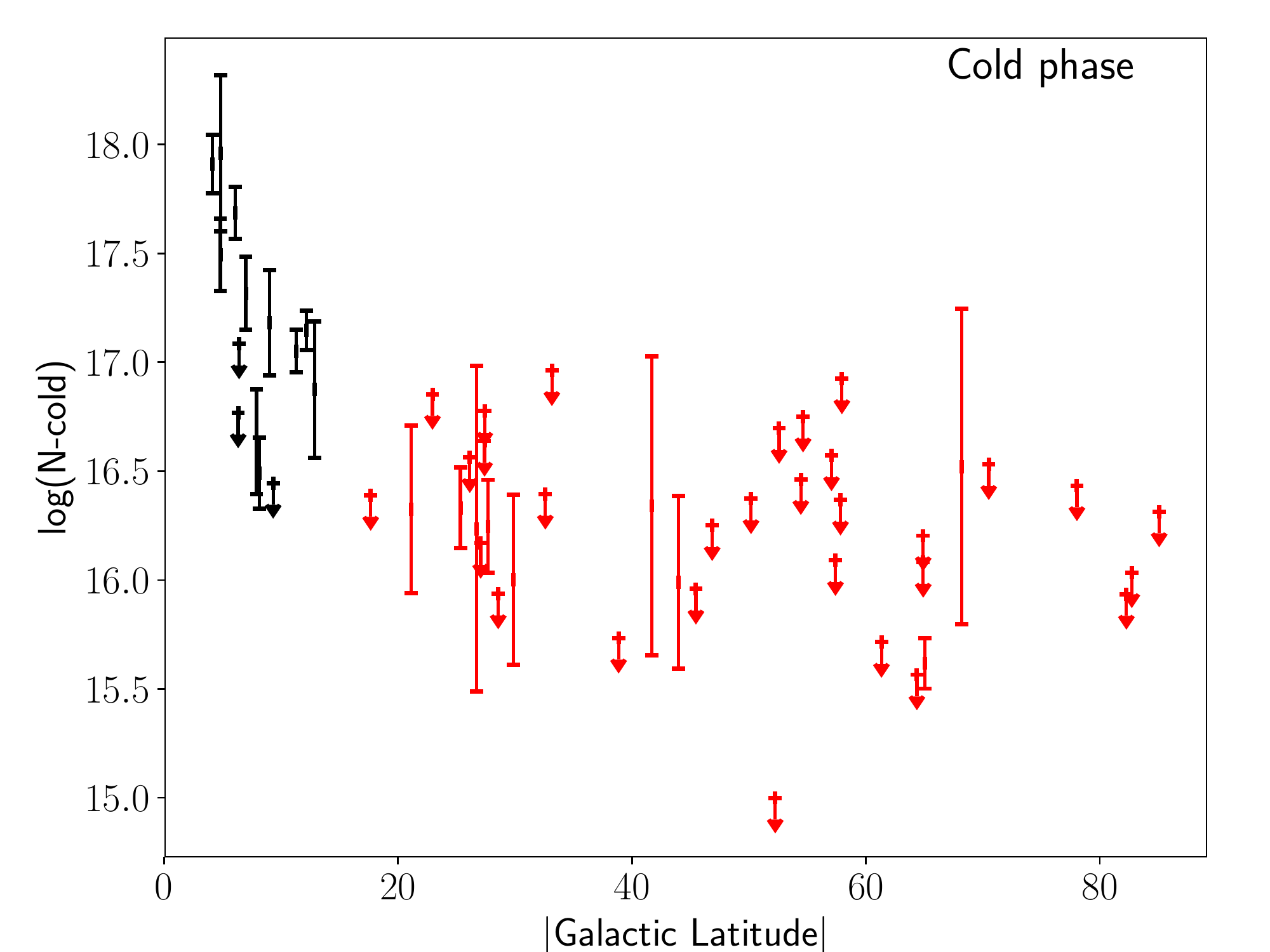}
\includegraphics[scale=0.43]{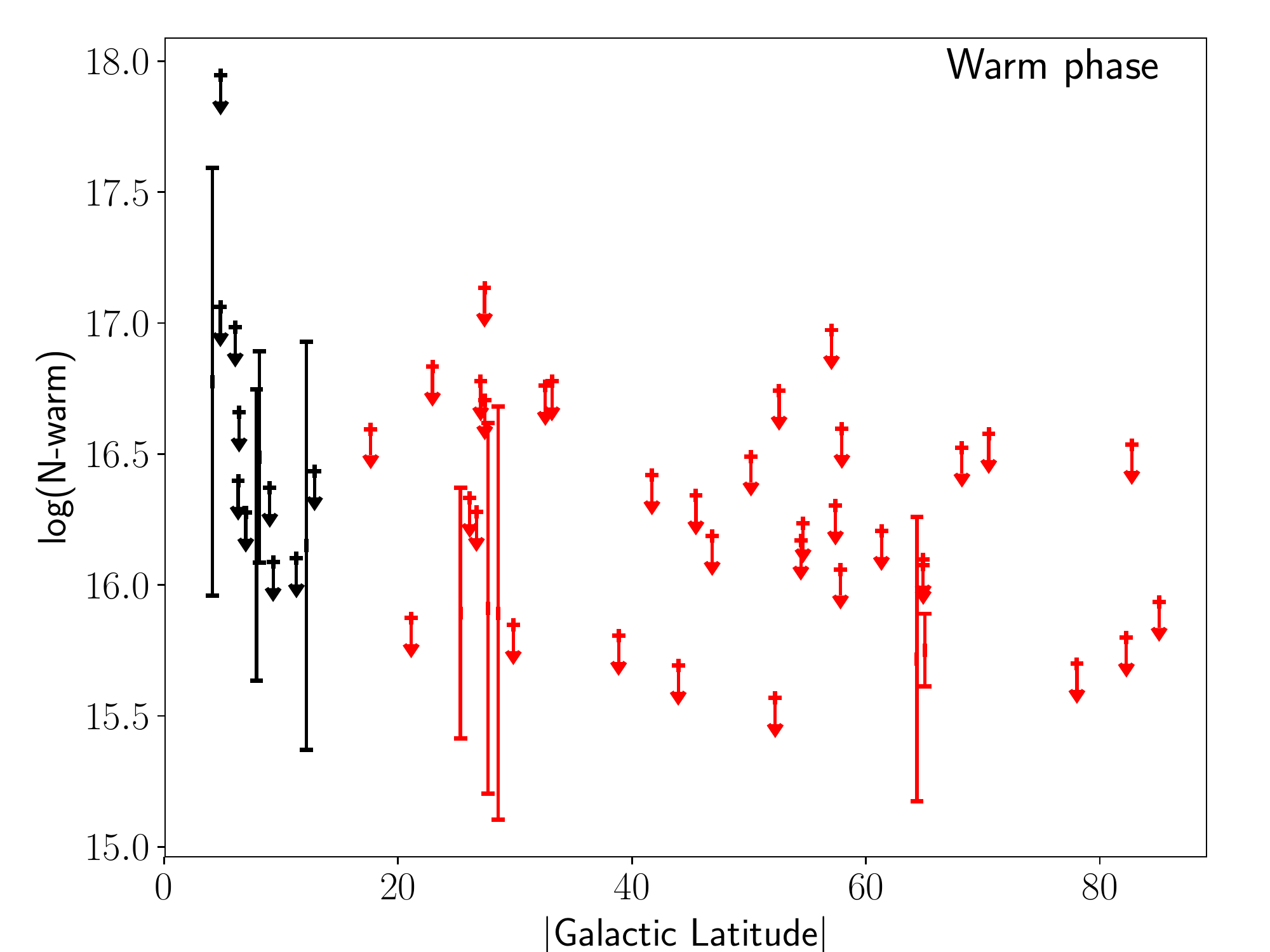}
\includegraphics[scale=0.43]{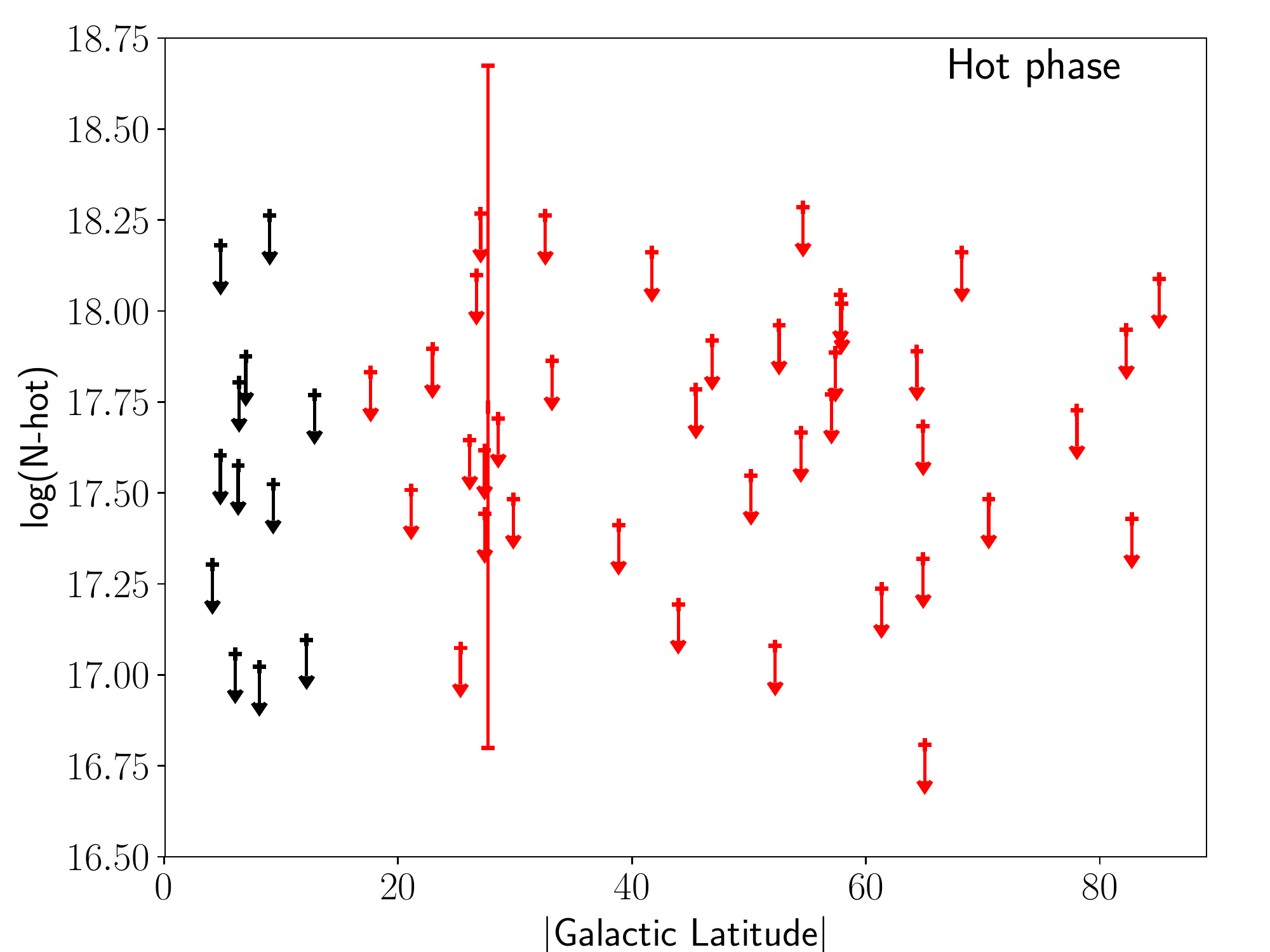}
      \caption{Best fit column densities for the cold ({\rm N}~{\sc i}), warm ({\rm N}~{\sc ii}+{\rm N}~{\sc iii}) and hot ({\rm N}~{\sc v}+{\rm N}~{\sc vi}+{\rm N}~{\sc vii}) ISM phases as function of the Galactic latitue. Black data points correspond to LMXB while red points correspond to extragalactic sources.}\label{fig_columns_latittude}
   \end{figure}

 The best-fit results are listed in Table~\ref{tab_ismabs}. We have found acceptable fits, from the statistical point of view, although for most of the sources we have obtained upper limits for the relevant parameters. We labeled the multiple phases of the gaseous ISM as cold ({\rm N}~{\sc i}), warm ({\rm N}~{\sc ii}+{\rm N}~{\sc iii}) and hot ({\rm N}~{\sc v}+{\rm N}~{\sc vi}+{\rm N}~{\sc vii}). Figure~\ref{fig_data_density_fractions} shows the best-fit column densities obtained. We note that for the cold component the column densities are systematically larger for the Galactic sources (black points) than for the extragalactic sources (red points). For the warm-hot components, in the other hand, the column densities tend to be similar between both types of sources. Figure~\ref{fig_columns_latittude} shows the distribution of the cold, warm and hot column densities as a function of the Galactic latitude. The plot shows that the cold-warm column densities decreases with the Galactic latitude, while the hot component does not appear to show any correlation, although for this case most of the results correspond to upper limits. This implies that the cold component is mostly concentrated in the Galactic disc, while the hot component is more homogeneously distributed. Previous studies suggest that a single disc model cannot fit the hot component of the ISM but requires a spherically symmetric profile, to account for the Galactic halo contribution \citep{mil13,mil15,nic16c,gat18a}.
 
 The presence of absorption features due to material intrinsic to the source (e.g. warm absorbers in AGNs) may lead to misidentification of certain absorption lines. To study such intrinsic absorber we used the {\tt warmabs} model, which is computed with the XSTAR photoionization code \citep{kal01}. This model considers the physical conditions for an ionizing source surrounded by a gas and takes into account physical processes such as photoionization, dielectronic and radiative recombination, excitation and electron impact collisional ionization. The model assumes ionization equilibrium conditions, a Maxwellian electron velocity distribution, and that the gas responsible for emission and absorption has an uniform temperature and ionization throughout. The parameters of the model includes the column density of the absorber ($N{\rm HI}$), the ionization parameter ($\log\xi$), elemental abundances (A$x$), broadening turbulence ($v_{turb}$), and redshift ($z$).

We have tested the effects in the {\tt ISMabs} column densities when including a {\tt warmabs} component in addition to the {\tt ISMabs} components, for Ark 564. In order to account for variations in the spectral energy distribution (SED), which affects the photoionization rate, we fitted the {\tt warmabs} $N({\rm HI})$ and $\log\xi$ parameters independently for each observation, while the {\tt ISMabs} column densities were linked between the different observations (i.e. to account for the ISM contribution).  We have found that the {\tt warmabs} best-fit prefers a highly ionized component, but the {\tt warmabs} column densities for such component are poorly constrained.  Moreover, while the cold and warm {\tt ISMabs} column densities obtained are not affected by the inclusion of {\tt warmabs}, the uncertainties in {\tt ISMabs} column densities for the hot component increase significantly. Similar results were obtained for other sources with large number of counts, including 1ES1028+511, 1ES1553+113, 1H0414+009 and 1H0707-495. It is important to note that the fits here presented are done within a small wavelength region (24-32 \AA). In order to perform a more detailed analysis of the absorber associated to the sources, a complete analysis of the RGS data is desired to identified absorption features in the spectra due to different ions, apart from nitrogen. However, such analysis is beyond the scope of this work. Finally, it is important to note that when fitting independently the intrinsic absorber for different observations, the computation time to perform error calculation increases exponentially (e.g. there are 37 observations for Mrk~421).

  \begin{table*}
\caption{\label{tab_ismabs}Best-fit nitrogen column densities obtained. }
\centering
\begin{tabular}{lccccccccc}
\hline
Source  & N\,{\sc i} &  N\,{\sc ii}  &  N\,{\sc iii} &  N\,{\sc v} &  N\,{\sc vi}  &  N\,{\sc vii}    &{\tt cash}/d.o.f.\\
 \hline
\hline 
\multicolumn{8}{c}{ Galactic sources }\\
\hline
4U~1254--690	&$	<12.2$&$<3.8$&$	<0.7$&$<0.4$&$<40.9$&$<22.3$&$	4355/3974$\\
4U~1543--62	&$<5.9 $&$<1.7 $&$<0.0$&$<1.7 $&$<16.0 $&$<19.9 $&$	859/790	$\\
4U~1636--53	&$	31.2^{+	12.7}_{-10.8}$&$3.9^{+5.6}_{-3.0}$&$<2.1 $&$<0.2 $&$<29.5 $&$< 10.5 $&$ 9010/7969	$\\
4U~1735--44	&$	20.7^{+7.7}_{-7.6}$&$<1.3$&$<0.6$&$<0.4 $&$<56.7 $&$<17.9 $&$	1684/1587	$\\ 
4U~1957+11	&$	<2.8$&$<1.1$&$<0.2 $&$<1.3 $&$<31.5 $&$<0.5 $&$1693/1587	$\\
4U~1820-30  &$ 4.3\pm 2.2$&$ 0.87_{-0.46}^{+0.56} $&$ 0.67_{-0.36}^{+0.48} $&$<0.2 $&$<0.1 $&$ <0.02  $&$2367/1587	$\\   
Aql~X--1	&$	81.2^{+22.0}_{-25.5}$&$3.9^{+5.6}_{-2.9}$&$2.1\pm 1.3$&$<2.1$&$<17.4$&$<0.7$&$2694/2383	$\\ 
Cygnus~X--2 &$11.3\pm 2.4 $&$0.58_{-0.34}^{+0.40} $&$ <0.3 $&$ <0.8 $&$ <0.5 $&$ <0.1 $&$3971/2382	$\\  
GS1826--238	&$	48.4^{+11.8}_{-13.7}$&$4.1^{+3.9}_{-2.8}$&$<1.6$&$<1.1$&$<10$&$<0.3$&$4577/3986	$\\  
GX9+9	&$	15.2^{+7.8}_{-7.6}$&$<1.8$&$<0.6$&$<1.5$&$<14.2$&$121.0^{+46.3}_{-46.1}$&$1711/1588	$\\
HETEJ1900.1--2455	&$7.5^{+5.0}_{-4.5}$&$<0.8$&$0.8^{+1.1}_{-0.7}$&$<1.6$&$<29.6$&$<27.7$&$	849/790	$\\
SAXJ1808.4--3658	&$3.1^{+1.2}_{-1.1}$&$2.8^{+0.6}_{-0.5}$&$0.26^{+0.19}_{-0.16}$&$<0.1$&$<3.9$&$<6.4$&$	2334/1589	$\\ 
Serpens~X--1	&$9.1_{-5.6}^{+8.9} $&$<4.7 $&$<4.1 $&$<2.4 $&$<1.2 $&$<0.9 $&$	2555/2384	$\\   
Swift~J1753.5--0127	&$	14.0^{+3.7}_{-2.4}$&$1.1\pm 0.6$&$	0.28^{+0.28}_{-0.23}$&$<0.2$&$<2.1$&$<10.3$&$8031/5556	$\\ 
\hline
\multicolumn{8}{c}{ Extragalactic sources }\\
\hline
1ES1028+511	&$<2.9$&$<0.9$&$<0.5$&$<1.2$&$<39.0$&$<6.1$&$2373/2381$\\
1ES1553+113	&$0.9^{+0.9}_{-0.7}$&$<0.3$&$<0.2$&$0.8^{+0.6}_{-0.6}$&$<4.2$&$<9.9$&$18237/17547	$\\
1H0414+009	&$<9.2$&$<1.9$&$<4.0$&$<4.5$&$<45.3$&$<23.6$&$1681/1586	$\\
1H0707-495	&$<2.5$&$2.0^{+1.4}_{-1.1}$&$<0.5$&$<3.1$&$<26.3$&$<38.5$&$	8939/8763	$\\
1H1219+301	&$<1.1$&$<0.8$&$<2.6$&$<0.2$&$<7.1$&$<19.5$&$790/789	$\\ 
1H1426+428	&$<1.2$&$<0.5$&$0.32^{+0.43}_{-0.30}$&$<0.4$&$<17.2$&$<3.2$&$4727/4770	$\\
3C120	&$	<4.3$&$5.8^{+6.3}_{-4.2}$&$<1.6$&$<0.1$&$<39.4$&$<1.9$&$828/788	$\\  
3C273	&$<0.4$&$0.21\pm 0.14$&$0.31\pm 0.12$&$<0.4$&$30.6^{+8.2}_{-7.9}$&$22.9^{+15.5}_{-13.2}$&$33390/29457	$\\
3C279	&$<3.7$&$<3.7$&$<5.7$&$<1.1$&$<57.8$&$<0.1$&$870/791	$\\  
3C390.3	&$<1.5$&$<2.3$&$1.4^{+2.3}_{-1.3}$&$<4.1$&$76.9^{+55.8}_{-53.9}$&$<48.8$&$1555/1585	$\\
Ark120	&$2.1^{+2.1}_{-1.4}$&$<0.5$&$<0.3$&$<0.8$&$<17.5$&$<13.9$&$	5036/4779	$\\
Ark564	&$2.14^{+0.98}_{-0.81}$&$0.51^{+0.24}_{-0.22}$&$0.26^{+	0.16}_{-0.14}$&$0.9\pm 0.4$&$<1.1$&$<9.4$&$12412/10357	$\\
ESO141-G055	&$1.72^{+2.63}_{-1.58}$&$<0.7$&$<1.2$&$<0.7$&$38.7^{+30.2}_{-27.6}$&$<55.7$&$3467/3183	$\\
ESO198-G24	&$<8.4$&$<1.7$&$<2.2$&$<5.5$&$<55.4$&$<43.8$&$1575/1584	$\\ 
Fairall9	&$<2.3$&$<0.4$&$<0.7$&$<0.8$&$38.2^{+38.6}_{-34.8}$&$<33.2$&$2522/2385	$\\
H1426+428	&$<1.6$&$<0.5$&$<0.7$&$<0.4$&$18.5^{+18.8}_{-15.4}$&$	<10.7$&$5078/4771	$\\
H1821+643	&$<5.9$&$<2.1$&$<3.0$&$<3.3$&$<16.3$&$<8.2$&$7958/7174	$\\
H2356-309	&$<2.7$&$<0.3$&$<0.2$&$<1.3$&$<22.9$&$<29.2$&$6799/6373	$\\
HE1143-1810	&$2.2^{+3.4}_{-1.9}$&$1.1^{+1.2}_{-0.9}$&$<0.3$&$<1.5$&$47.5^{+35.1}_{-33.0}$&$<60.8$&$4983/4779	$\\ 
IZw1	&$<2.4$&$<0.9 $&$	0.88^{+1.33}_{-0.76}$&$<1.5$&$<3.0$&$<30.8 $&$3595/3181	$\\
MCG-6-30-15	&$1.8^{+0.9}_{-0.8}$&$0.45^{+0.36}_{-0.31}$&$0.36^{+0.27}_{-0.22}$&$2.3^{+0.9}_{-1.2}$&$1.5^{+2.9}_{-1.2}$&$50.6^{+19.4}_{-15.6}$&$	9528/7165	$\\
MR2251-178	&$<0.5$&$<0.5 $&$0.53^{+0.57}_{-0.38}$&$<0.2 $&$<3.4 $&$ 13.7 $&$ 7112/6375	$\\
Mrk279	&$<1.8 $&$<0.8$&$<0.7 $&$<1.7 $&$<30.3 $&$<51.0 $&$	3290/3178	$\\
Mrk421	&$0.41\pm 0.11$&$0.36\pm 0.05$&$0.20\pm 0.03$&$<0.02$&$2.2^{+1.6}_{-1.5}$&$1.4^{+1.2}_{-0.8}$&$60741/29471$\\ 
Mrk501	&$	<0.5 $&$<0.3$&$<0.4 $&$<1.3 $&$<24.2 $&$<0.2 $&$5044/4777	$\\
Mrk509	&$1.01^{+0.97}_{-0.67}$&$0.34^{+0.23}_{-0.21}$&$<0.13 $&$ 0.40 ^{+0.36}_{-0.24}$&$14.1^{+8.5}_{-7.5}$&$<7.0$&$12739/10361	$\\
Mrk766	&$	<0.9 $&$<0.4 $&$<0.3 $&$<2.3$&$ 3.11^{+7.87}_{-3.05}$&$38.8^{+	36.6}_{-35.6}$&$	7194/6359	$\\
Mrk841	&$<5.6$&$<1.4 $&$<0.4$&$<1.6 $&$ 83.7^{+48.0}_{-47.3}$&$<59.5 $&$ 5003/4773	$\\ 
NGC3783	&$	<7.1  $&$<0.5 $&$ 3.5^{+2.8}_{-1.9}$&$<2.5$&$	<37.3 $&$<38.9$&$	3049/2382	$\\
NGC4593	&$	<1.2 $&$<0.6 $&$ 0.74^{+0.70}_{-0.48}$&$<0.7 $&$ 34.9^{+24.1}_{-23.2}$&$<17.2 $&$ 5333/4778	$\\
NGC5548	&$	<3.4$&$ 1.8^{+1.6}_{-1.1}$&$<0.3$&$<0.8 $&$<29.2 $&$ <0.5$&$	2806/2381	$\\
NGC7213	&$	<4.9 $&$<1.5 $&$ 1.42	^{+	2.60}_{-1.27}$&$<2.1 $&$<72.9 $&$<16.5 $&$ 1762/1587	$\\
NGC7469	&$<0.9 $&$ 1.3^{+	0.6}_{-0.5}$&$<0.3 $&$<0.2 $&$ 31.2^{+	15.1}_{-14.3}$&$<14.4 $&$9105/7967	$\\
PG1116+215	&$3.3^{+5.3}_{-3.0}$&$<1.8 $&$<1.6 $&$<0.3 $&$<48.5 $&$ <96.2$&$	3411/3182	$\\
PKS0548-32	&$<3.7 $&$<1.4$&$<0.8 $&$<1.2$&$<29.9 $&$<13.0 $&$2536/2381	$\\
PKS0558-504	&$<0.9 $&$	0.39^{+0.35}_{-0.30}$&$	0.39^{+	0.30}_{	-0.24}$&$<0.3$&$<10.1 $&$ 17.7 ^{+22.6}_{-15.1}$&$13007/11945	$\\
PKS2005-489	&$<2.5 $&$ 2.17^{+2.50}_{-1.56}$&$<1.1 $&$<0.8 $&$<49.7 $&$ 53.3^{+	79.2}_{-49.3}$&$2608/2382	$\\
PKS2155-304	&$<0.1$&$	<0.1$&$	<0.3$&$	<0.1$&$	<11.8$&$	<0.1$&$	39762	/	27898	$\\ 
Tons180	&$	<2.1 $&$<0.5$&$<0.4$&$<1.1$&$<39.2	$&$<82.3$&$3308/3181	$\\
 \hline
\multicolumn{8}{l}{ Column densities in units of $10^{16}$cm$^{-2}$.}
 \end{tabular}
\end{table*}

 \subsection{Comparison with previous works}\label{sec}
We have collected {\rm N}~{\sc i} column density values from previous studies. \citet[][MEY+97]{mey97} analyzed the interstellar {\rm N}~{\sc i} $\lambda\lambda$ 1160,1161 absorption doublet toward 8 stars using the Hubble Space Telescope Goddard High Resolution Spectrograph (GHRS). \citet[][MOO+02]{moo02} determined {\rm N}~{\sc i} column densities along 7 lines-of-sight using {\rm N}~{\sc i} $\lambda$ 1160 observations obtained with the Far Ultraviolet Spectroscopic Explorer (FUSE). \citet[][KNA+06]{kna06}, reported on the analysis of 13 stars using FUSE observations.   Finally, \citet[][GUD+12]{gud12}, created a database of multiple interstellar column densities using absorption line data toward 3008 stars, 164 of them with {\rm N}~{\sc i} values.

\begin{figure}
          \centering
\includegraphics[scale=0.48]{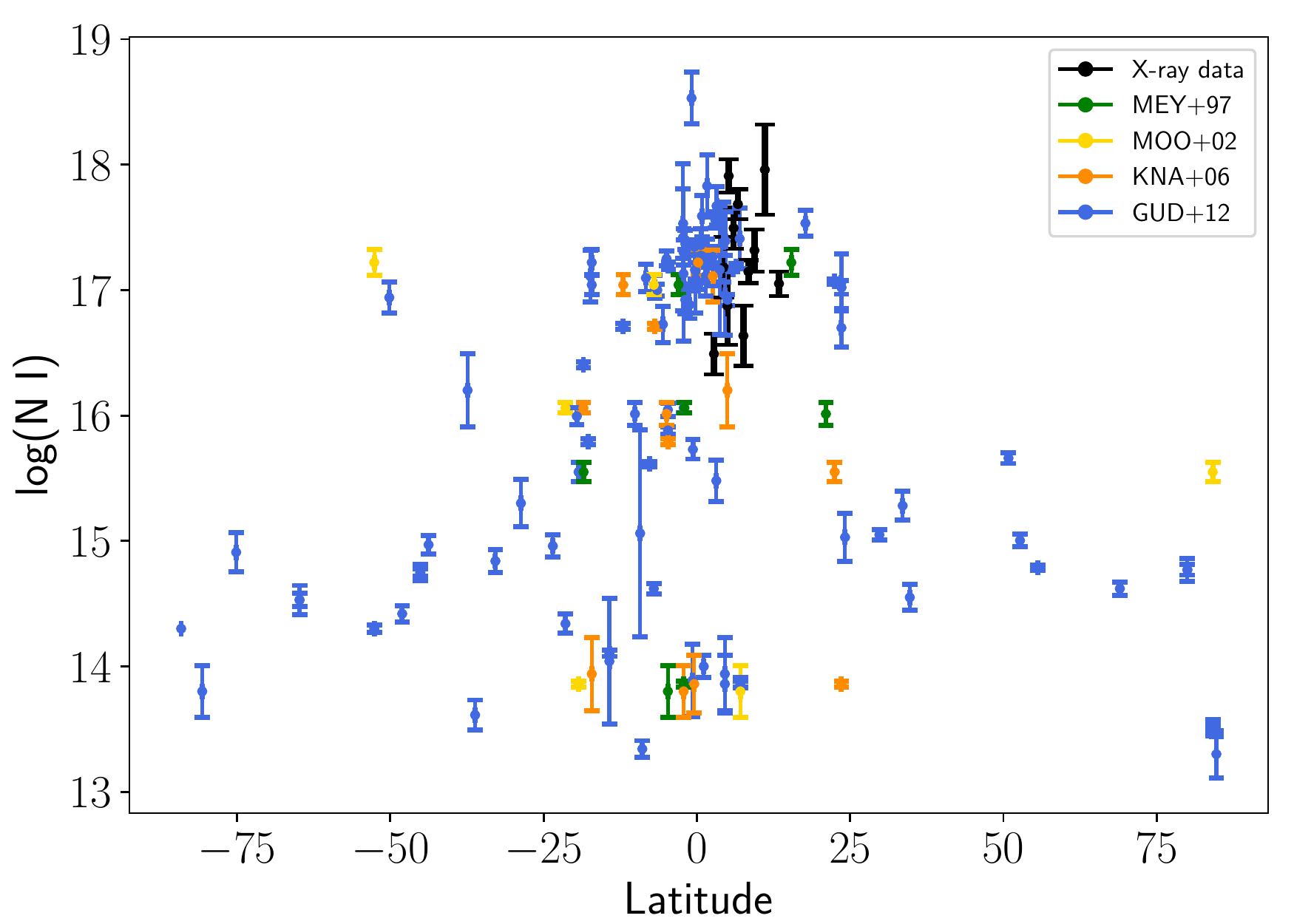} \\
\includegraphics[scale=0.48]{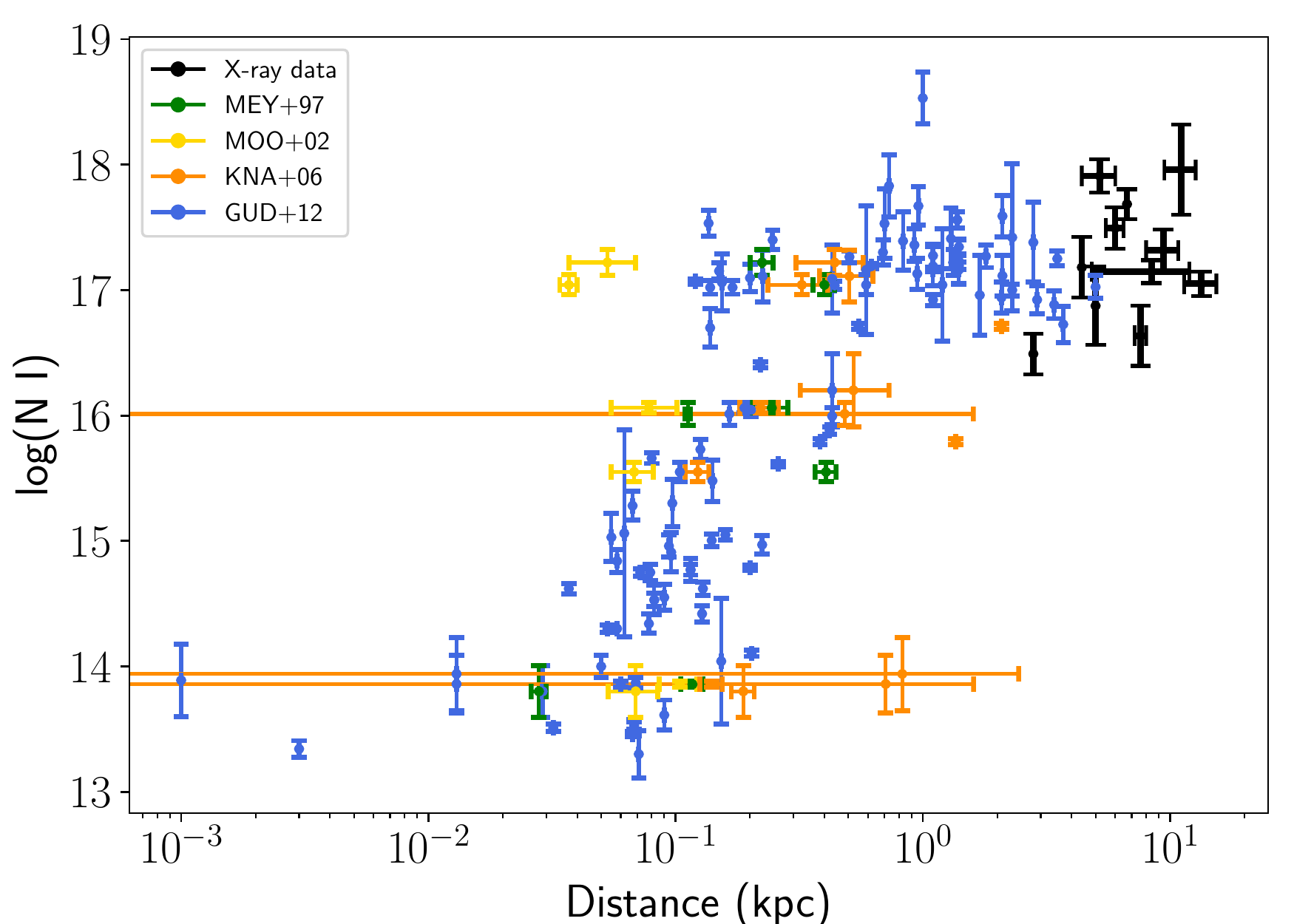}
      \caption{\emph{Top panel:} {\rm N}~{\sc i} column densities distribution as function of Galactic latitude for different Galactic samples. \emph{Bottom panel:} {\rm N}~{\sc i} column densities distribution as function of the distance. }\label{fig_columns_latittude_comparison}
   \end{figure}

Figure~\ref{fig_columns_latittude_comparison} shows the distribution of {\rm N}~{\sc i} values as function of Galactic latitude (top panel) and distance (bottom panel) including the collected sample. In the case of the distances, we have not included the LMXB 4U1957+11 due to its large uncertainty. For the cold ISM component, it is expected to show a strong correlation between the column densities and the Galactic latitude \citep[see for example][]{gat18a}.  For the cold component of the ISM, it is commonly assumed in the literature to be exponentially decreasing perpendicular to the Galactic plane, with larger column densities near the Galactic center \citep[see e.g.][]{rob03,kalb09}.  For the X-ray data we have included only those Galactic sources for which {\rm N}~{\sc i} is well constrained (i.e. no upper limits. See Table~\ref{tab_ismabs}).  We found that the X-ray column densities derived from the spectral fitting are in good agreement with previous results, with a clear tendency to decrease as we move away from the Galactic plane ($|Latitude|>50$ $^{\circ}$) with variations in regions that form stars more actively than others (e.g., spiral arm regions such as Vela or Orion). However, the number of lines of sight does not allow to cover small scale structure (i.e. AU) but rather large-scale. Finally, it is important to note that the hot component is only accessible through UV/X-ray observations.

 \subsection{ISM structure}\label{sec_ism}  
Previous analysis of the ISM multiphase structure using X-ray absorption technique have shown that the gas physical state is dominated by the cold component, with mass fractions for the ISM phases in the Galactic disc of cold $\sim~ 90\%$, warm $\sim  8\%$ and hot $\sim  2\%$ components \citep[e.g.][]{yao06,pio07,pin13,gat18a}.  The uncertainties in the {\tt ISMabs} column densities obtained prevent us to compute accurate mass fractions for all sources. For example, we noted that our best-fit results show a very large contribution of the {\rm N}~{\sc vii} column density associated to the hot component (e.g. GX9+9, 3C273, MCG-6-30-15, Mrk766, PKS2005-489), while previous work indicates that the hot phase represent less than 1$\%$ of the total ISM \citep[e.g.][]{pin13}. However, for these sources our analysis show that the rest of ionic species are not well constrained. Moreover, our {\tt ISMabs} model includes the column densities as free parameters in the model and therefore we do not consider any ionization balance for the nitrogen ionic species. In this sense, a proper definition requires a spectral fitting with a more complete physical model that compute ionization balance depending on temperature. The contribution of the different ISM phases depends on the location of the absorber. While the density distribution for the cold-warm components is typically modeled with an exponential profile for the Galactic disc, the hot component includes the contribution of the Galactic halo \citep[e.g.][]{mil13,mil15,nic16c,gat18a}. However, the absence of sources in our sample near the Galactic center, a region heavily affected by the cold gas absorption, does not allow us to constrain the density profiles of the different ISM phases. More information about the physical properties of the gas (e.g. temperature and abundances) will be included in future works, with the inclusion of more complex models (e.g. {\tt warmabs}).

  \subsection{Future prospects}\label{sec_sim}
 
\begin{figure}
          \centering
\includegraphics[scale=0.45]{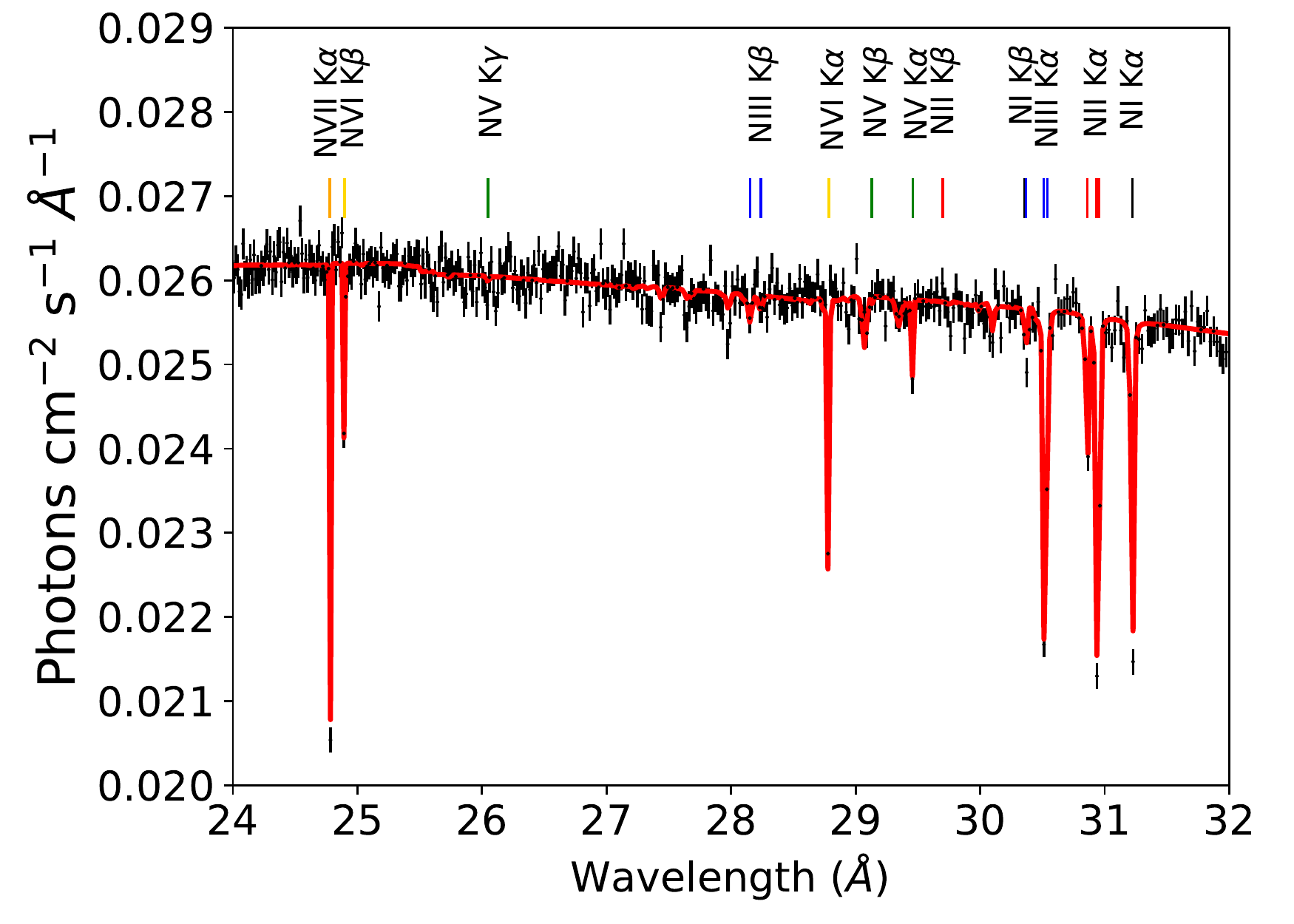} 
      \caption{{\it Athena} simulation of the N K-edge photoabsorption region for an extragalactic source (e.g. Mrk~421). The total exposure time is 10 ks and the flux is $\log F_{x}=-9.77$ erg cm$^{-2}$ s$^{-1}$ in the 24-32 \AA\ wavelength range.}\label{ath_sim}
   \end{figure}

 Future X-ray high-resolution spectra mission, such as {\it Arcus} \citep{smi16} and {\it Athena} \citep{nan13}, will benefit greatly from this atomic data benchmarking. For example, Figure~\ref{ath_sim} shows a 10 ks simulation for an extragalactic source (i.e. Mrk~421 in high-state) obtained with {\it Athena} for the baseline configuration\footnote{\url{http://x-ifu-resources.irap.omp.eu/PUBLIC/RESPONSES/CC_CONFIGURATION/}}. The position for the main absorption lines are indicated. The plot shows the oustanding capabilities of the instrument, with the main resonance absorption lines visible by-eye, compared with the RGS spectra analyzed in this paper. For such observation the line profiles can be studied with an accuracy at the percent level. More important, by measuring simultaneously multiple resonance lines (e.g. K$\alpha$ and K$\beta$) for the same ions, model-independent constraint on the broadening will be obtained.

\section{Conclusions}\label{sec_con}
We have carried out analysis of the ISM nitrogen K-edge absorption (24--32 \AA) using high-resolution {\it XMM-Newton} spectra. Our data sample consist of 12 LMXBs and 40 extragalactic sources. For each source we fitted all observations simultaneously using a {\tt powerlaw*constant} model for the continuum and a modified version of the {\tt ISMabs} model for the spectral absorption features. We have found acceptable fits, from the statistical point of view, for most of sources. We have measured column densities for {\rm N}~{\sc i}, {\rm N}~{\sc ii}, {\rm N}~{\sc iii}, {\rm N}~{\sc v}, {\rm N}~{\sc vi} and {\rm N}~{\sc vii} ionic species, which trace the cold, warm and hot phases of the ISM.  For the cold component we have found that the distribution of the column density distribution as function of Galactic latitude is in good agreement with UV measurements. For the hot component we have not found such correlation, most likely due to the contribution from the Galactic halo. We have tested the effects in the {\tt ISMabs} column densities when including a {\tt warmabs} component, to account for absorption intrinsic to the sources. For such test, we link the ISM column densities between different observations (to account for the ISM contribution) while the {\tt warmabs} $N({\rm HI})$ and $\log\xi$ parameters were fitted independently for each observation. We have found that both cold and warm {\tt ISMabs} column densities obtained are not affected by the inclusion of the {\tt warmabs} component. However, the uncertainties in the hot ISM phase increase significantly while the column densities for the {\tt warmabs} are not constrained. Such analysis point out the importance of modeling simultaneously multiple K-edge absorption regions (e.g. O, Ne, Mg) in order to study the contribution from both, the local ISM and intrinsic absorption. Future observations with new-generation instrumentation such as {\it Arcus} and {\it Athena} will allow a finer examination of the N K-edge structure.

\subsection*{Data availability}
Observations analyzed in this article are available in the {\it XMM-Newton} Science Archive (XSA) (\url{http://nxsa.esac.esa.int/nxsa-web/#search}). The {\tt ISMabs} model is included in the {\sc xspec} data analysis software (\url{https://heasarc.gsfc.nasa.gov/xanadu/xspec/})

 \subsection*{Acknowledgements}
We thank the anonymous referee for the careful reading of our manuscript and the valuable comments. J.A.G. acknowledges support from the Smithsonian Astrophysical Observatory grant AR0-21003X, and from the Alexander von Humboldt Foundation.

\bibliographystyle{mnras}

  \appendix
\label{sec_apx}

\section{Spectra of individual source and best-fitting models}\label{sec_apx_fit}
\begin{figure*}
          \centering
 
\includegraphics[scale=0.195]{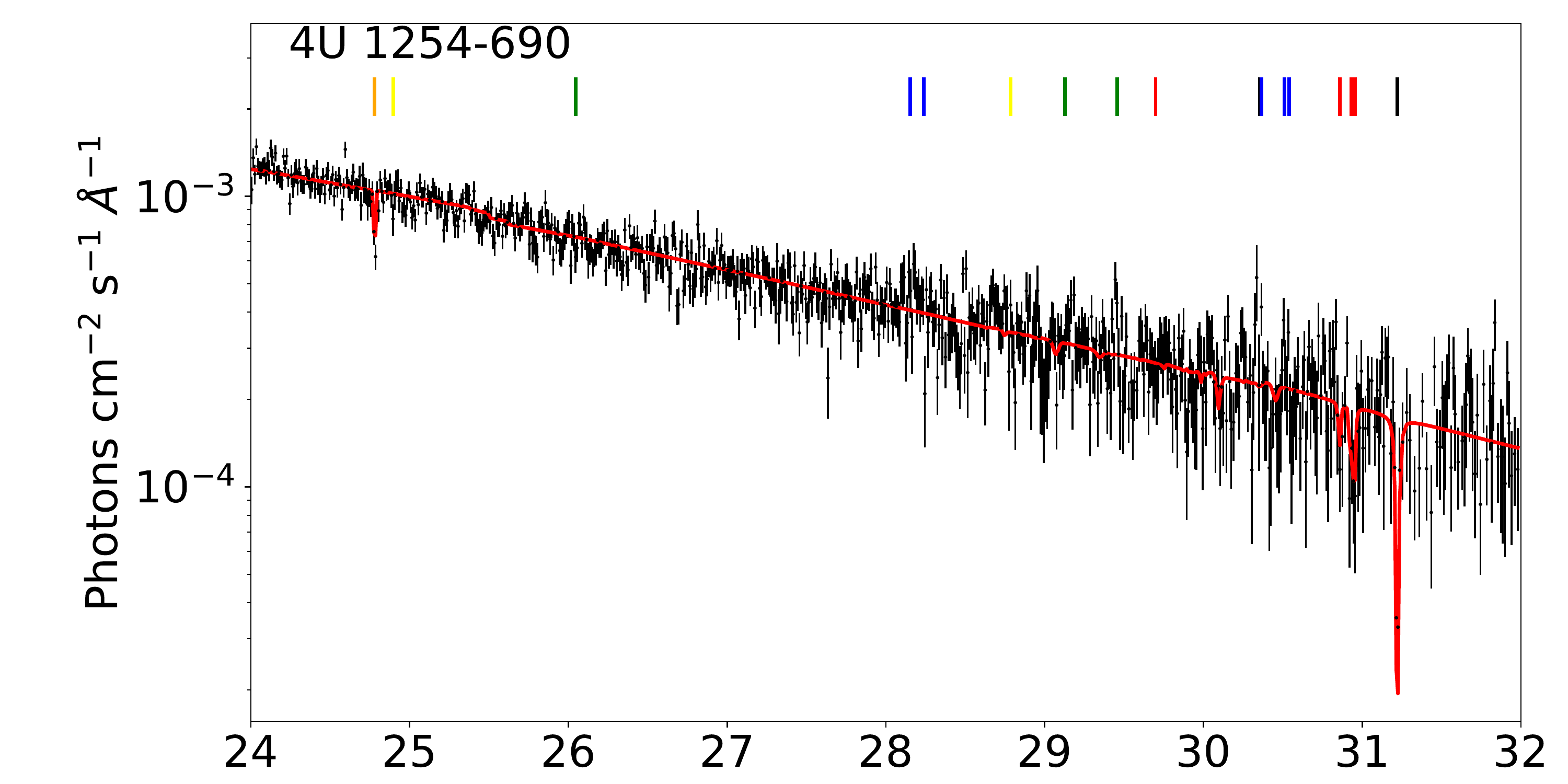}
\hspace{-2.5mm}
\includegraphics[scale=0.195]{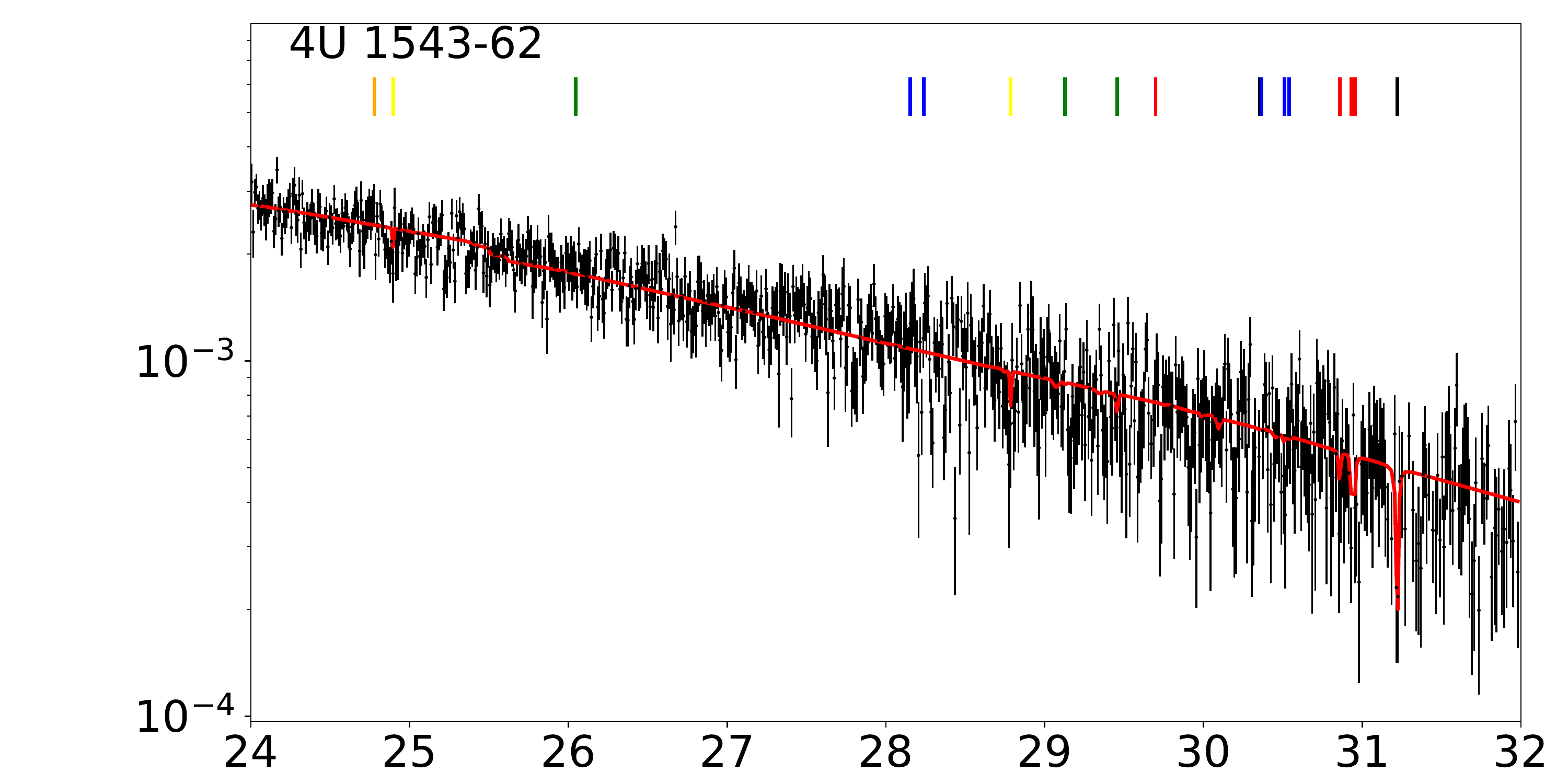} 
\hspace{-2.5mm}
\includegraphics[scale=0.195]{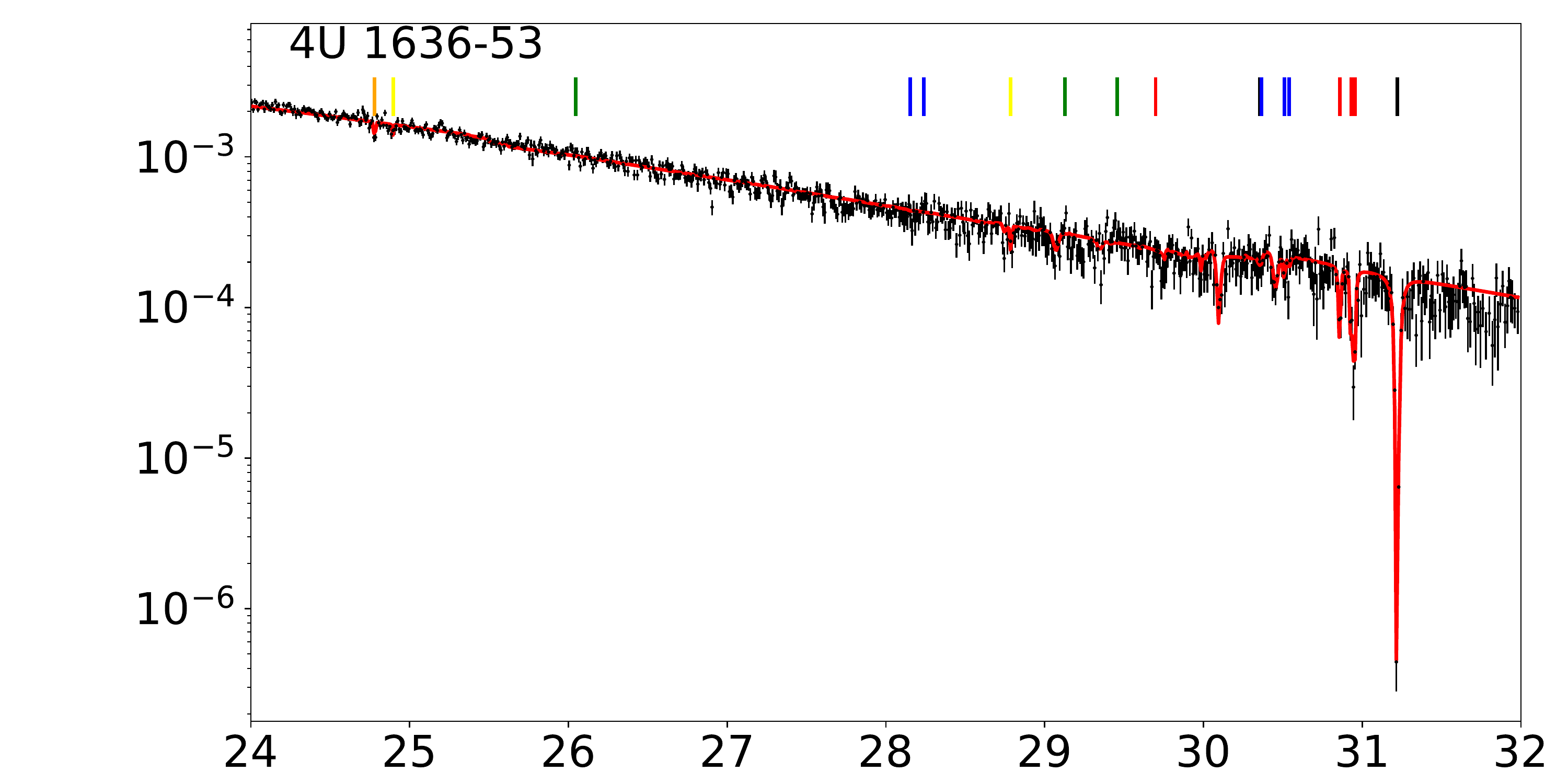}\\ 
\includegraphics[scale=0.195]{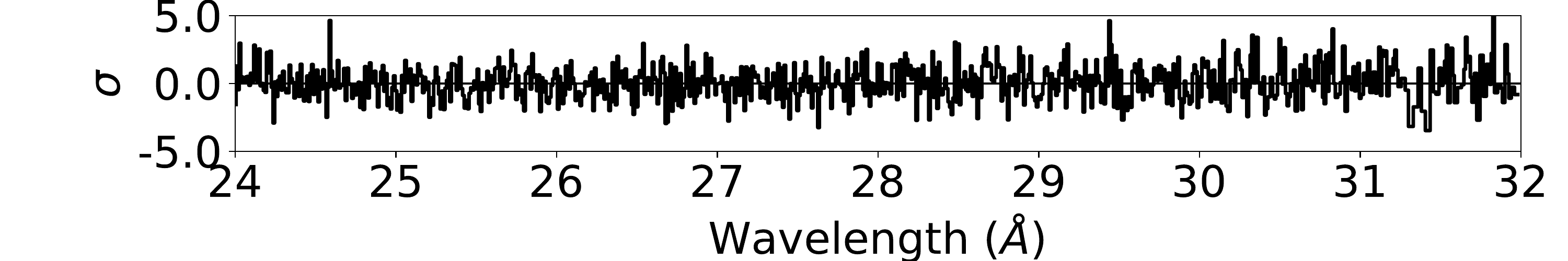} 
\hspace{-2.5mm}
\includegraphics[scale=0.195]{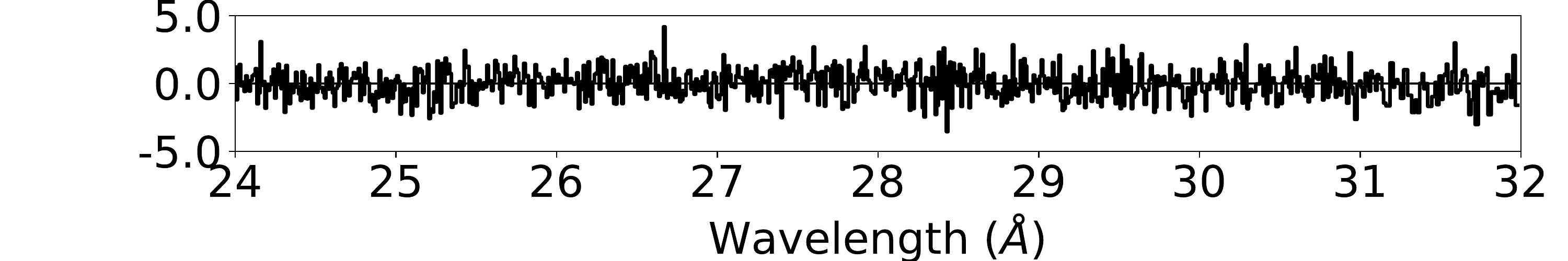} 
\hspace{-2.5mm}
\includegraphics[scale=0.195]{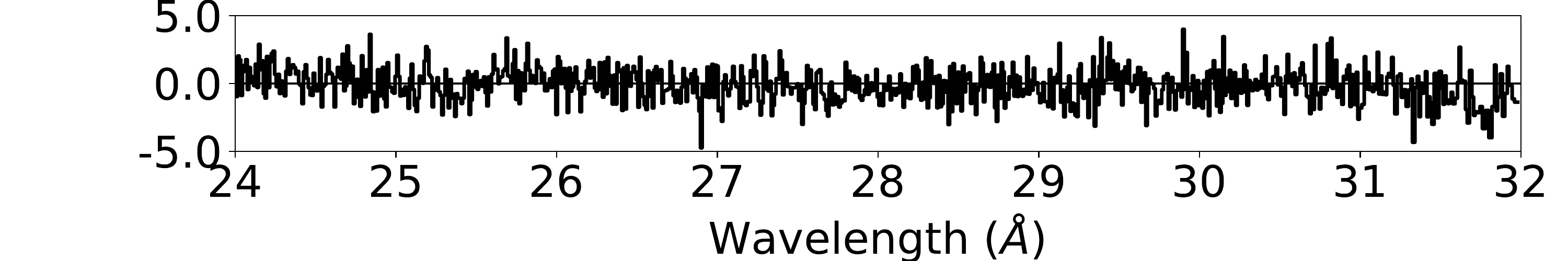}\\

\includegraphics[scale=0.195]{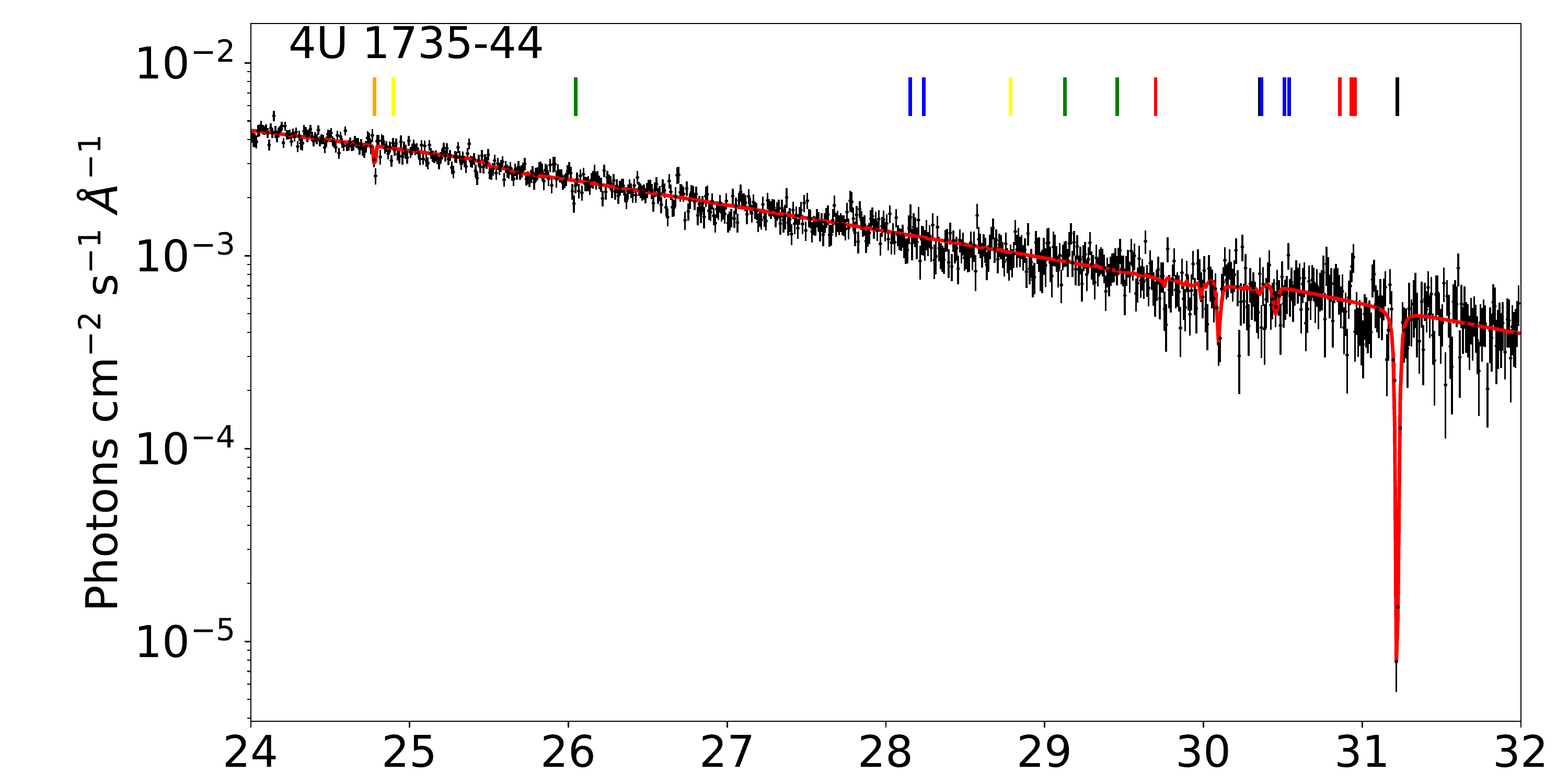}
\hspace{-2.5mm}
\includegraphics[scale=0.195]{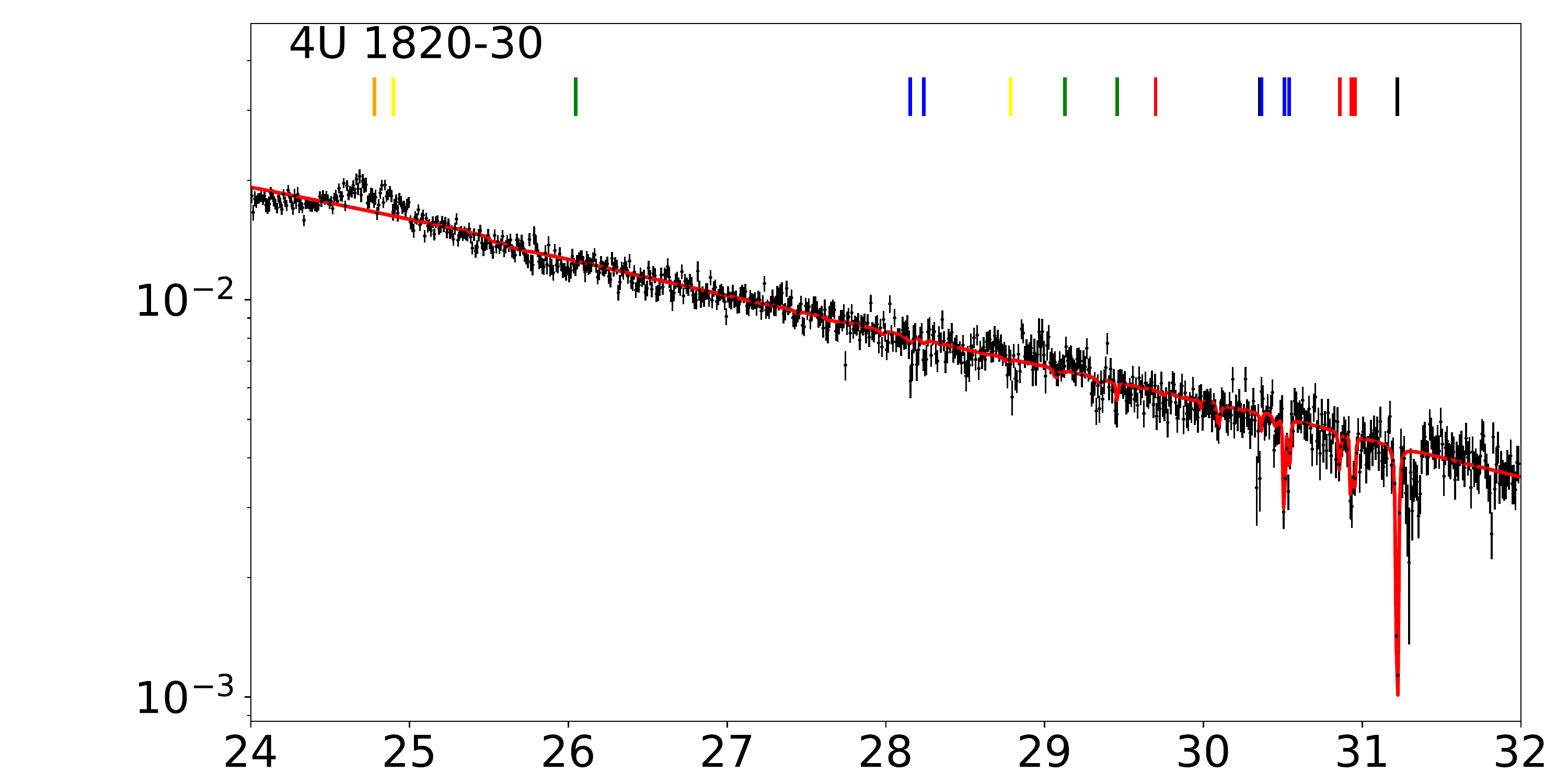}
\hspace{-2.5mm}
\includegraphics[scale=0.195]{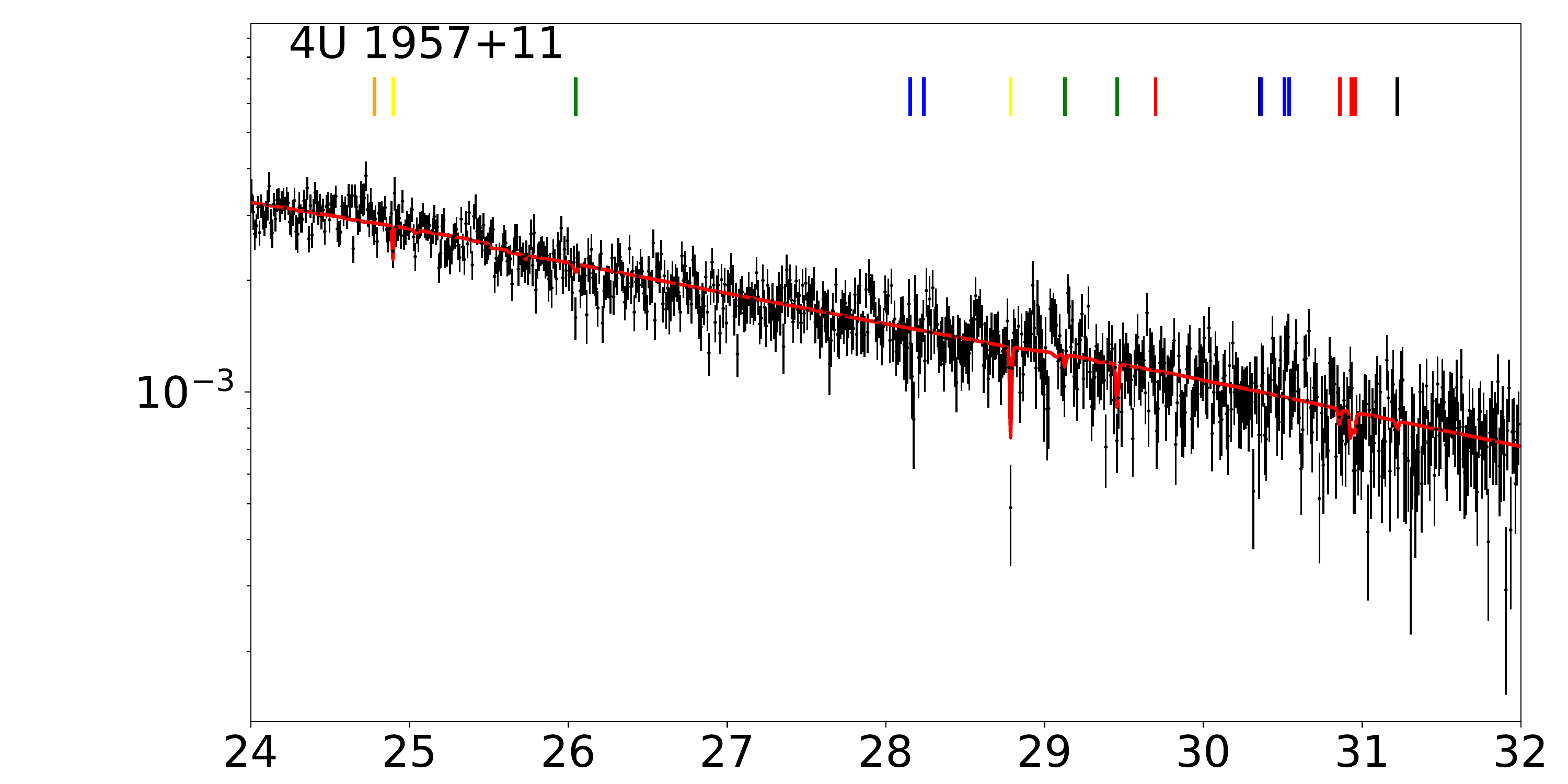}\\
\includegraphics[scale=0.195]{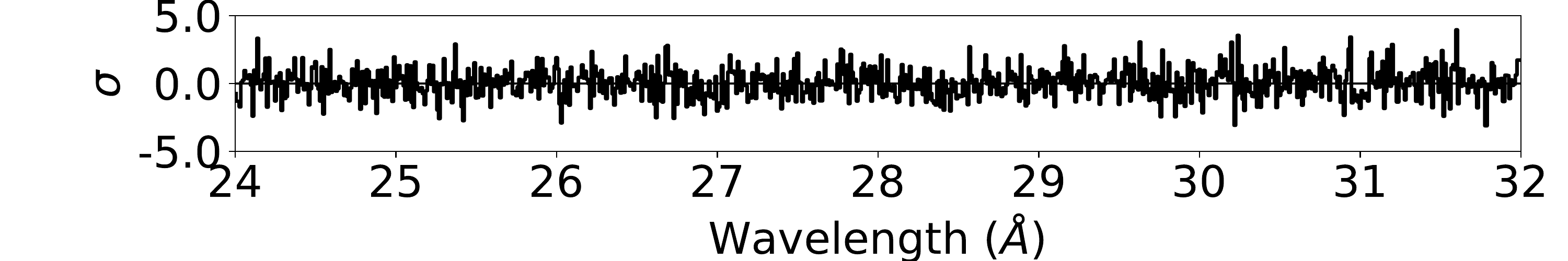}
\hspace{-2.5mm}
\includegraphics[scale=0.195]{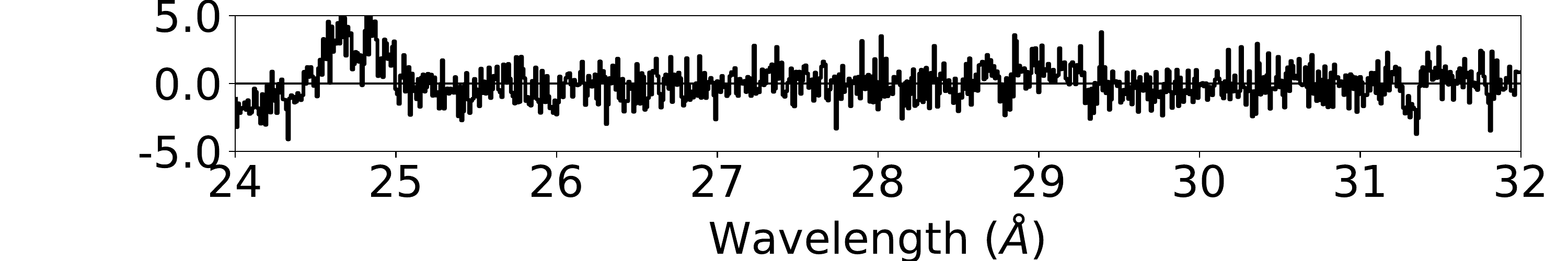}
\hspace{-2.5mm}
\includegraphics[scale=0.195]{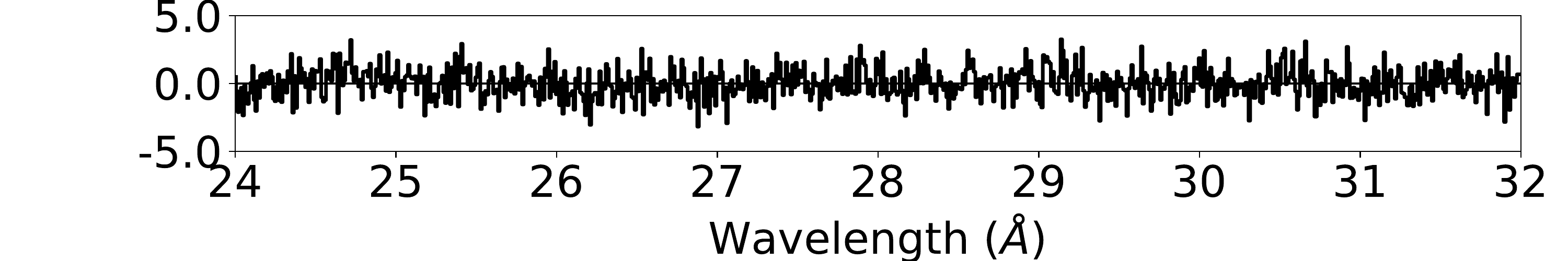}\\

\includegraphics[scale=0.195]{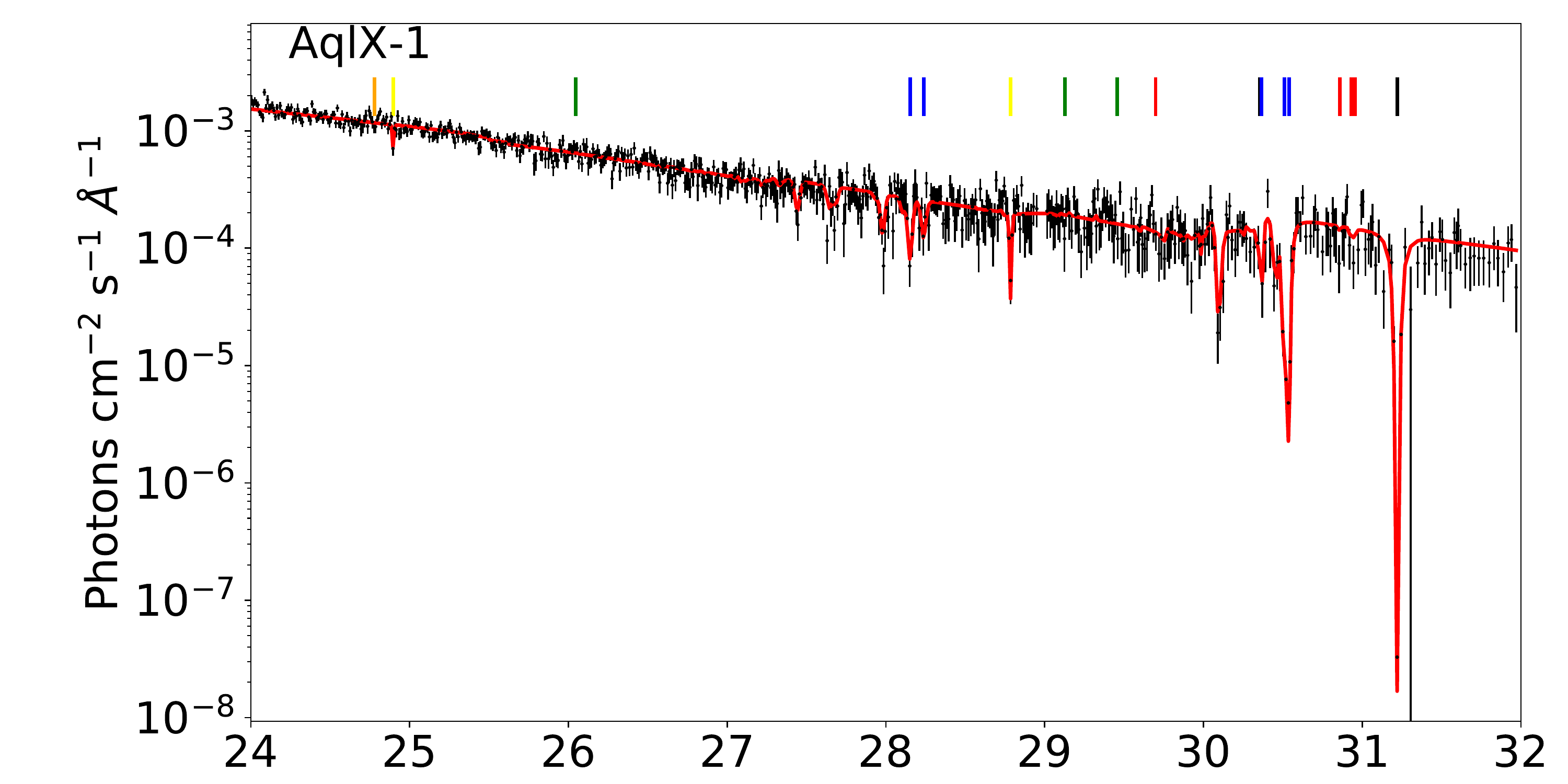}
\hspace{-2.5mm}
\includegraphics[scale=0.195]{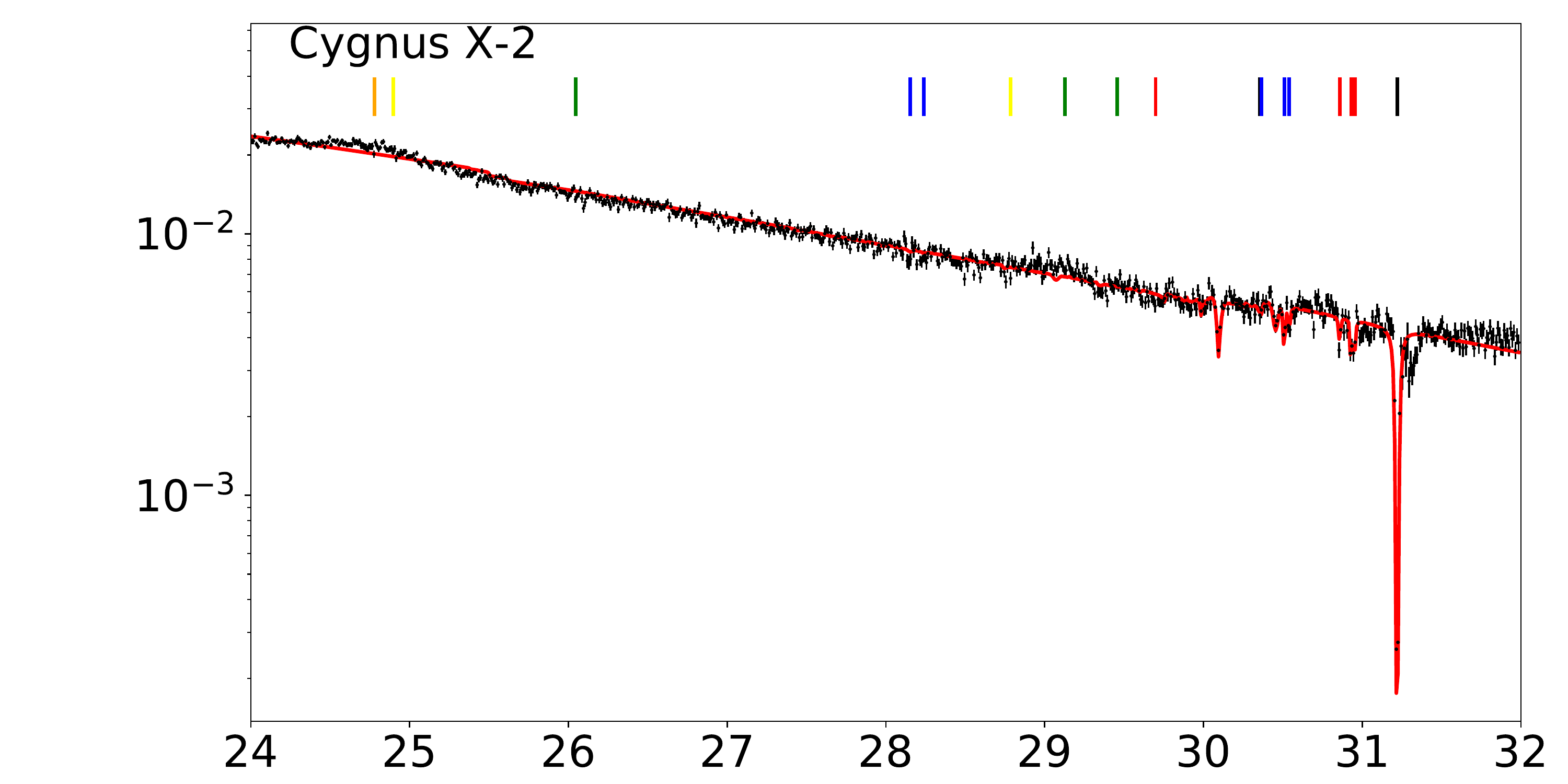}
\hspace{-2.5mm}
\includegraphics[scale=0.195]{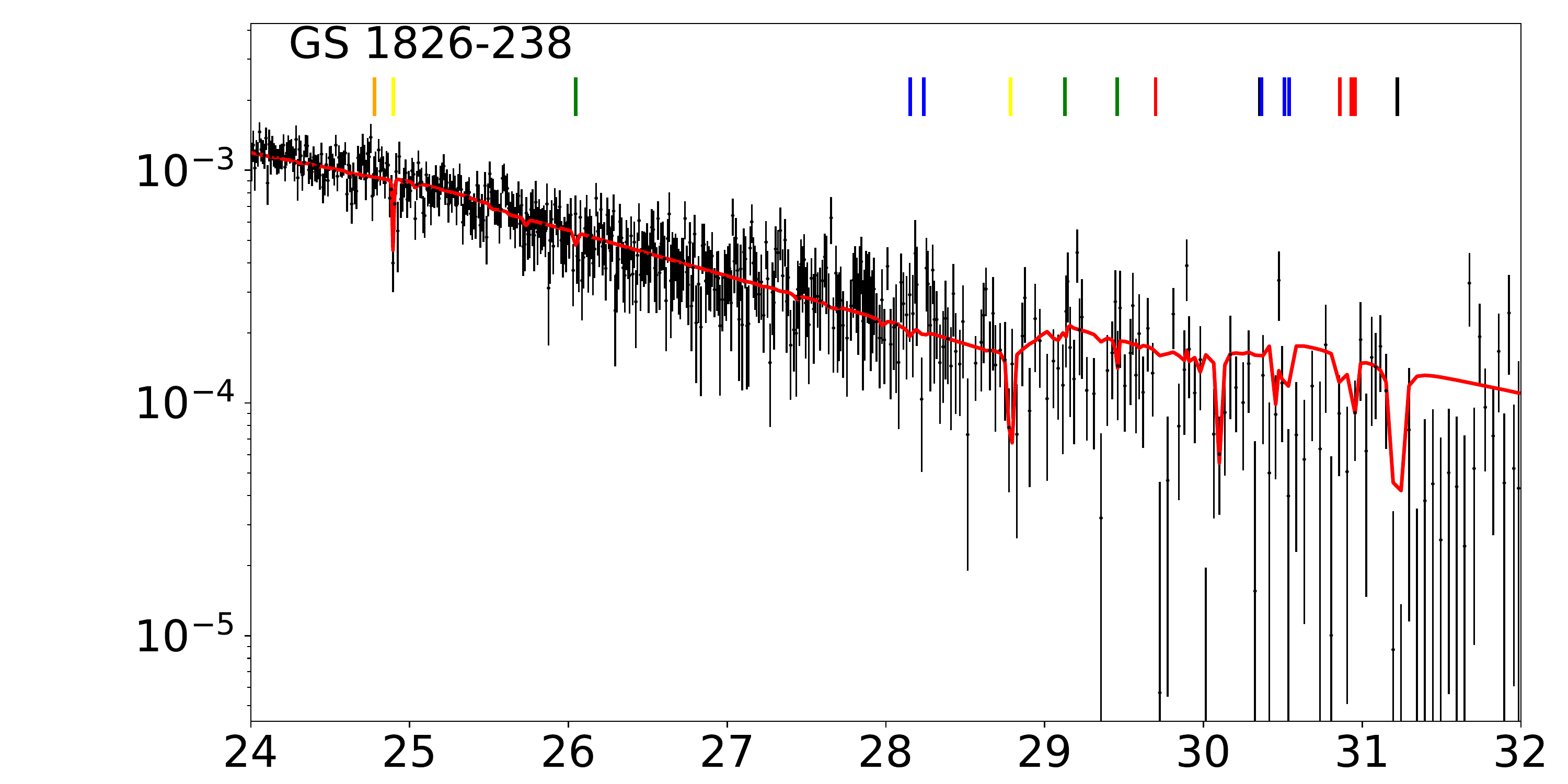}\\
\includegraphics[scale=0.195]{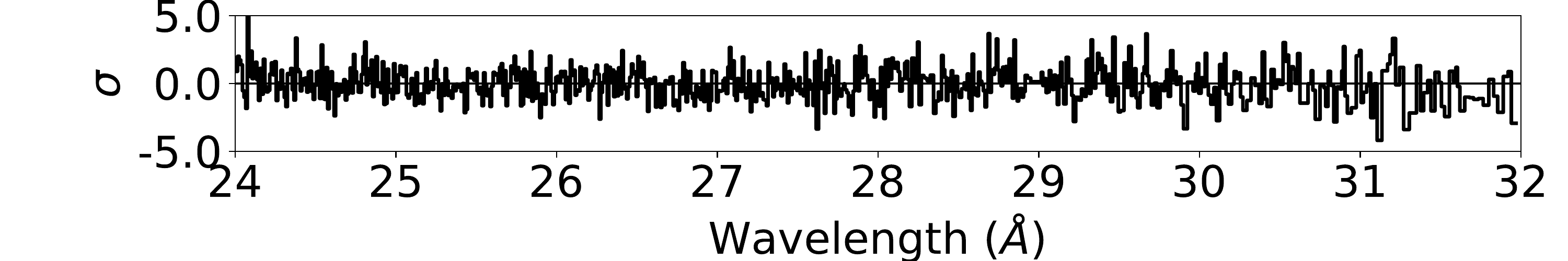}
\hspace{-2.5mm}
\includegraphics[scale=0.195]{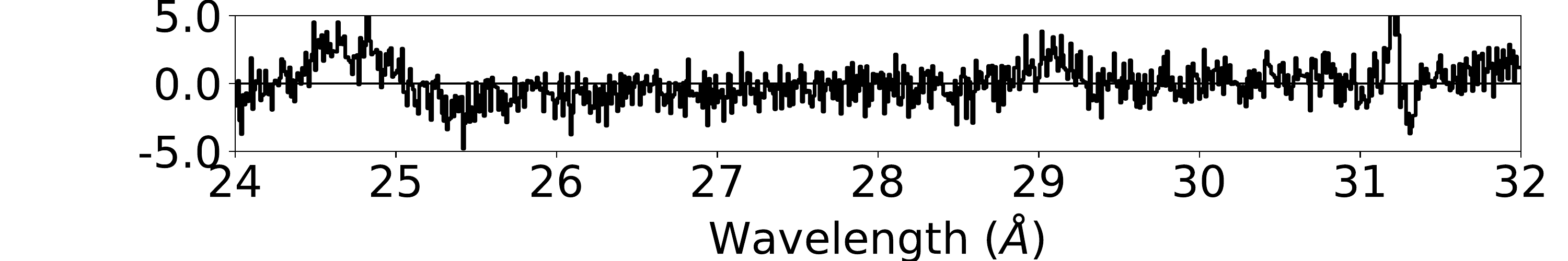}
\hspace{-2.5mm}
\includegraphics[scale=0.195]{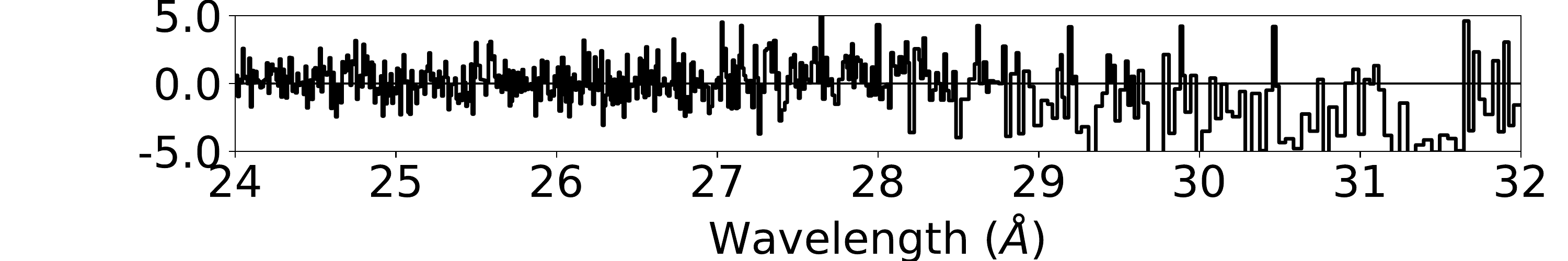}\\

\includegraphics[scale=0.195]{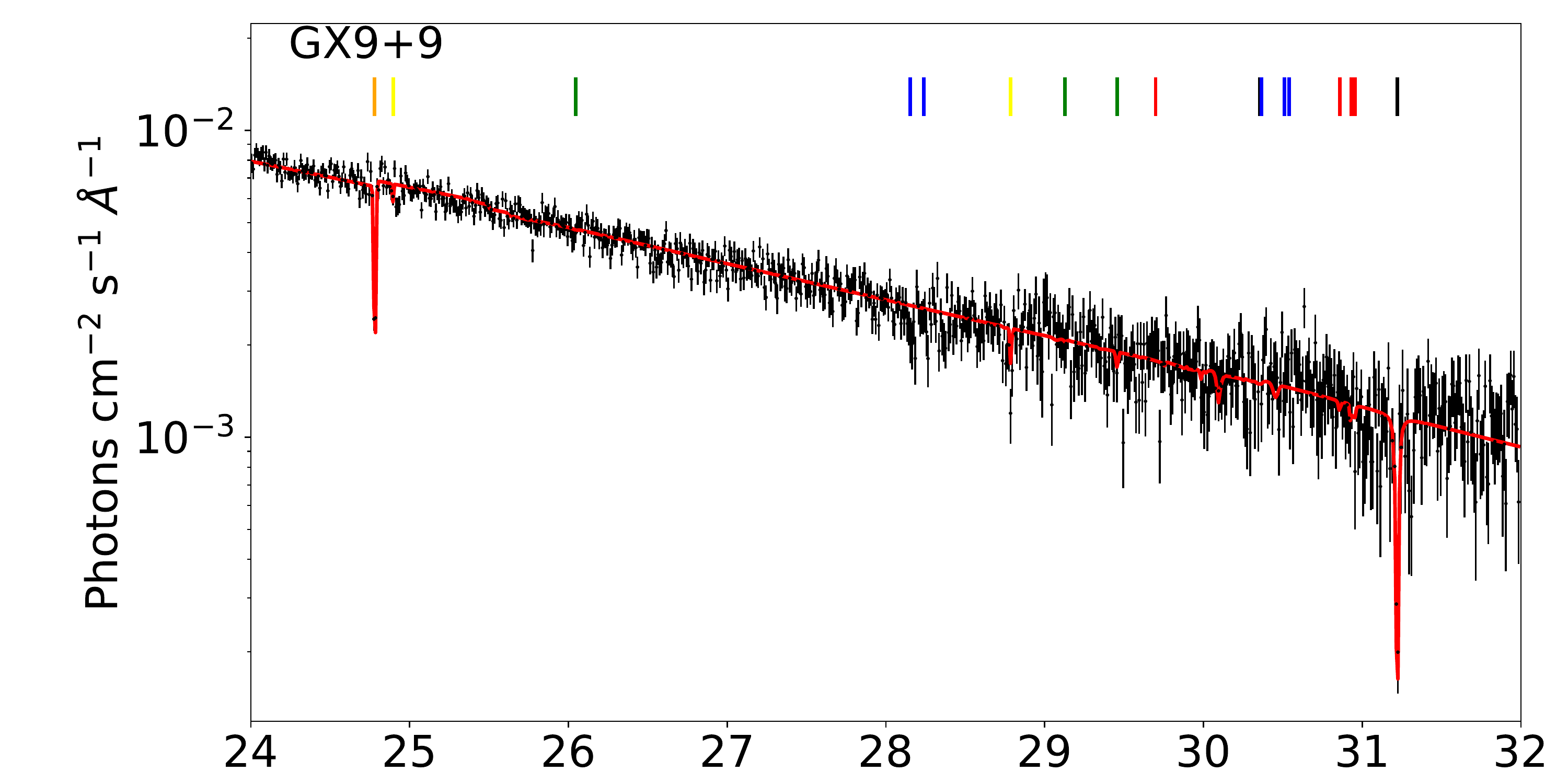}
\hspace{-2.5mm}
\includegraphics[scale=0.195]{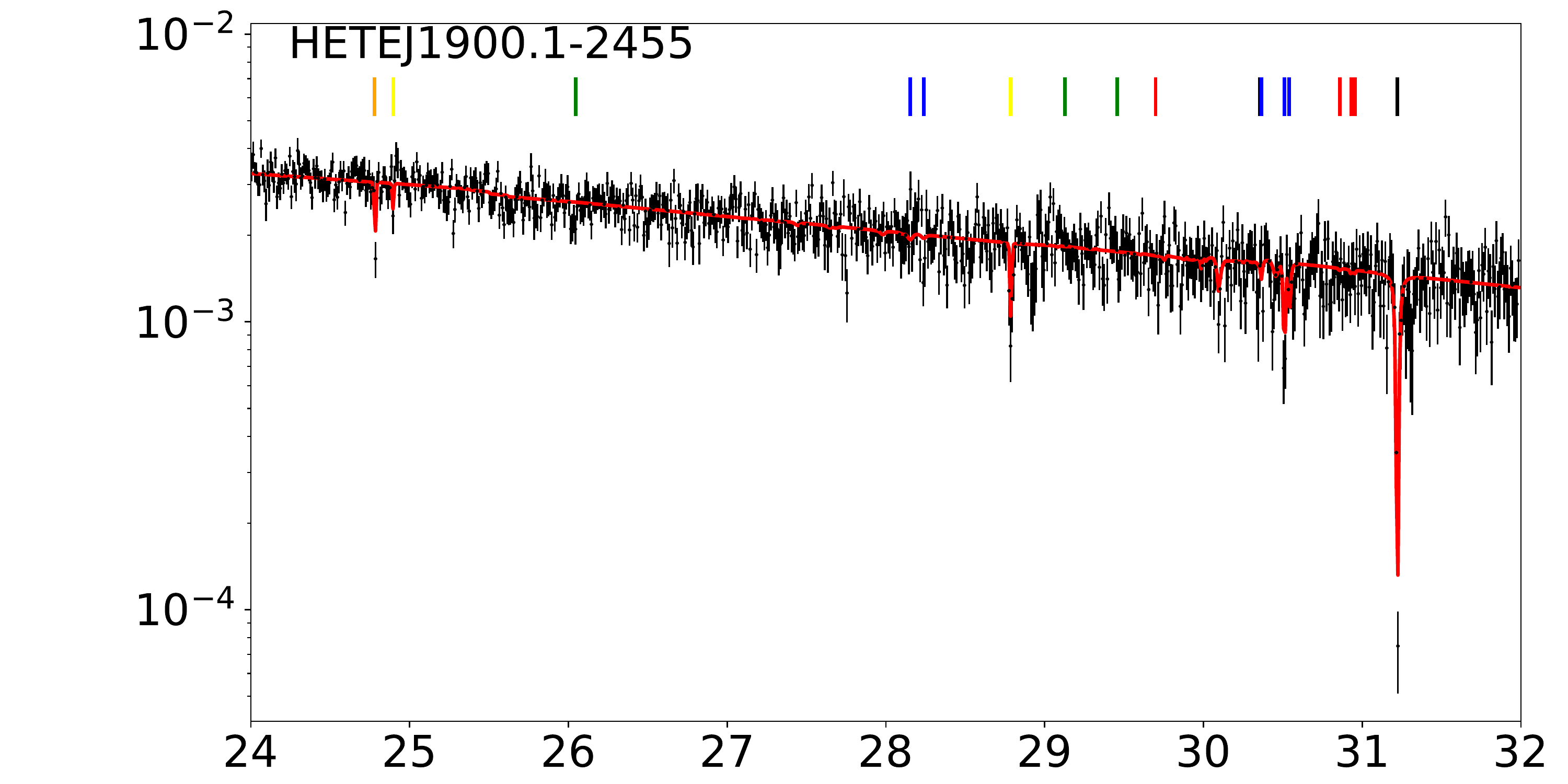}
\hspace{-2.5mm}
\includegraphics[scale=0.195]{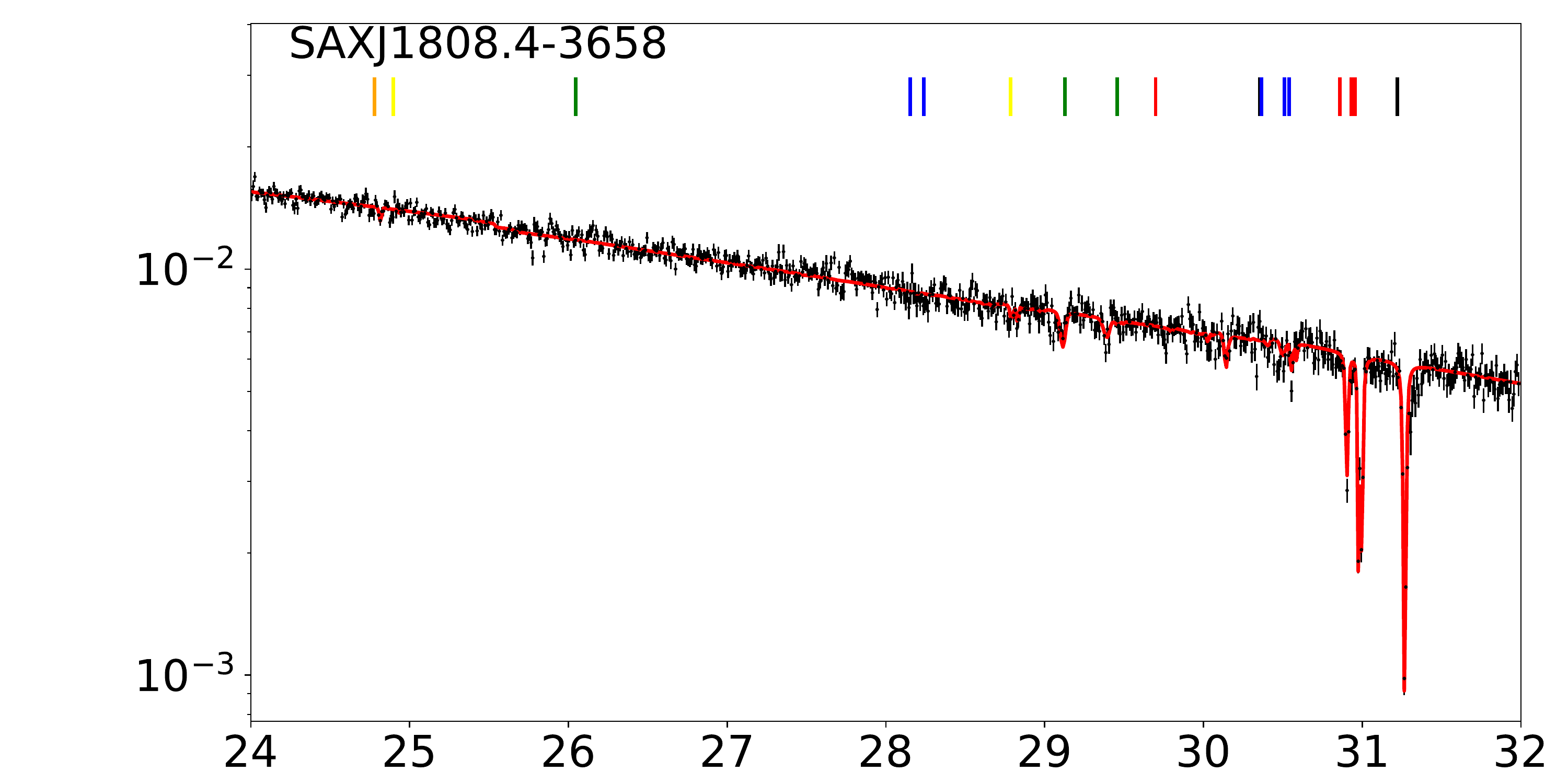}\\
\includegraphics[scale=0.195]{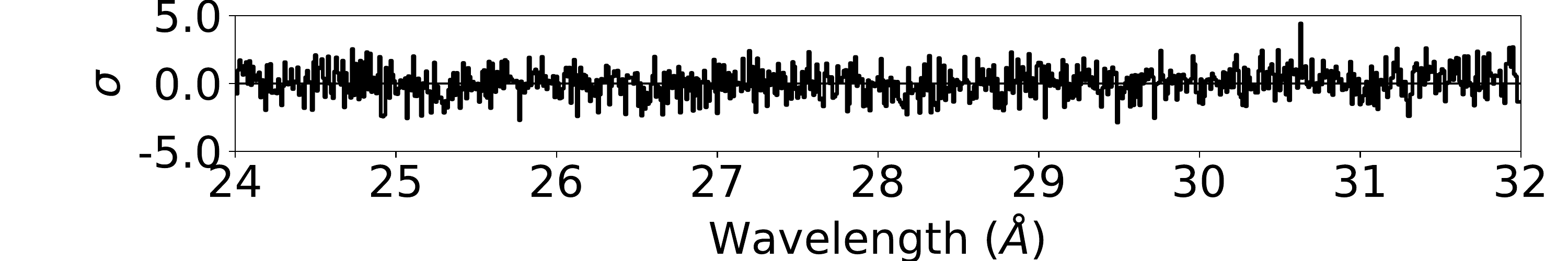}
\hspace{-2.5mm}
\includegraphics[scale=0.195]{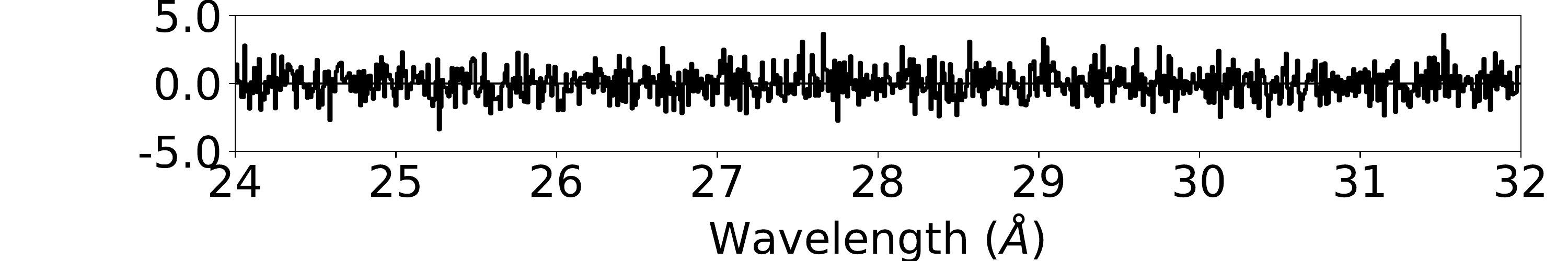}
\hspace{-2.5mm}
\includegraphics[scale=0.195]{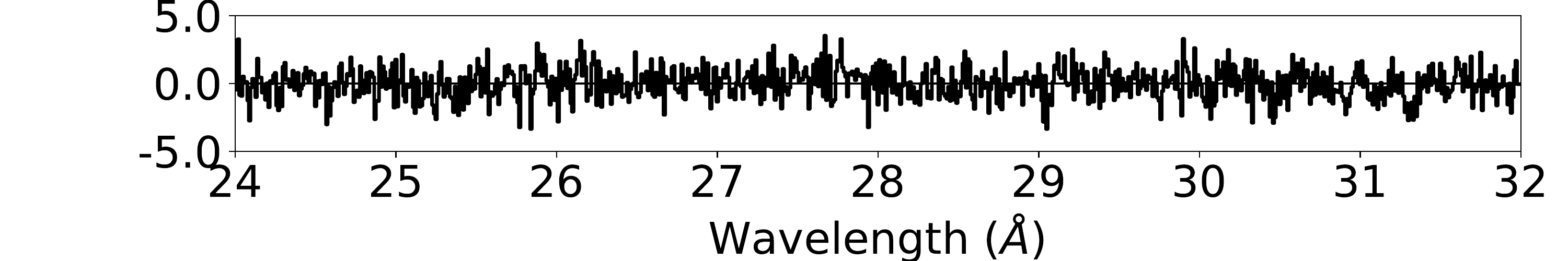}\\

\includegraphics[scale=0.195]{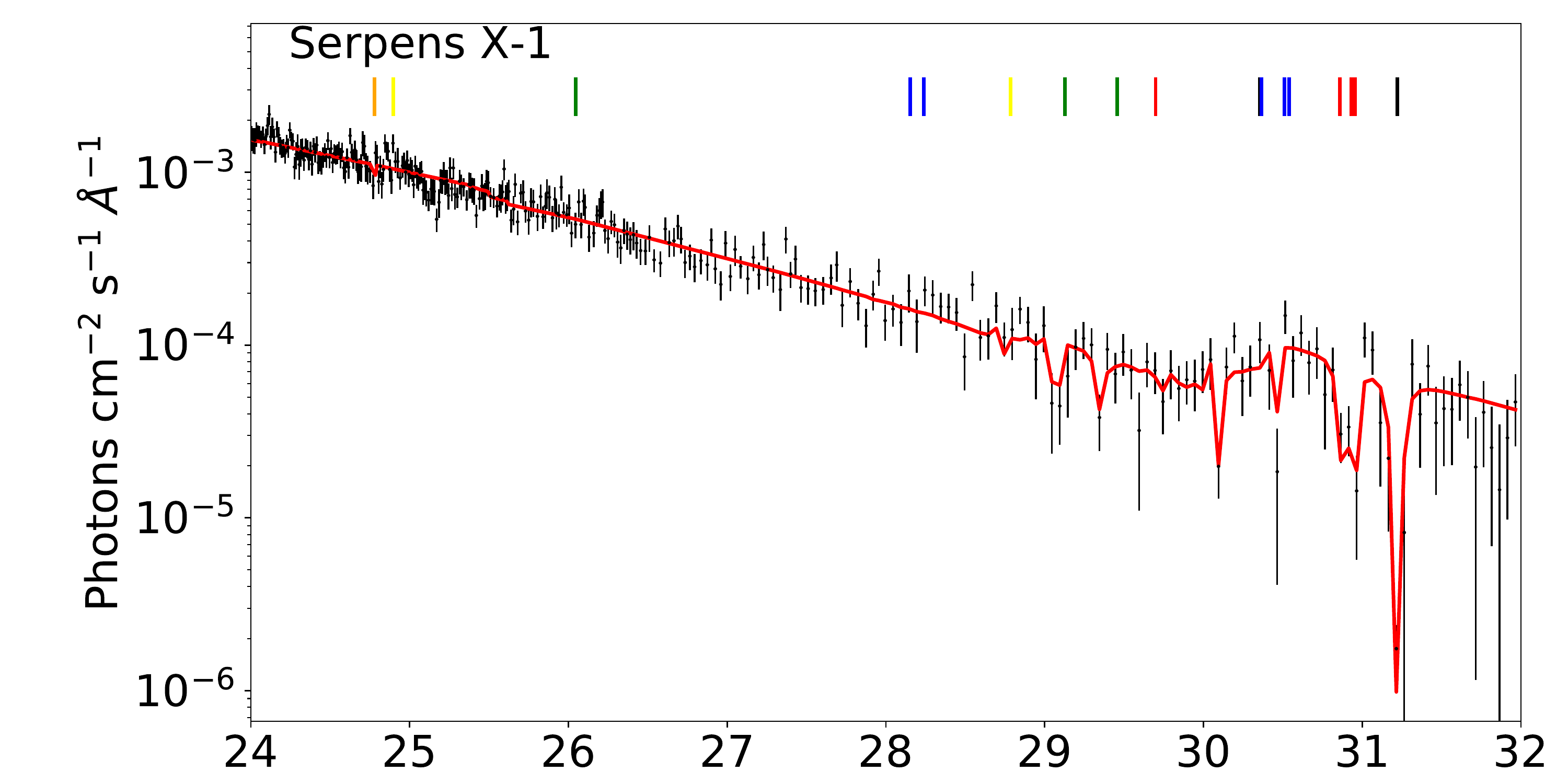}
\hspace{-2.5mm}
\includegraphics[scale=0.195]{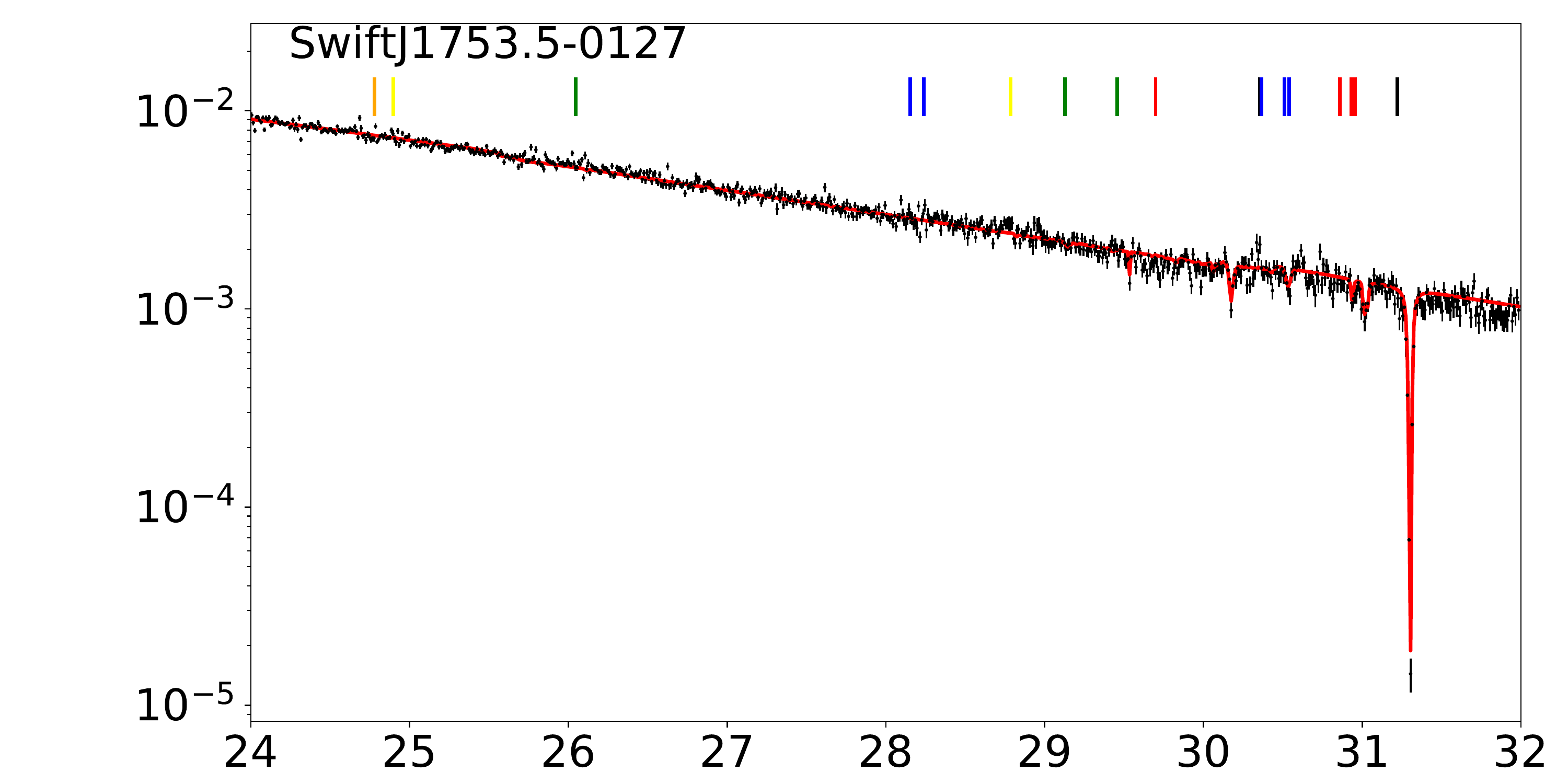} \\
\includegraphics[scale=0.195]{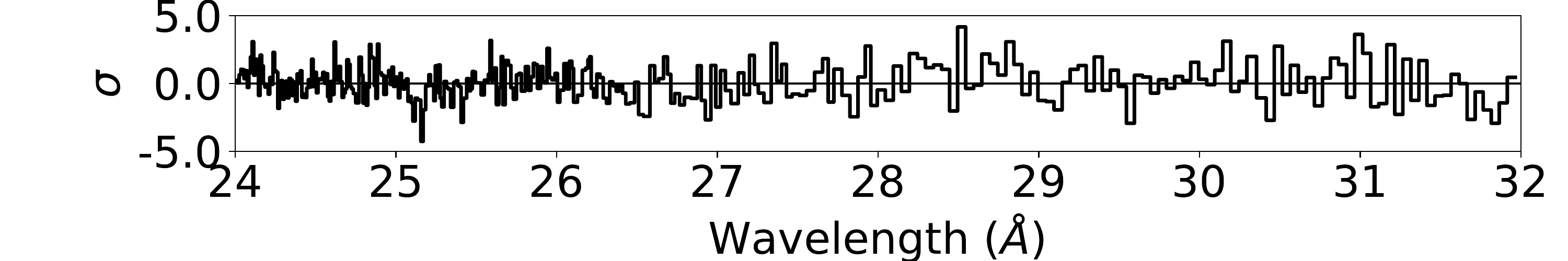}
\hspace{-2.5mm}
\includegraphics[scale=0.195]{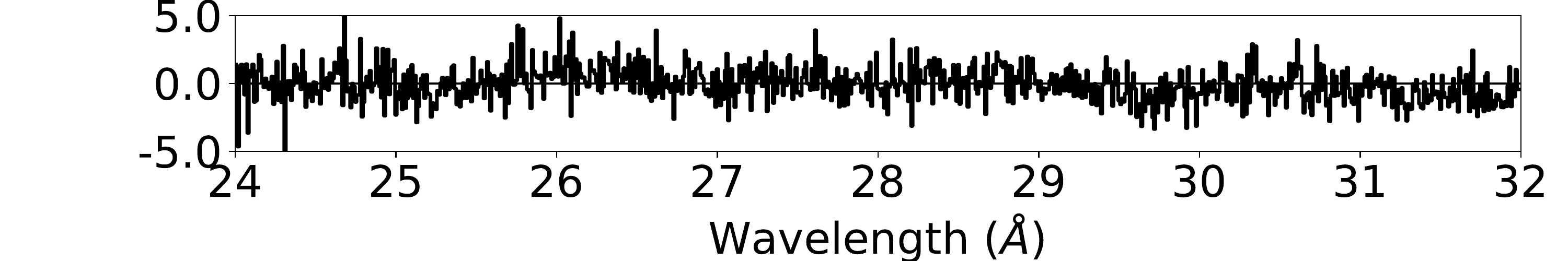} \\

      \caption{Best fit results in the N K-edge photoabsorption region for the LMXBs sample described in Table~\ref{tab_lmxbs}. In each panel, the black data points are the observations, while the solid red lines correspond to the best-fit models.}\label{fig_fits_lmxbs}
   \end{figure*}

Figures~\ref{fig_fits_lmxbs} and ~\ref{fig_fits_ex1} show the best-fit models to the spectra for the individual LMXBs and extragalactic sources, respectively.  For each source, all observations were combined for illustrative purposes. Black points correspond to the observations while the red lines indicate the best-fit model. Residuals, in units of $(data-model)/error$, are included. The positions of the K$\alpha$ and K$\beta$ resonance lines are indicated for each ion, following the color code used in Figure~\ref{fig_n_cross}.

         \begin{figure*}
          \centering
\includegraphics[scale=0.195]{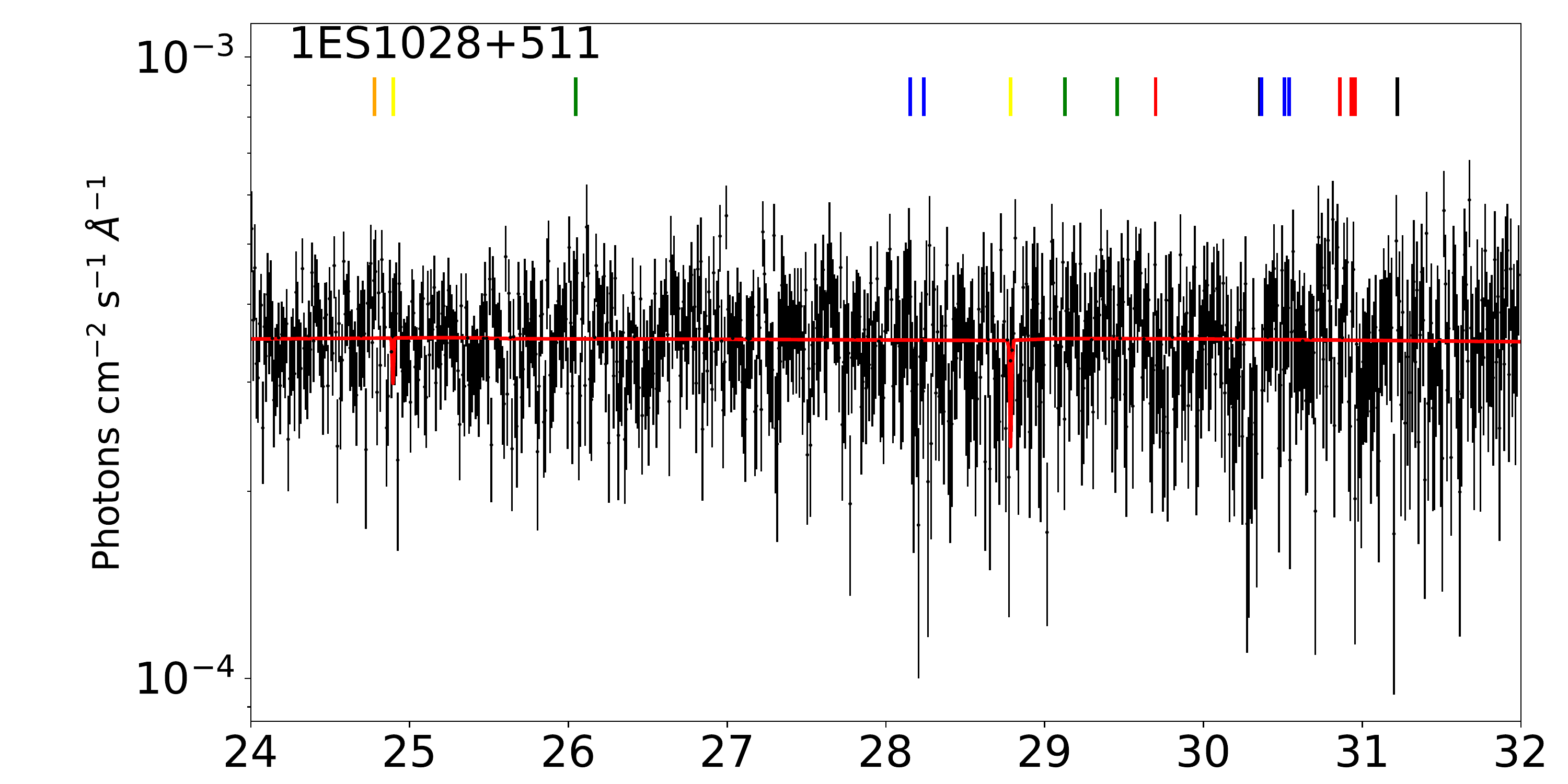}
\hspace{-2.5mm}
\includegraphics[scale=0.195]{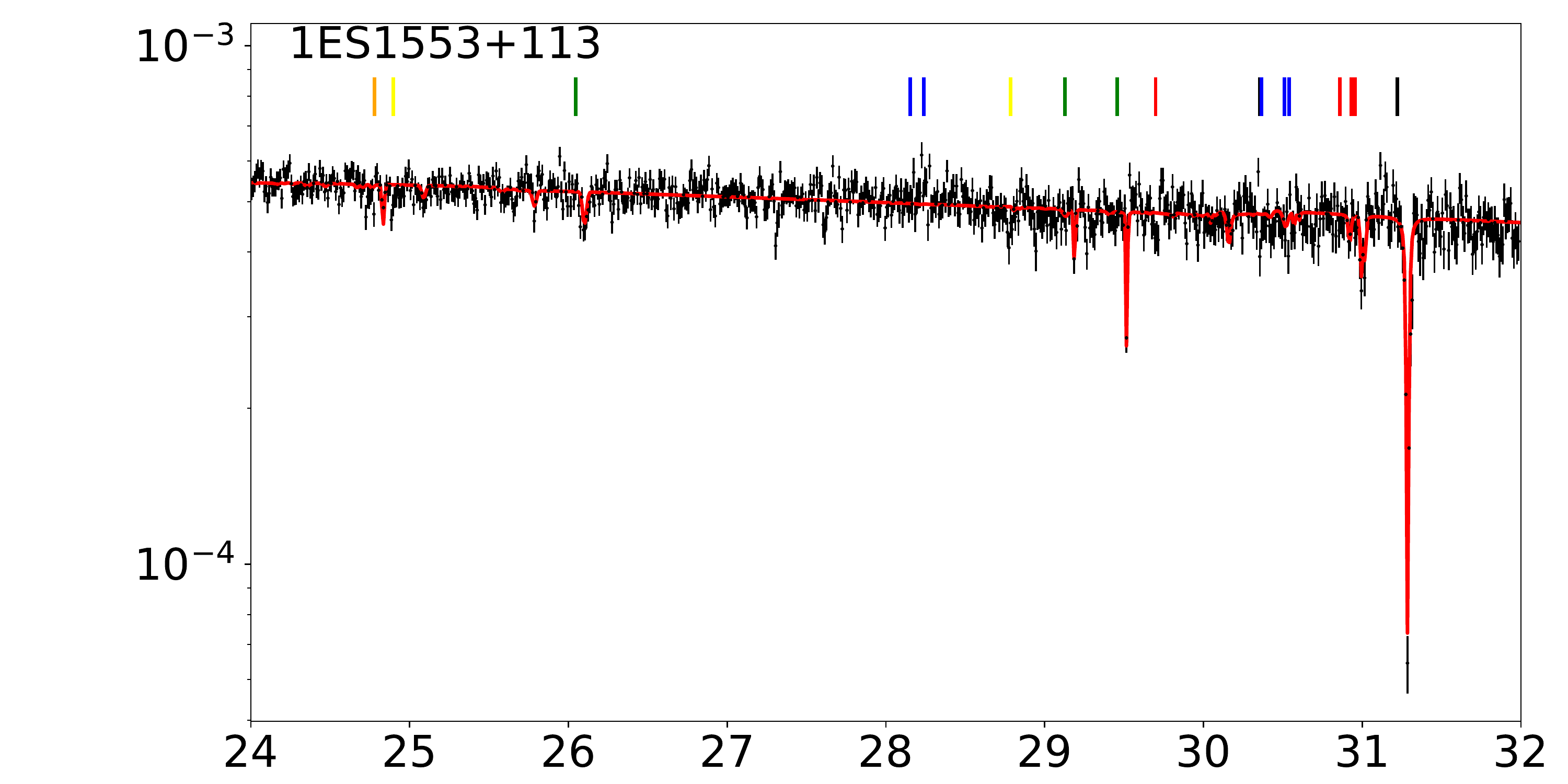} 
\hspace{-2.5mm}
\includegraphics[scale=0.195]{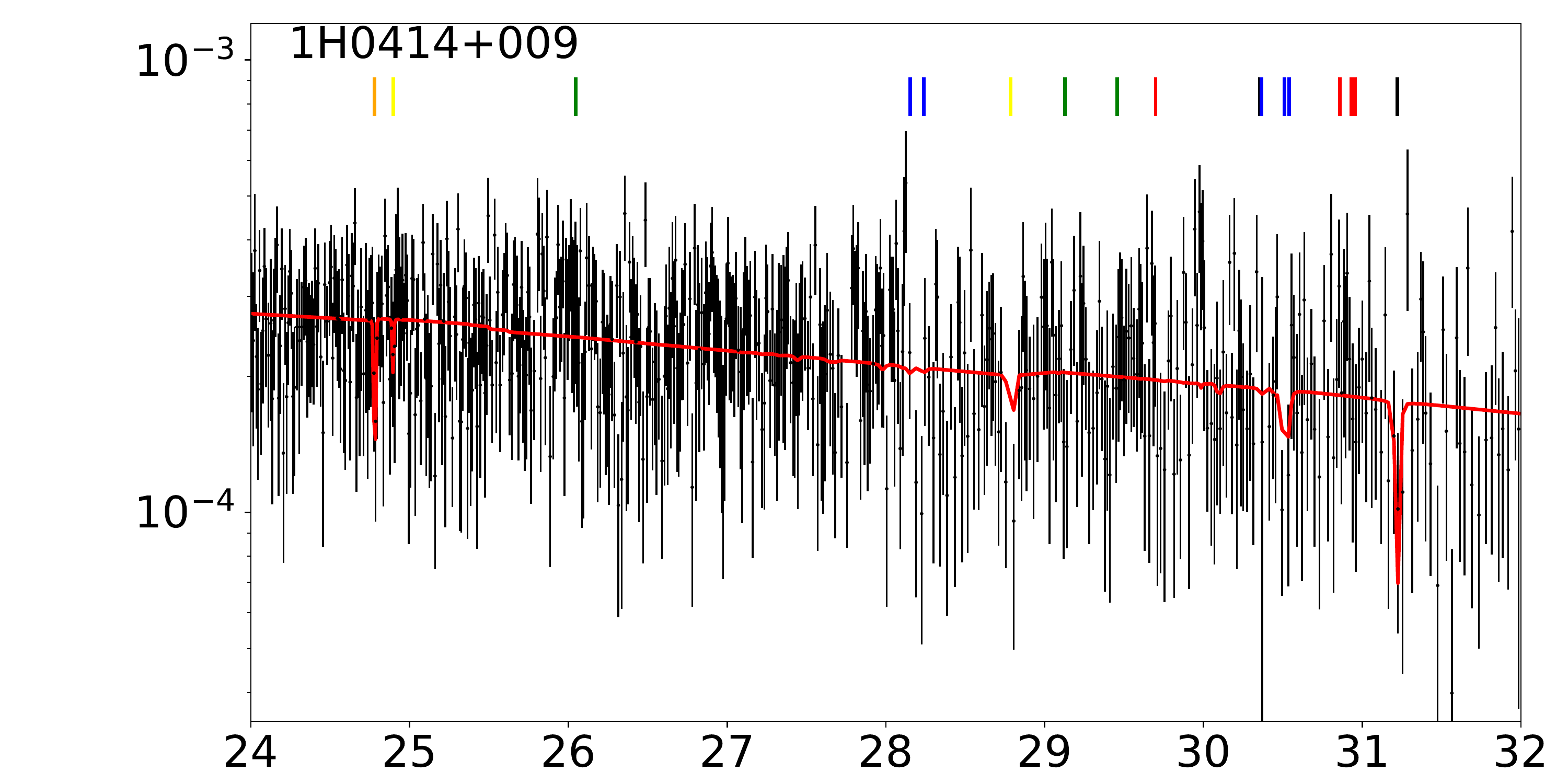}\\
\includegraphics[scale=0.195]{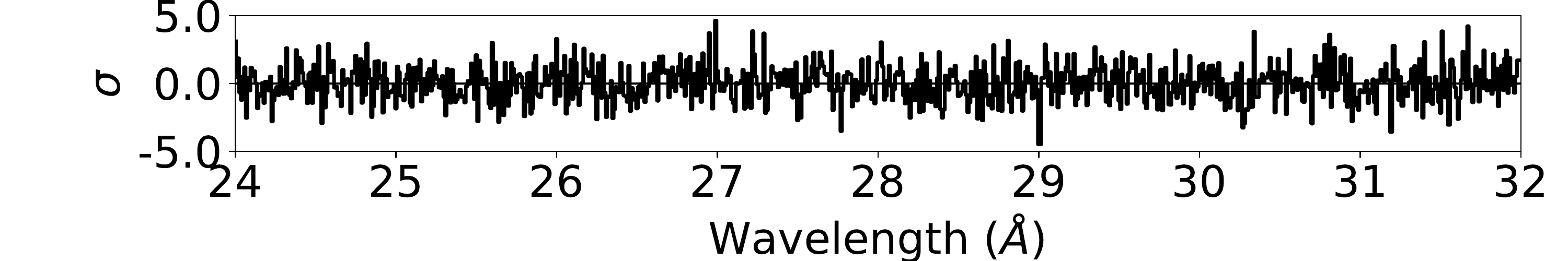}
\hspace{-2.5mm}
\includegraphics[scale=0.195]{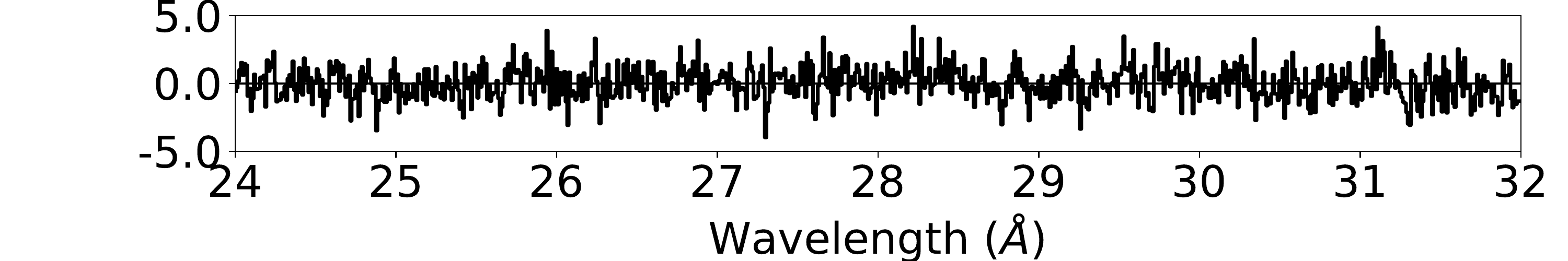} 
\hspace{-2.5mm}
\includegraphics[scale=0.195]{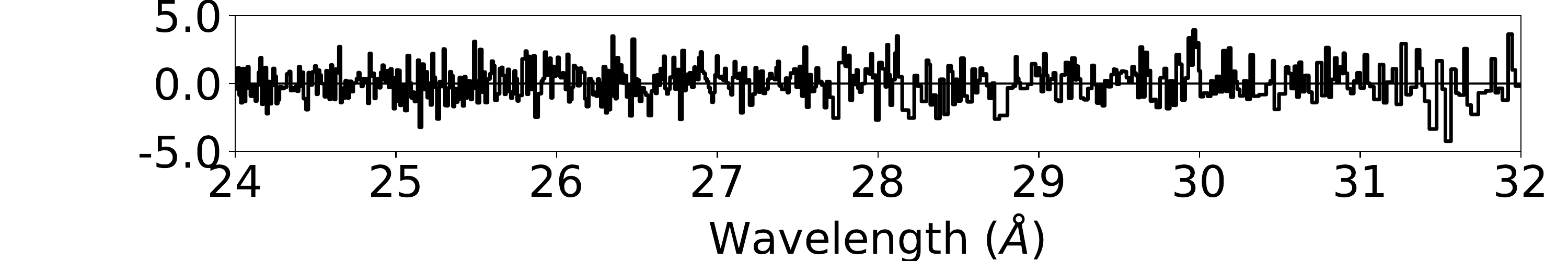}\\

\includegraphics[scale=0.195]{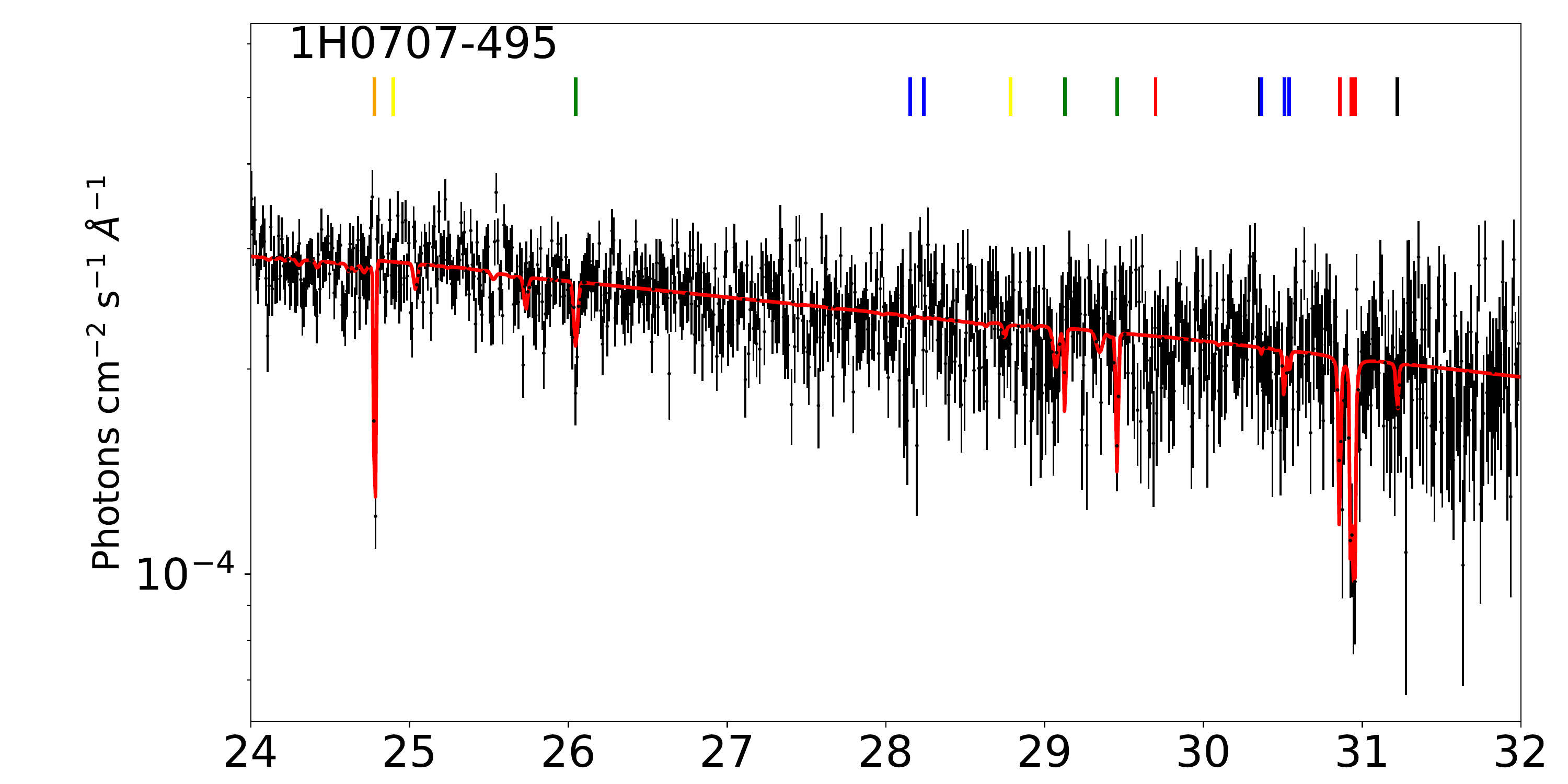}
\hspace{-2.5mm}
\includegraphics[scale=0.195]{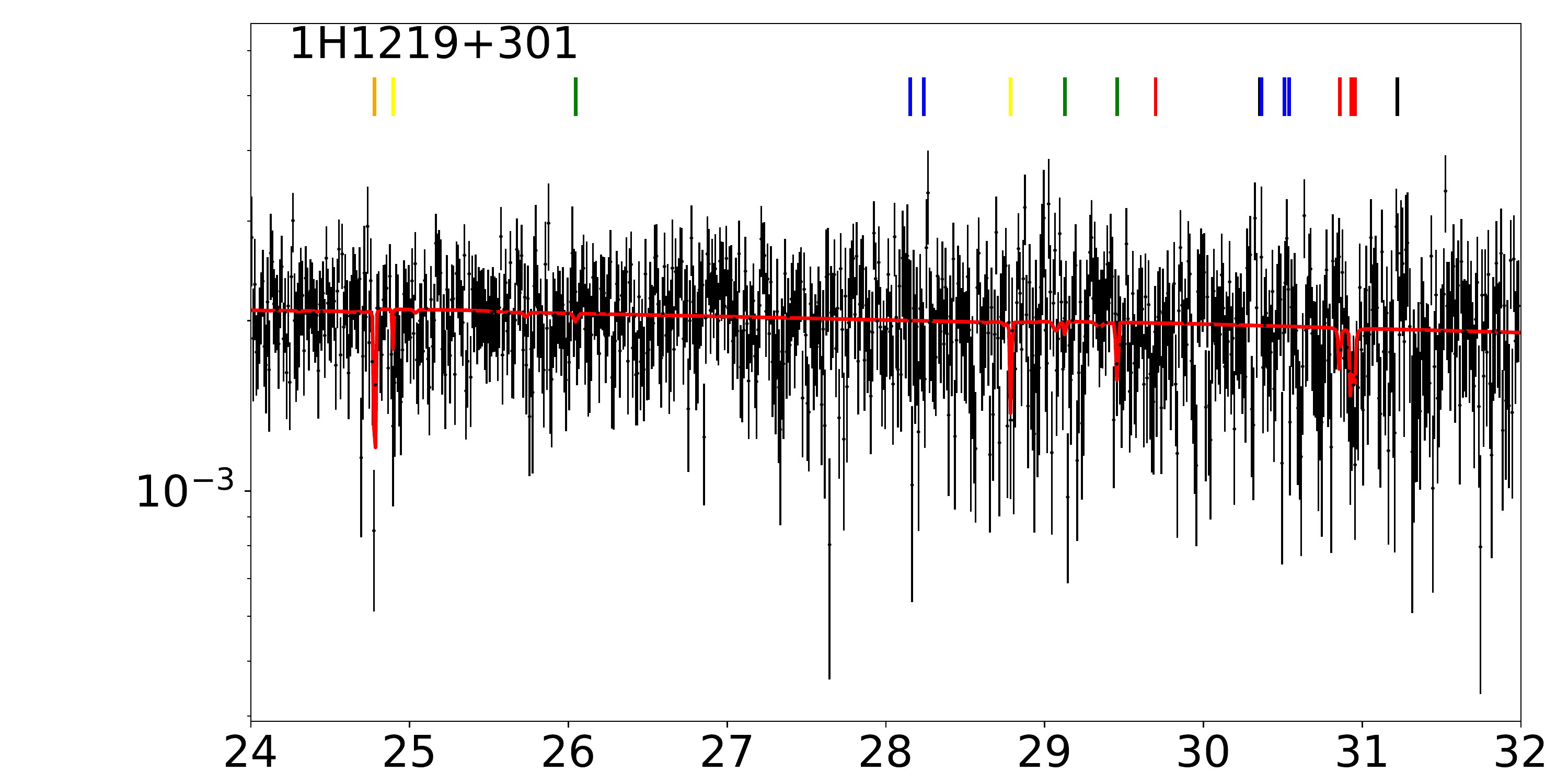} 
\hspace{-2.5mm}
\includegraphics[scale=0.195]{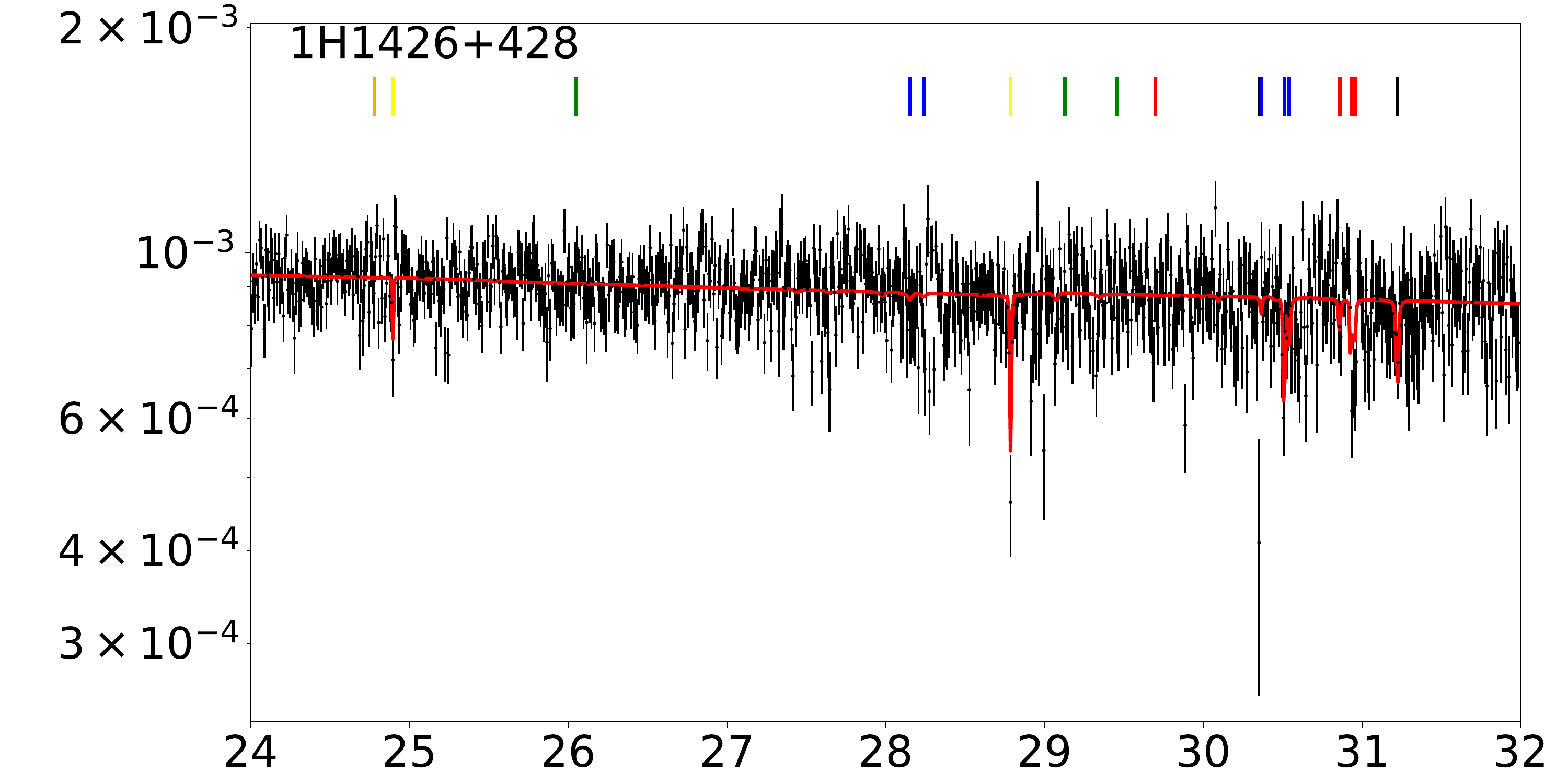}\\
\includegraphics[scale=0.195]{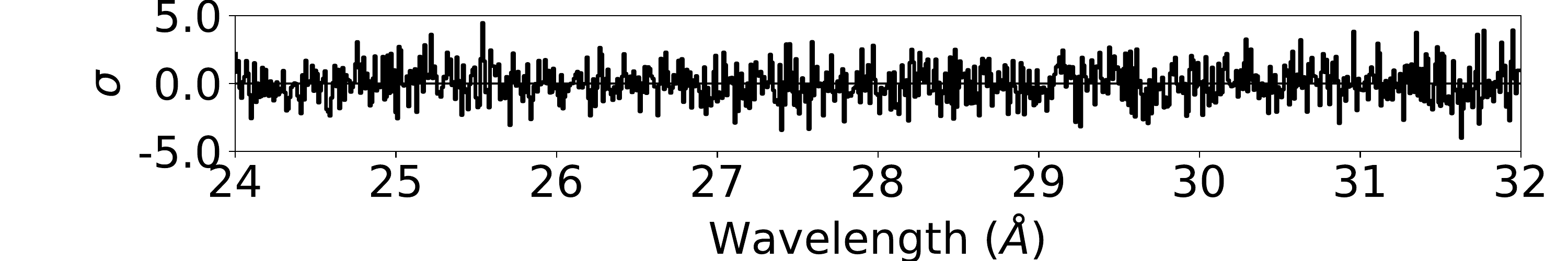}
\hspace{-2.5mm}
\includegraphics[scale=0.195]{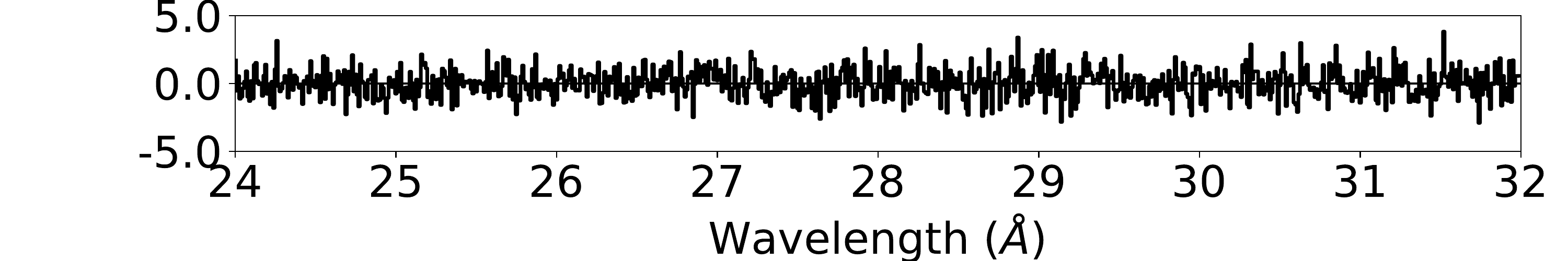} 
\hspace{-2.5mm}
\includegraphics[scale=0.195]{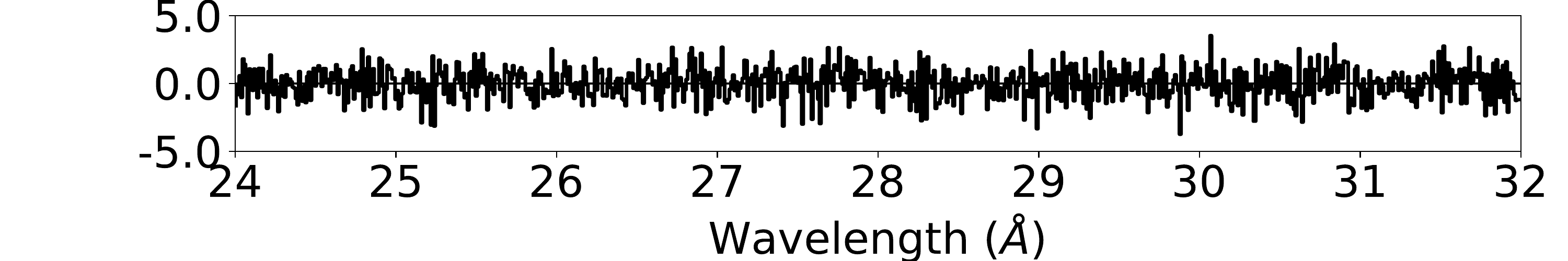}\\

\includegraphics[scale=0.195]{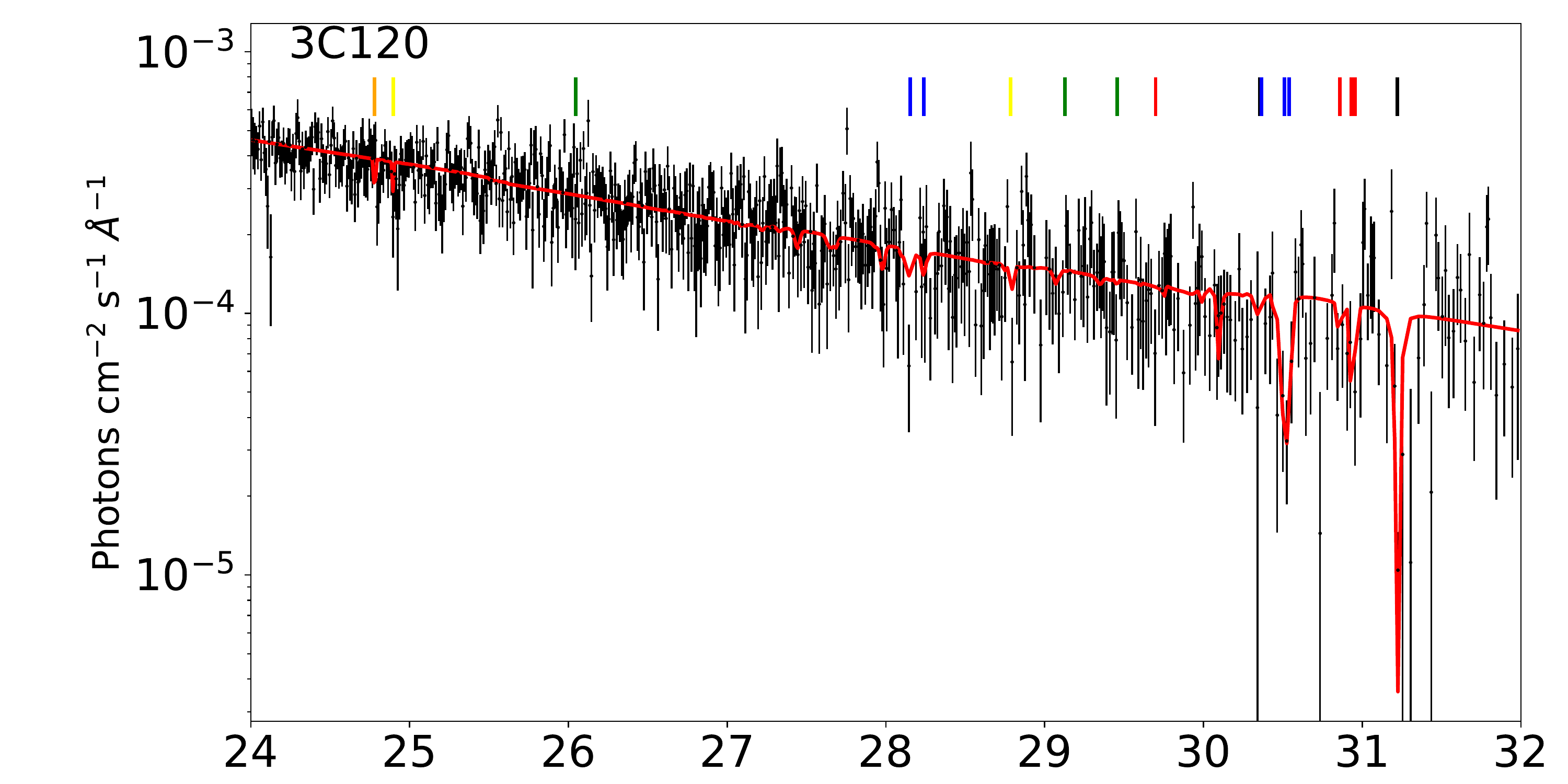}
\hspace{-2.5mm}
\includegraphics[scale=0.195]{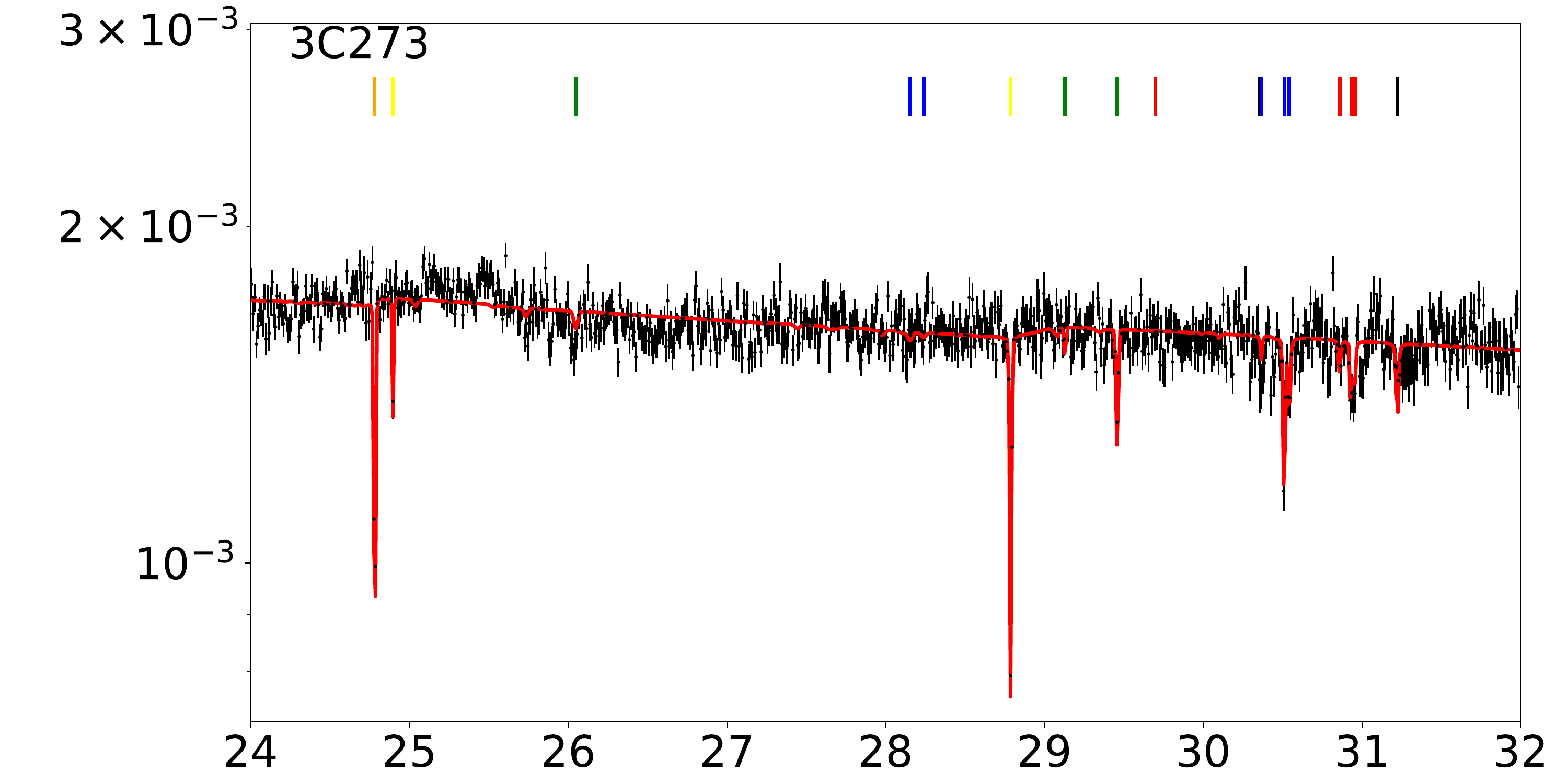} 
\hspace{-2.5mm}
\includegraphics[scale=0.195]{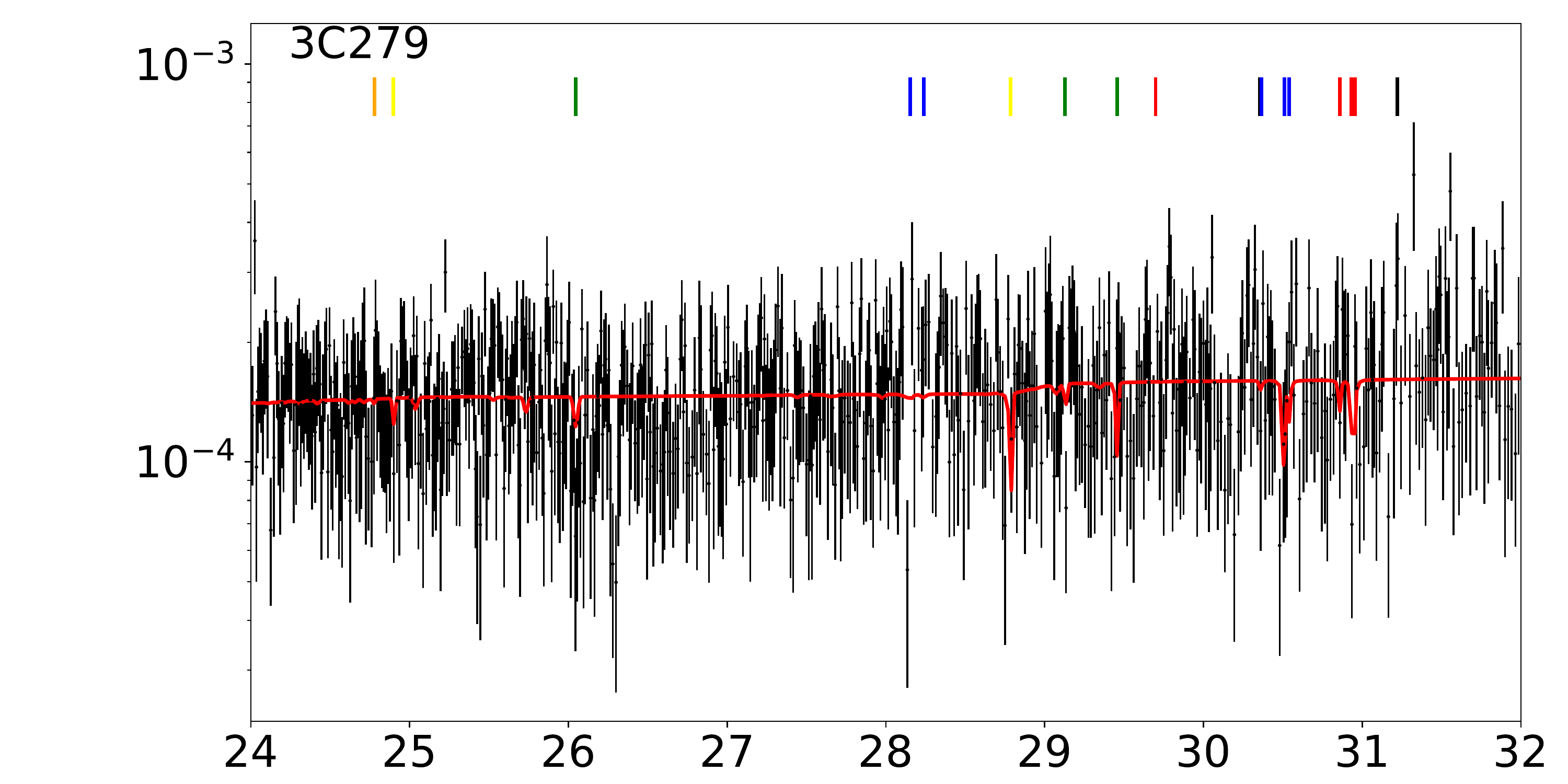}\\
\includegraphics[scale=0.195]{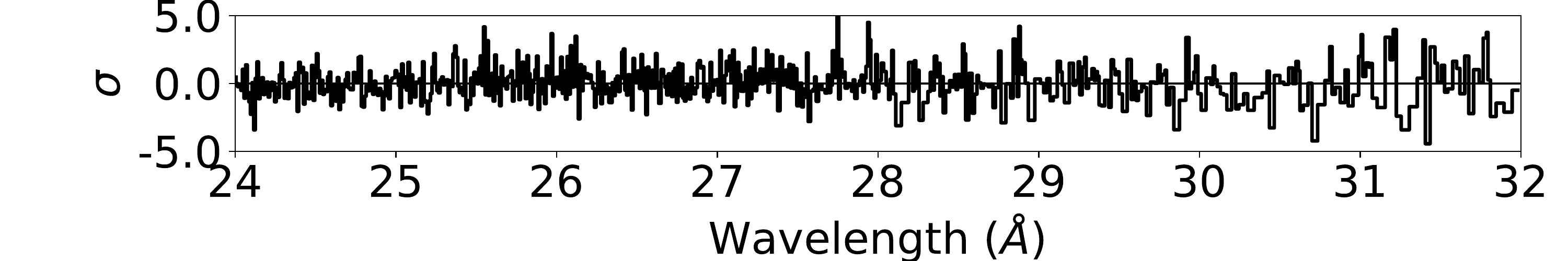}
\hspace{-2.5mm}
\includegraphics[scale=0.195]{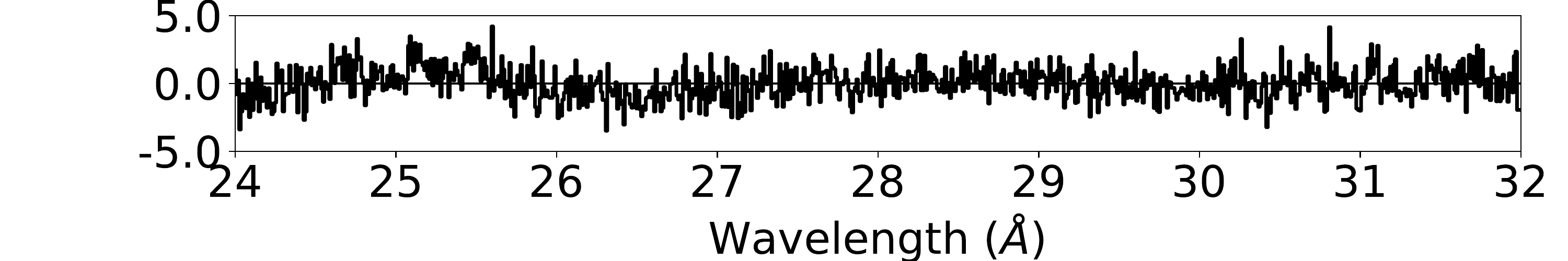} 
\hspace{-2.5mm}
\includegraphics[scale=0.195]{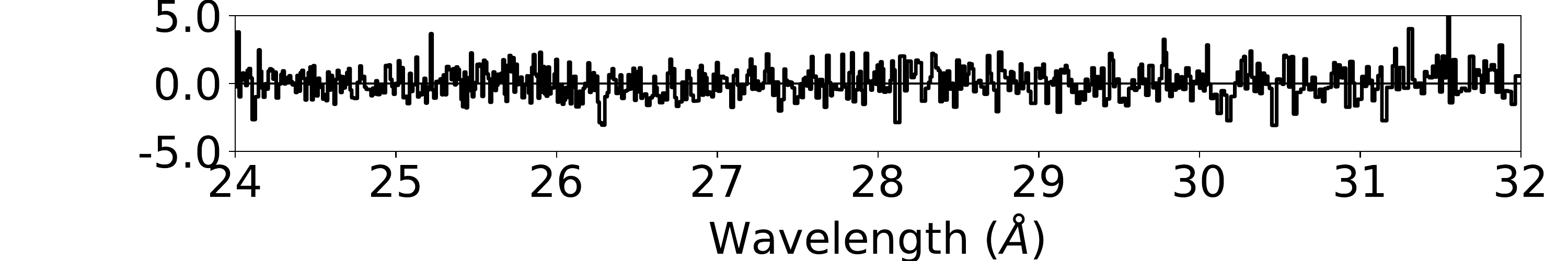}\\

\includegraphics[scale=0.195]{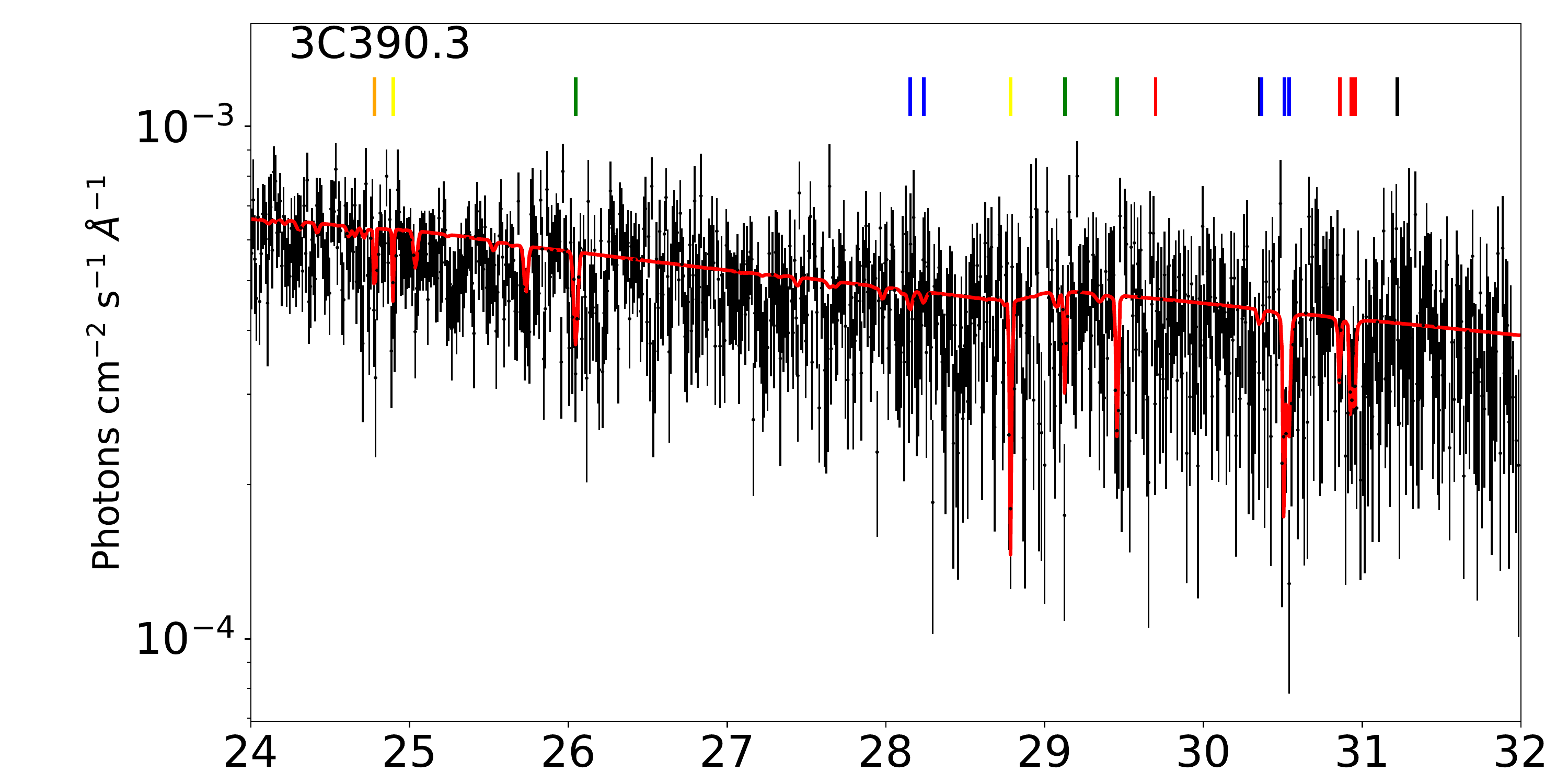}
\hspace{-2.5mm}
\includegraphics[scale=0.195]{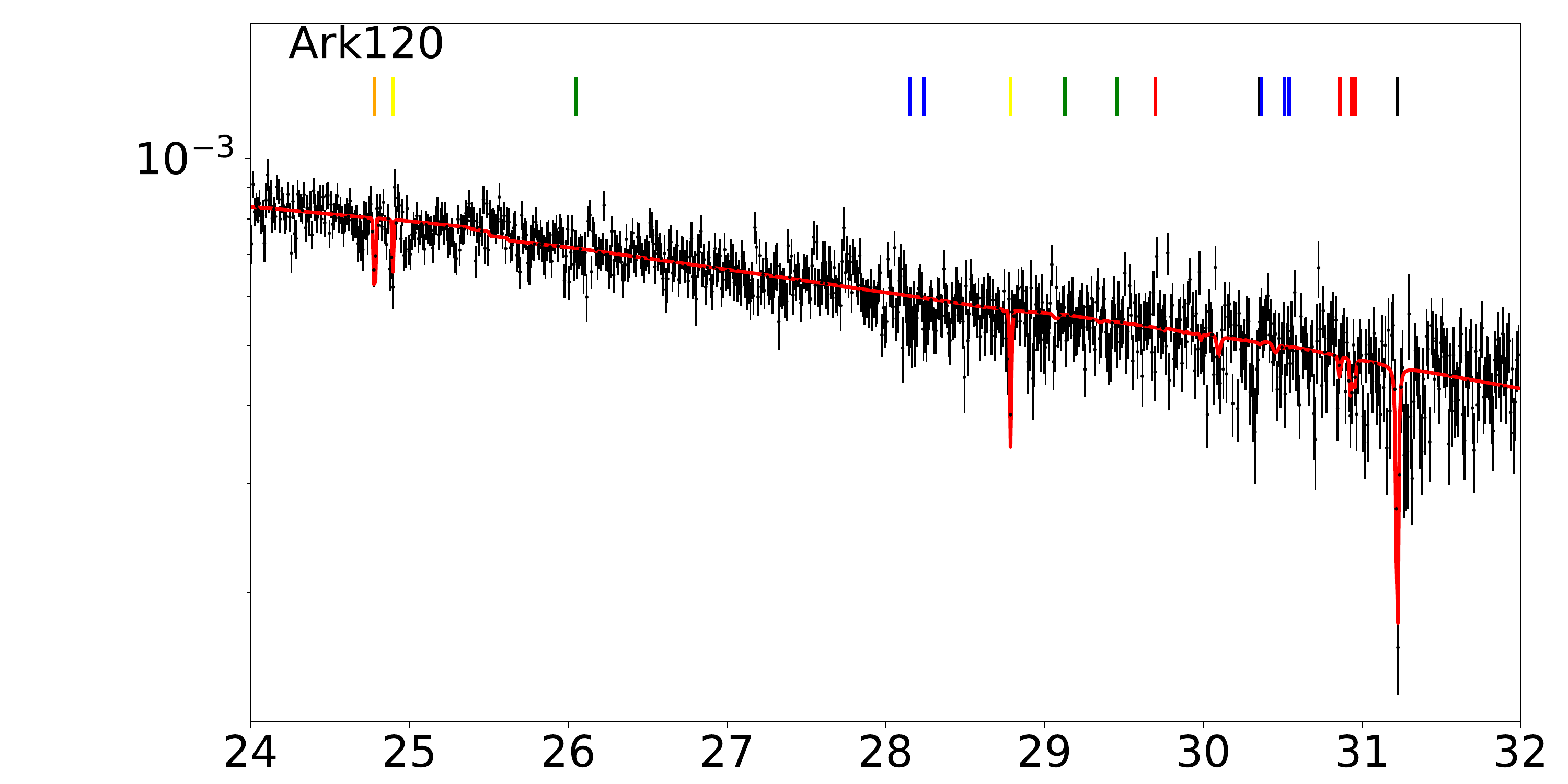} 
\hspace{-2.5mm}
\includegraphics[scale=0.195]{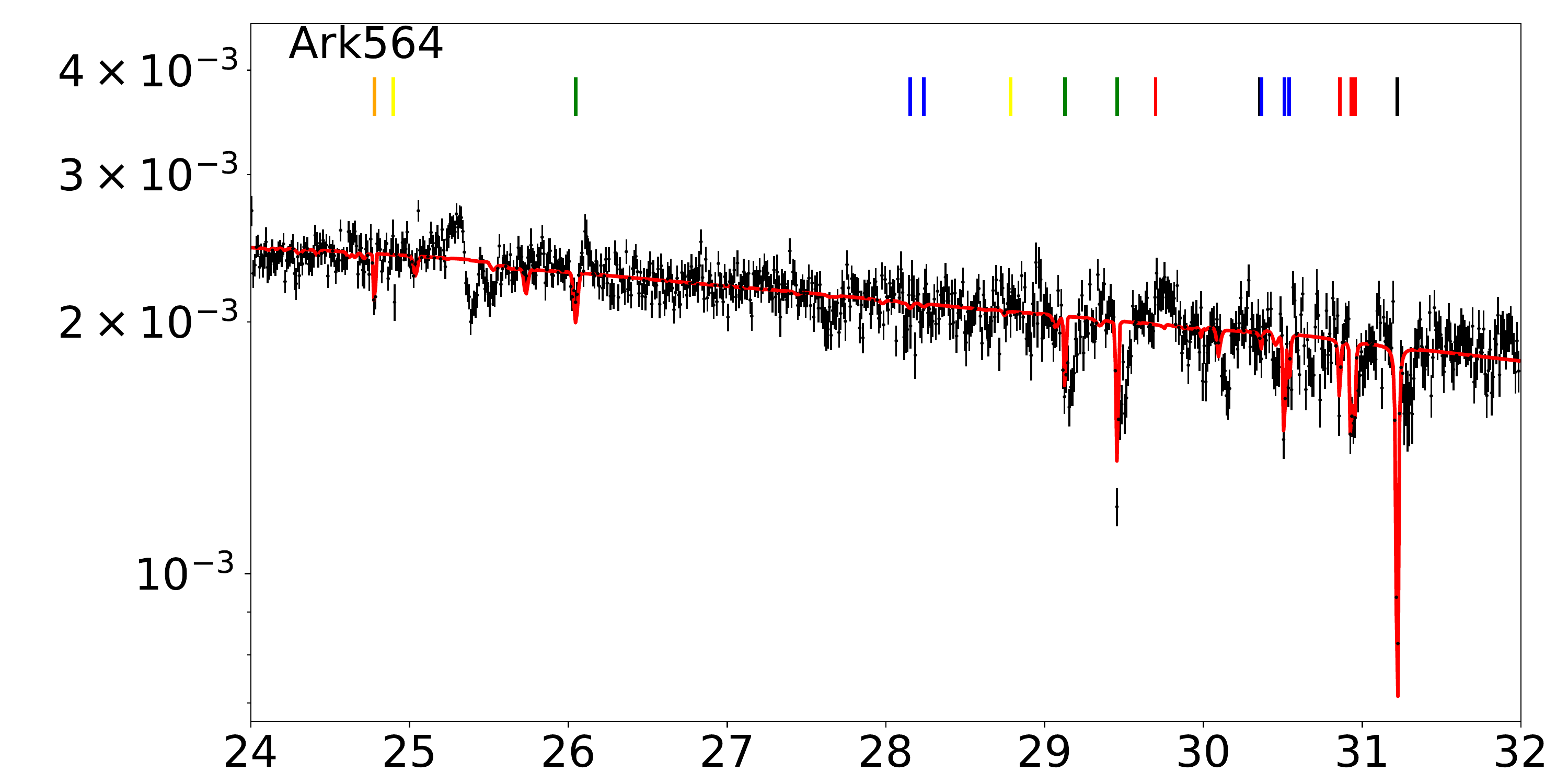}\\
\includegraphics[scale=0.195]{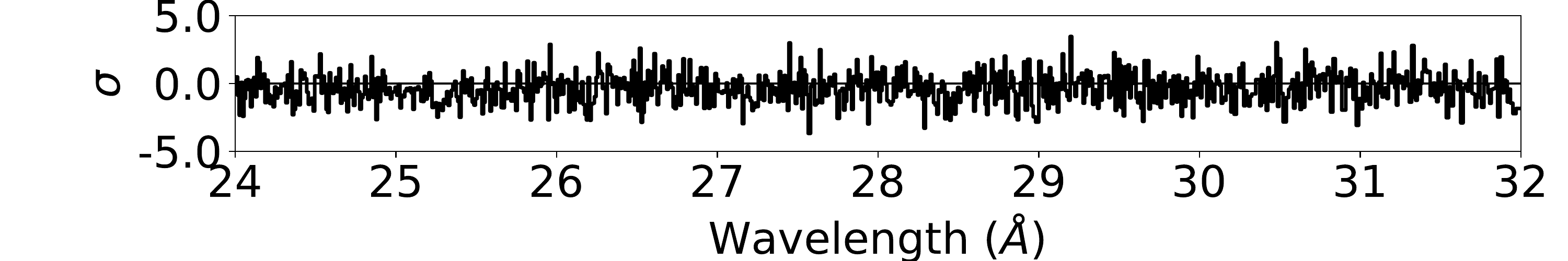}
\hspace{-2.5mm}
\includegraphics[scale=0.195]{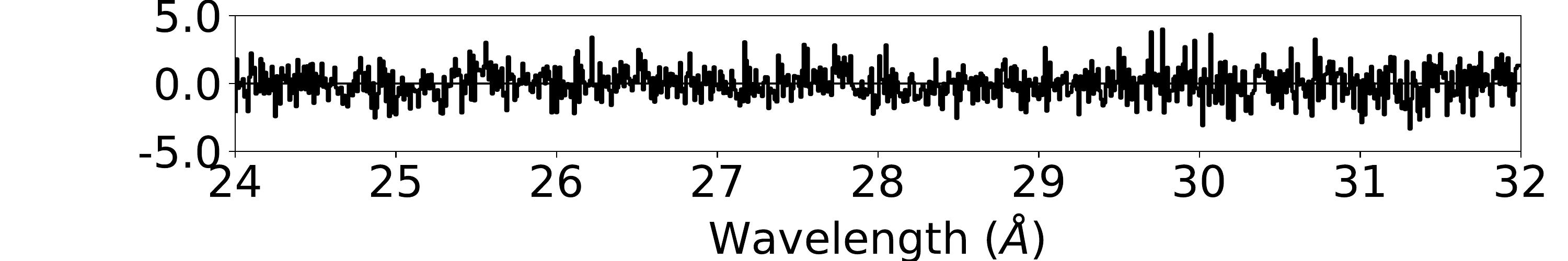} 
\hspace{-2.5mm}
\includegraphics[scale=0.195]{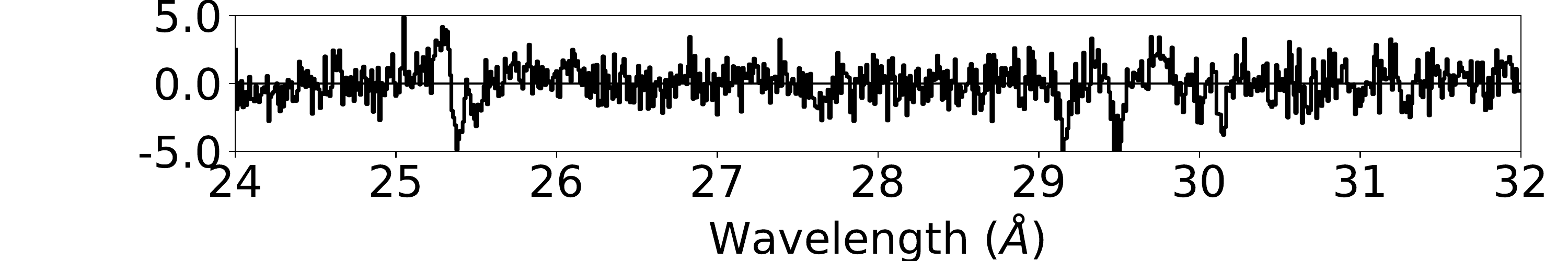}\\

\includegraphics[scale=0.195]{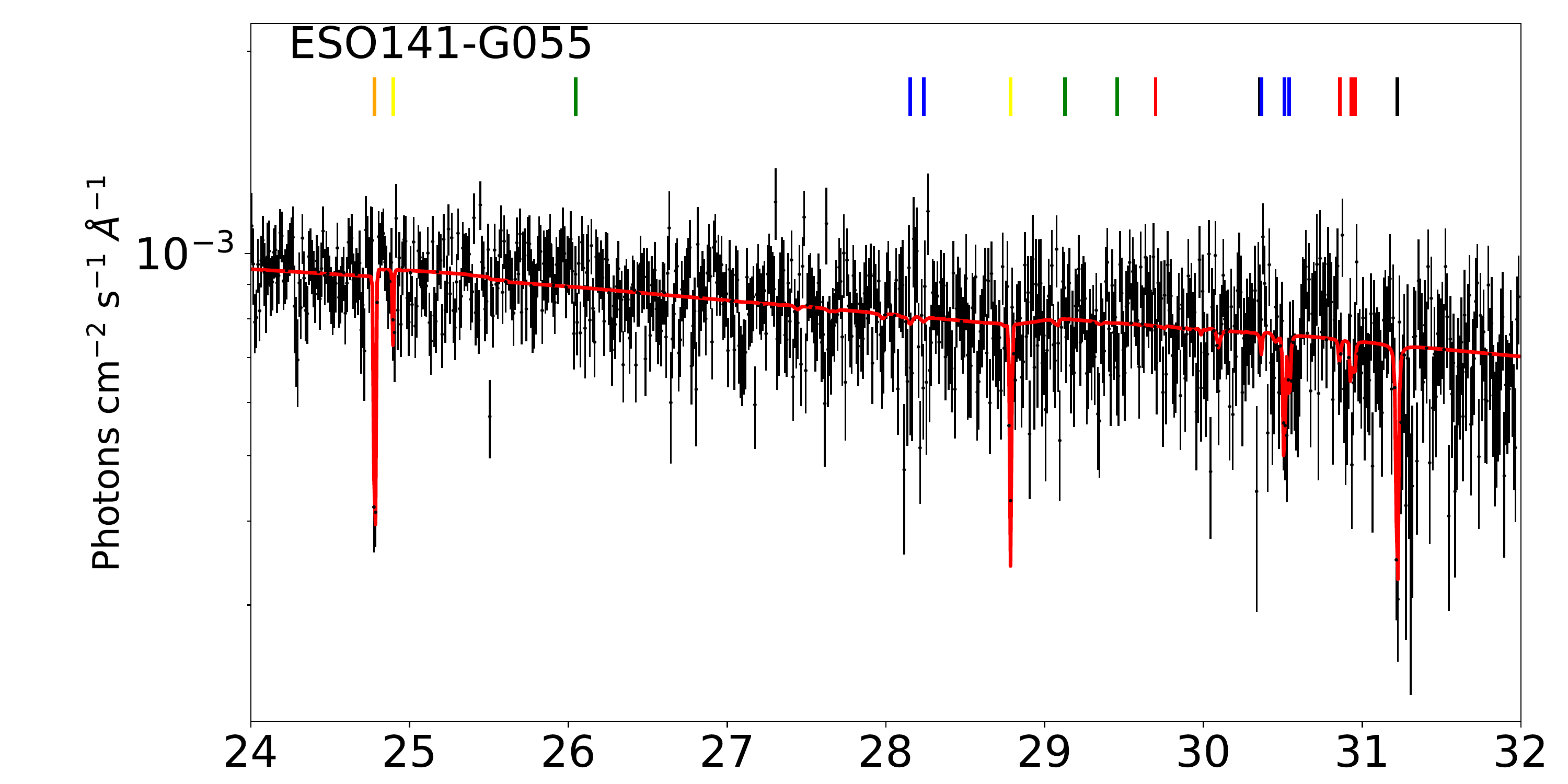}
\hspace{-2.5mm}
\includegraphics[scale=0.195]{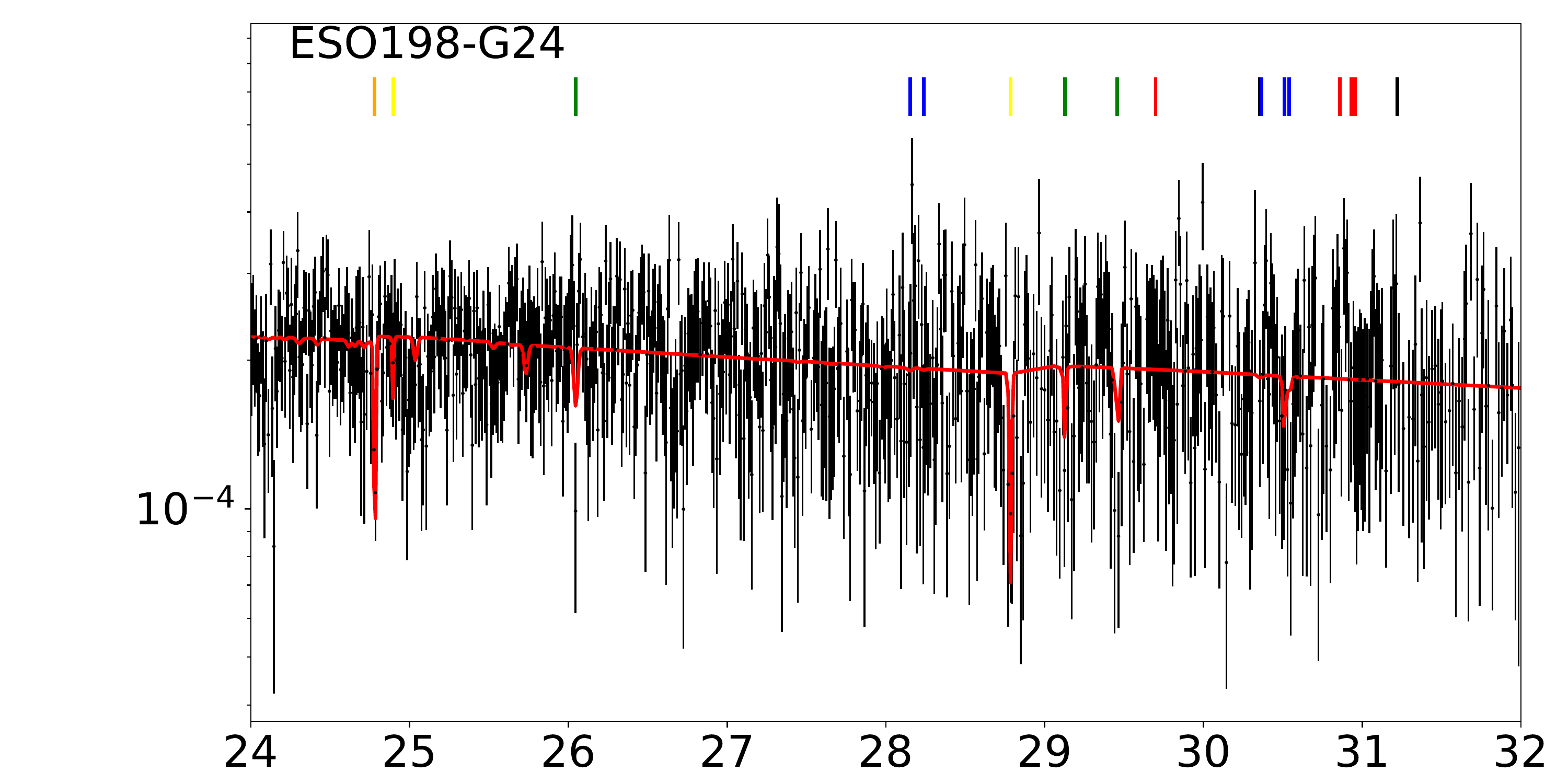} 
\hspace{-2.5mm}
\includegraphics[scale=0.195]{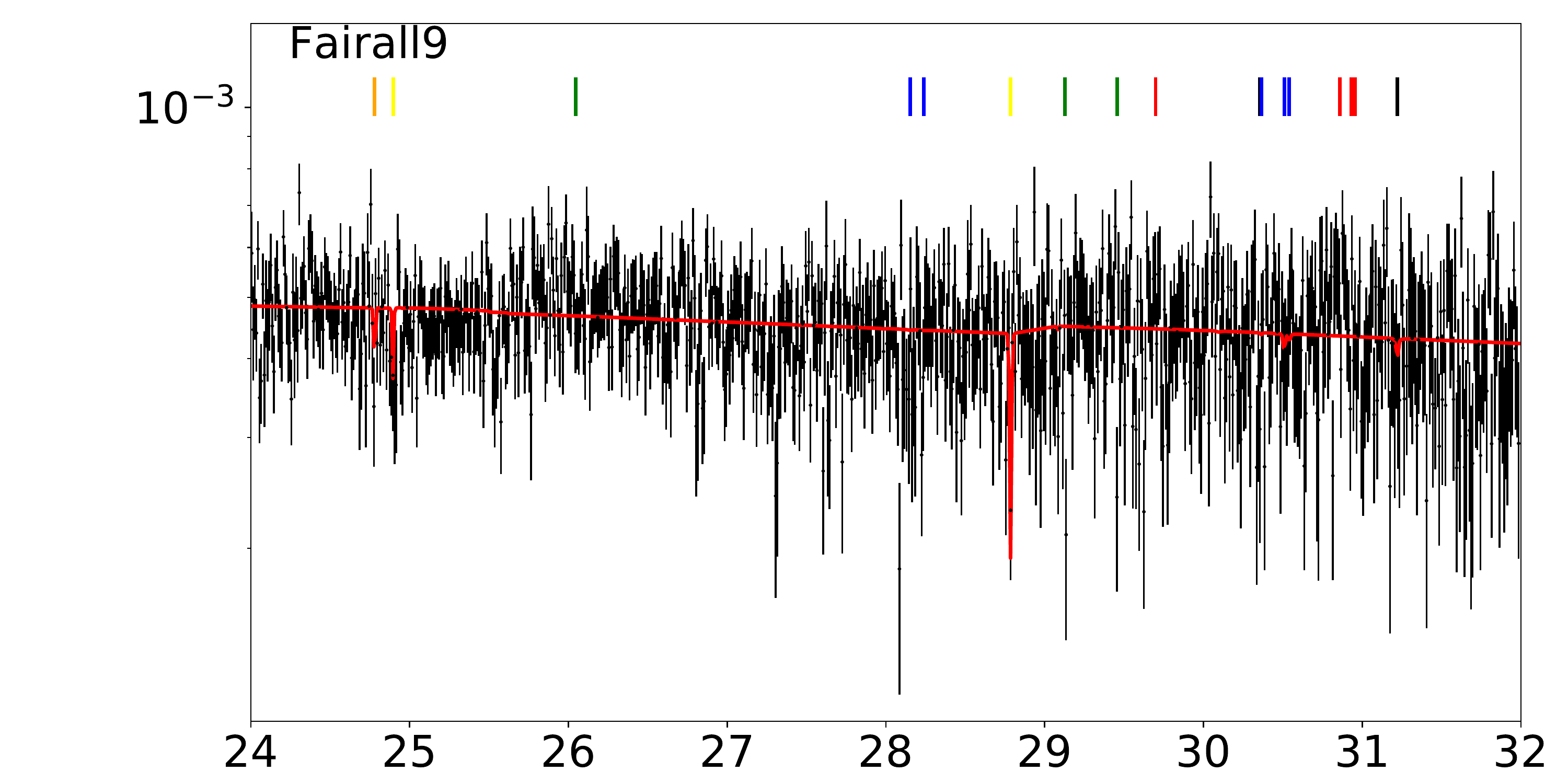}\\
\includegraphics[scale=0.195]{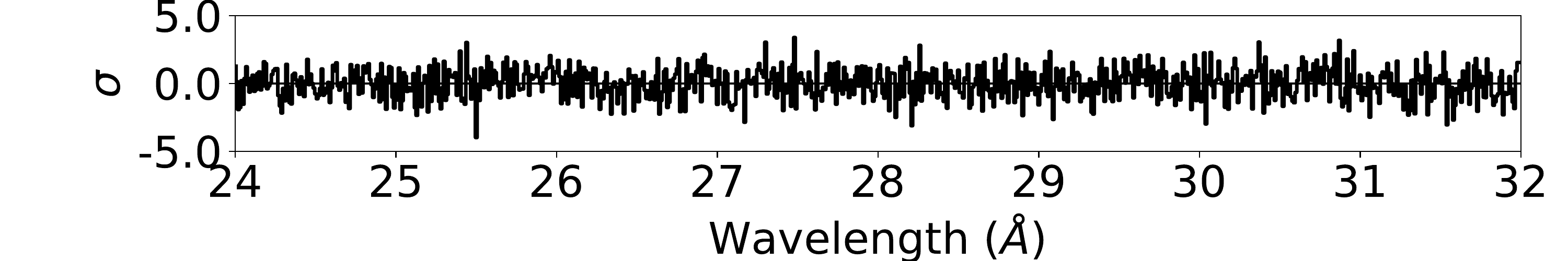}
\hspace{-2.5mm}
\includegraphics[scale=0.195]{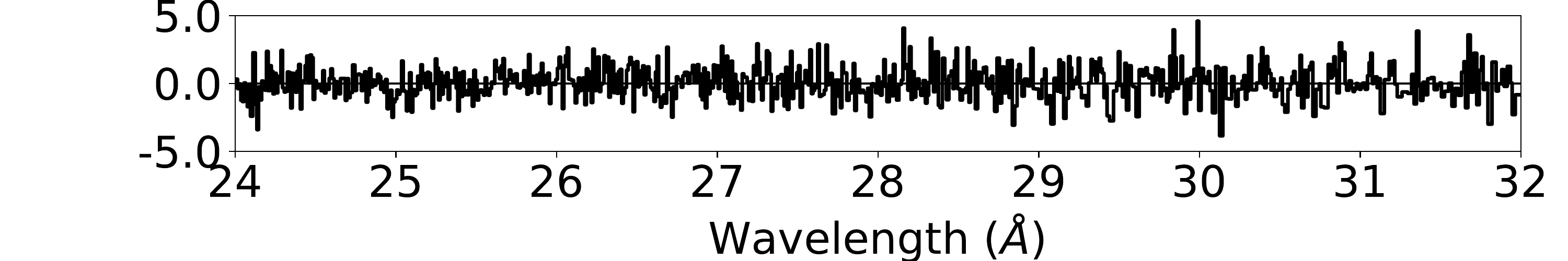} 
\hspace{-2.5mm}
\includegraphics[scale=0.195]{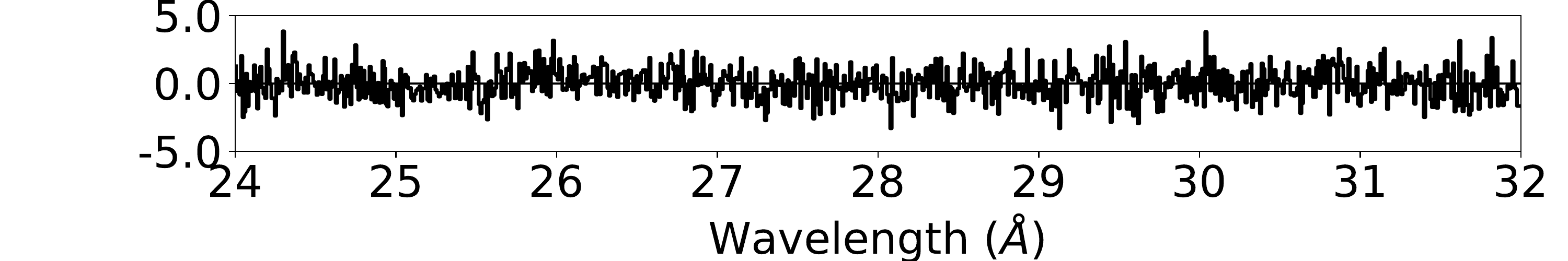}\\
      \caption{Best fit results in the N K-edge photoabsorption region for the Extragalactic sample described in Table~\ref{tab_ex}. In each panel, the black data points are the observations, while the solid red lines correspond to the best-fit models.}\label{fig_fits_ex1}
   \end{figure*}

         \begin{figure*}
          \centering
\includegraphics[scale=0.195]{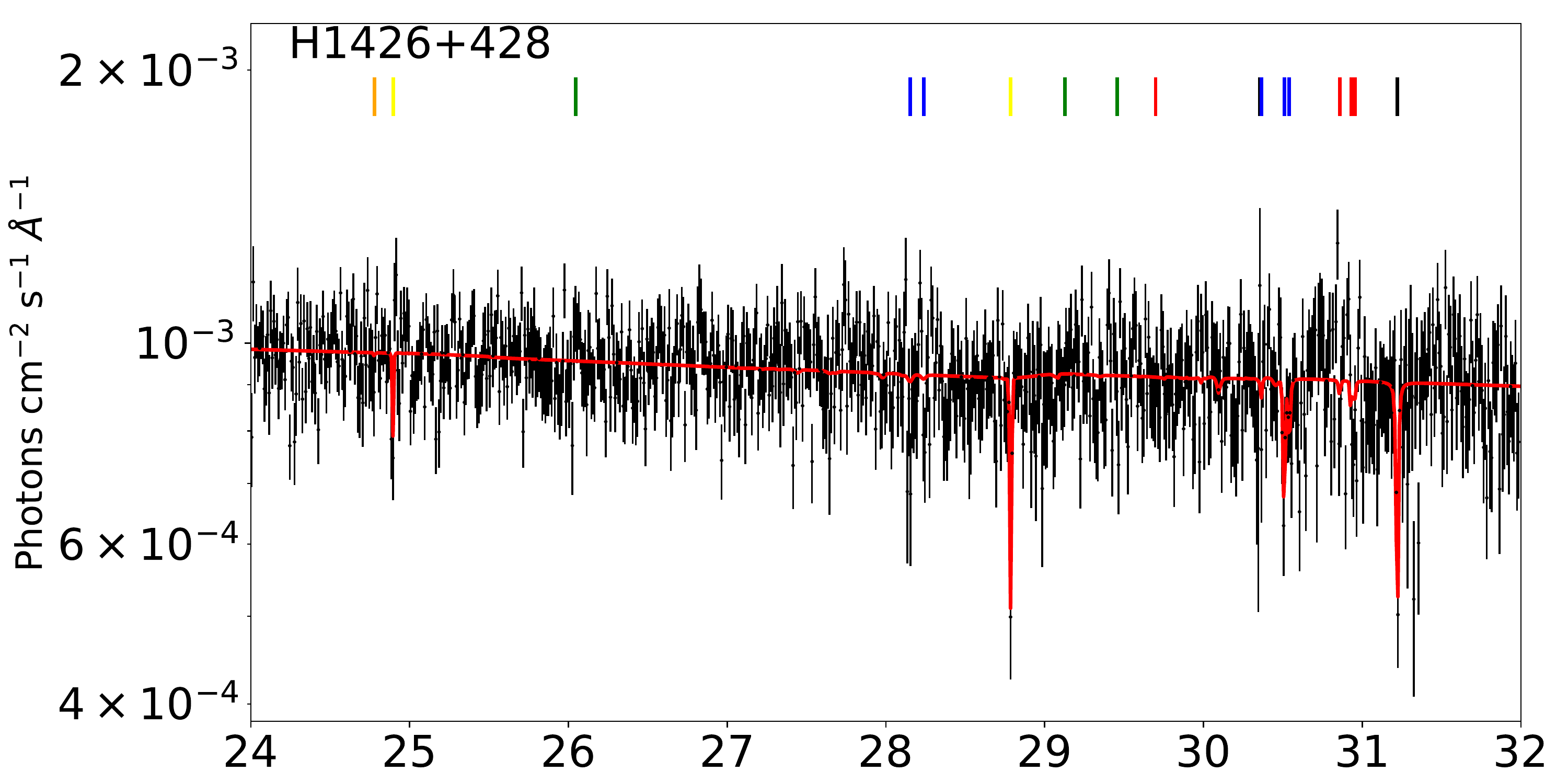}
\hspace{-2.5mm}
\includegraphics[scale=0.195]{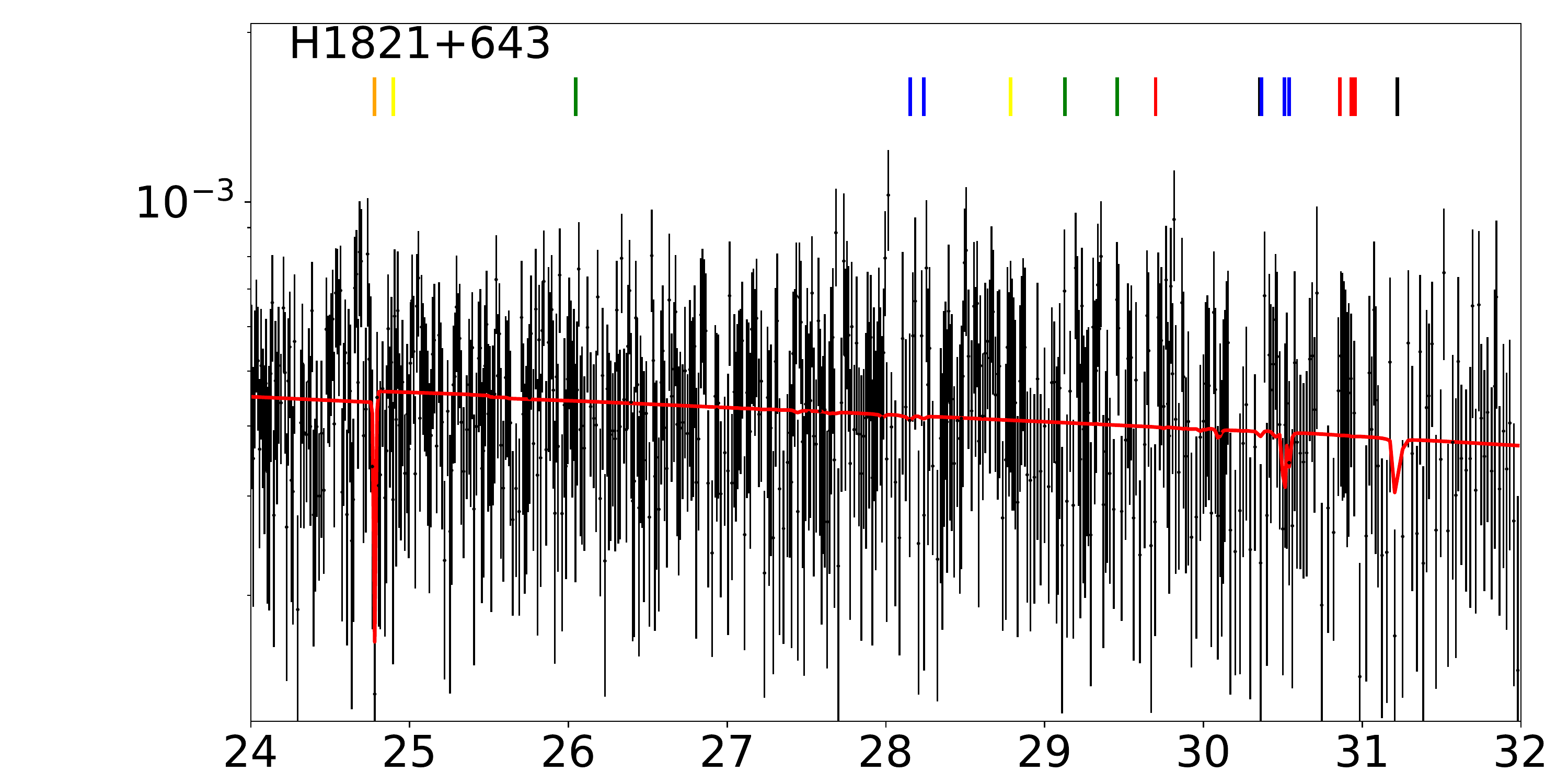} 
\hspace{-2.5mm}
\includegraphics[scale=0.195]{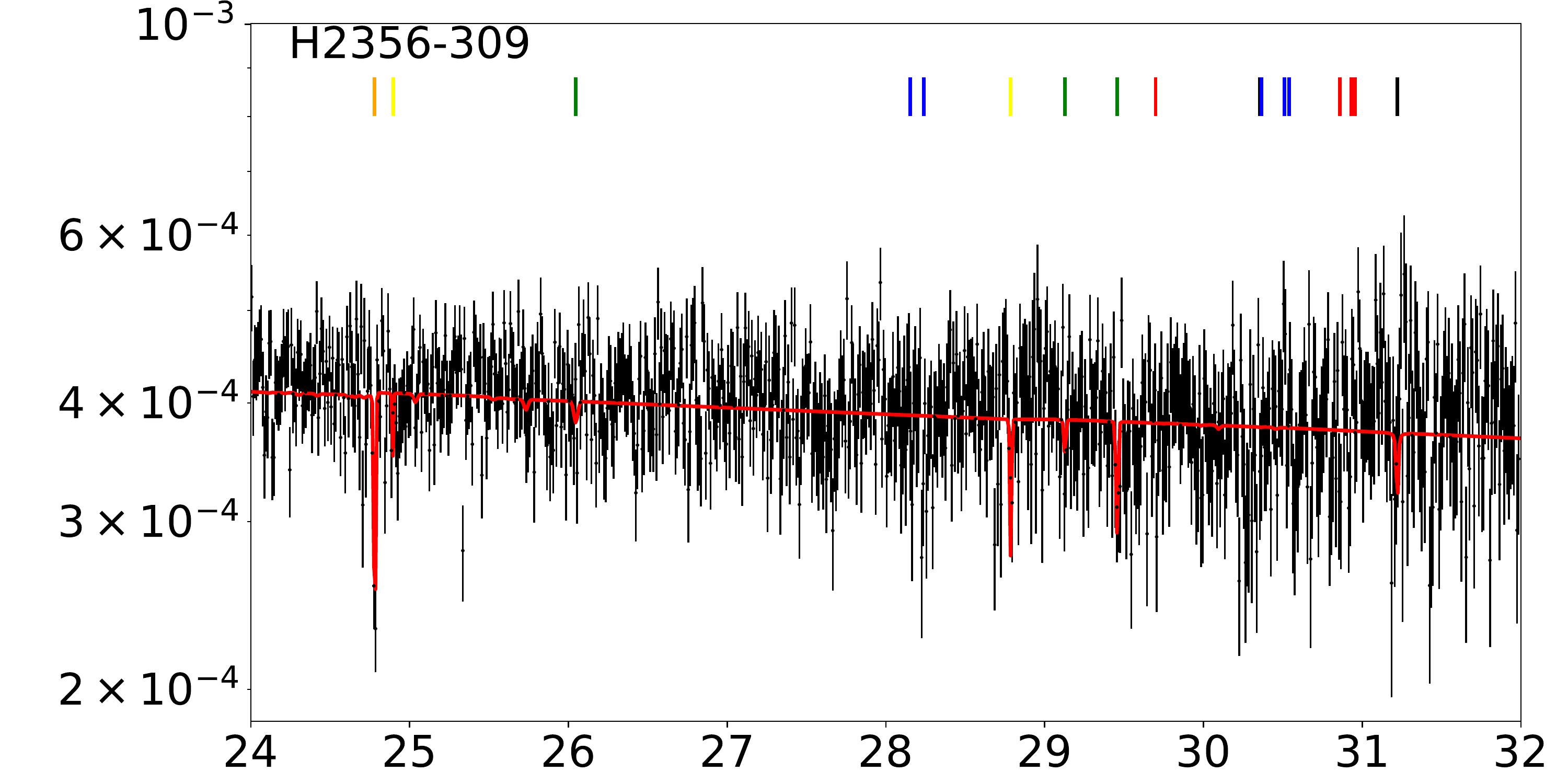}\\
\includegraphics[scale=0.195]{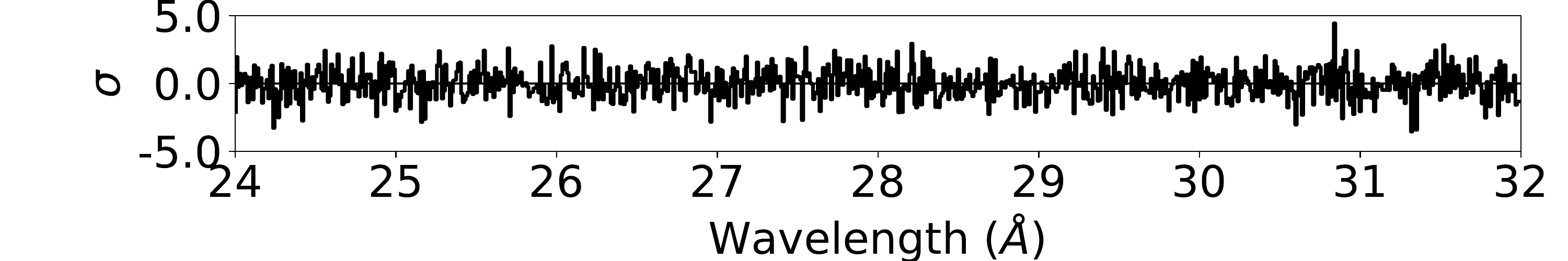}
\hspace{-2.5mm}
\includegraphics[scale=0.195]{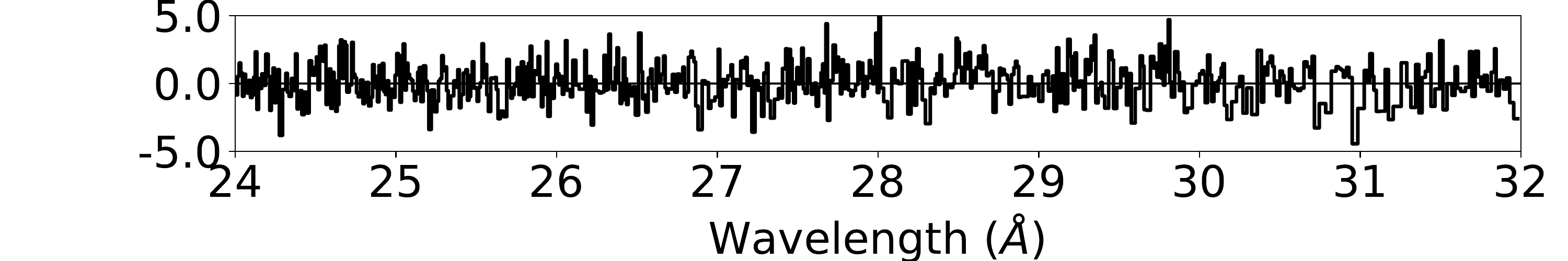} 
\hspace{-2.5mm}
\includegraphics[scale=0.195]{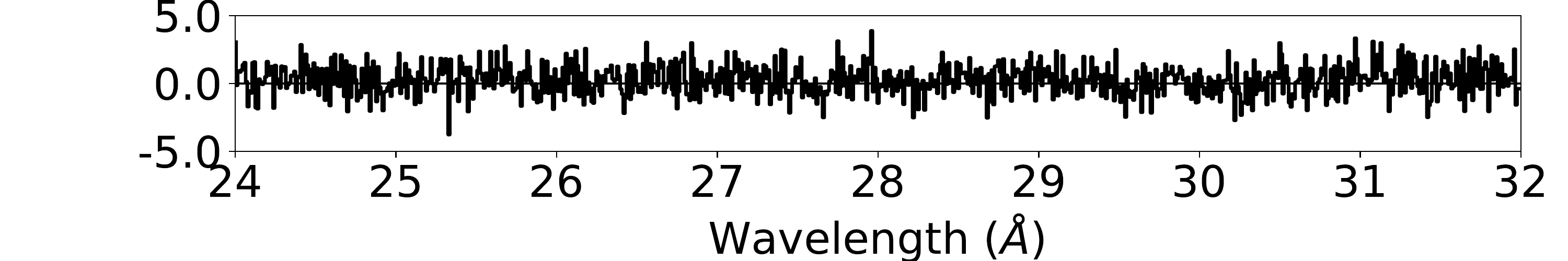}\\

\includegraphics[scale=0.195]{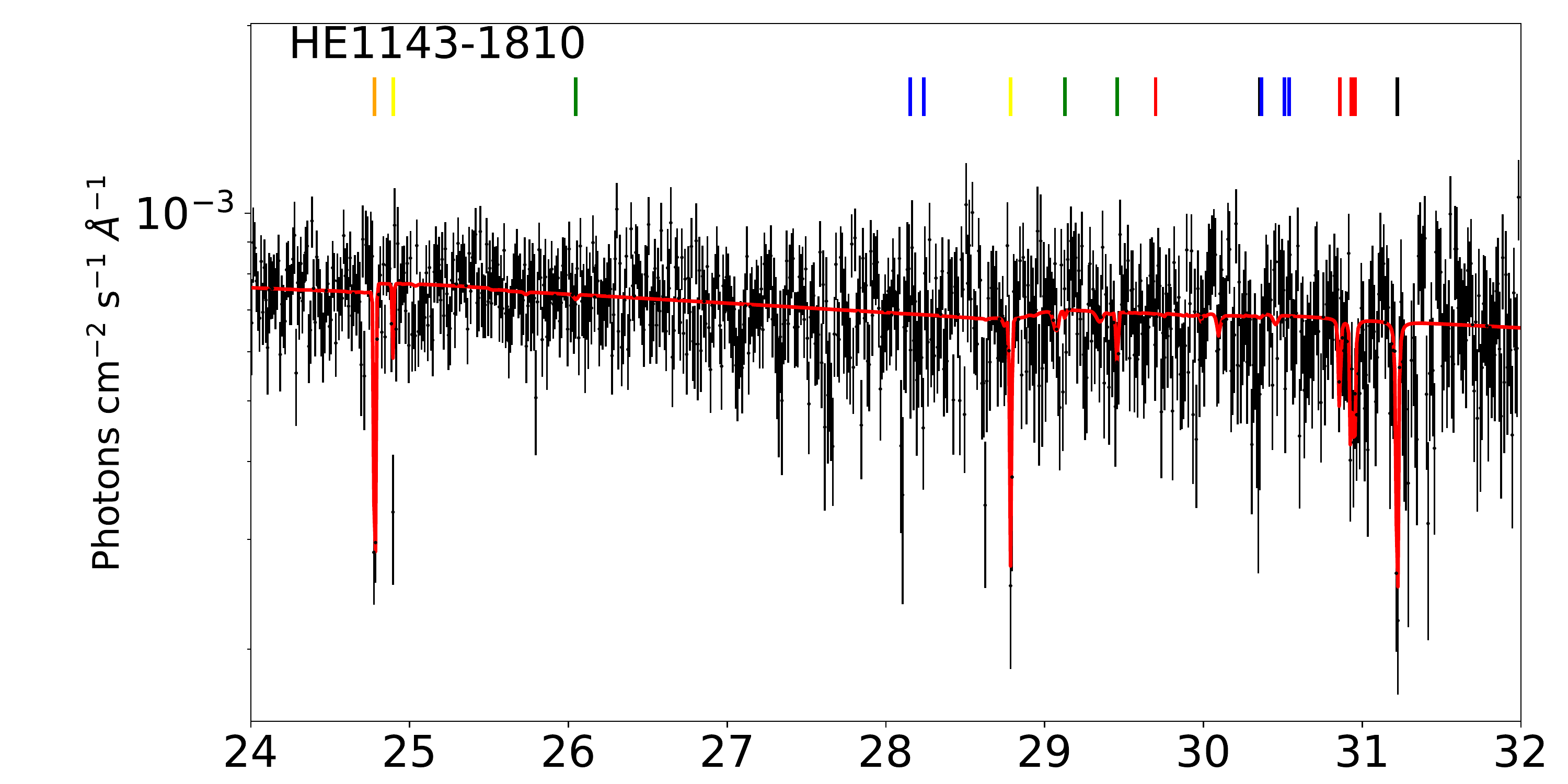}
\hspace{-2.5mm}
\includegraphics[scale=0.195]{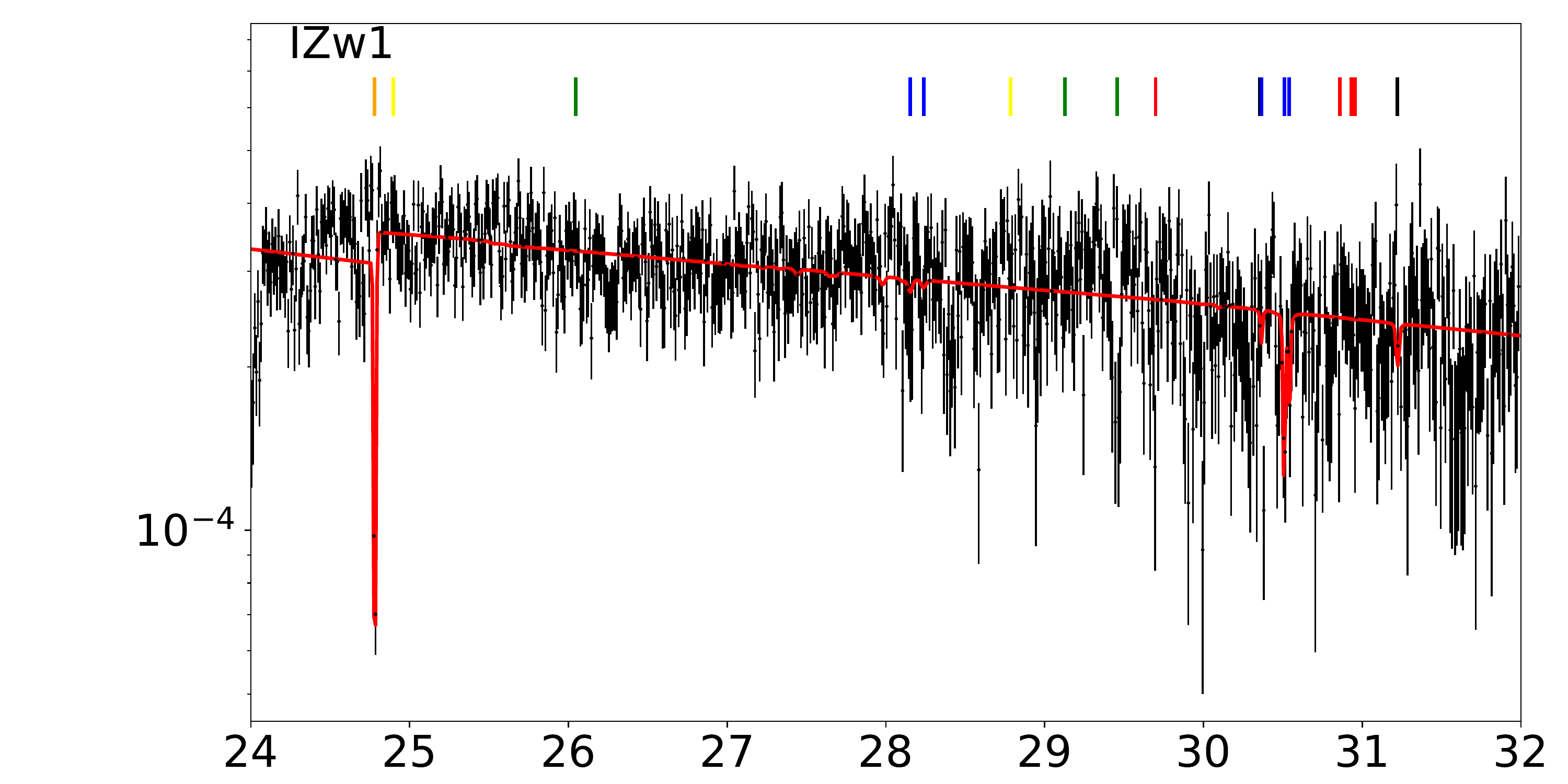} 
\hspace{-2.5mm}
\includegraphics[scale=0.195]{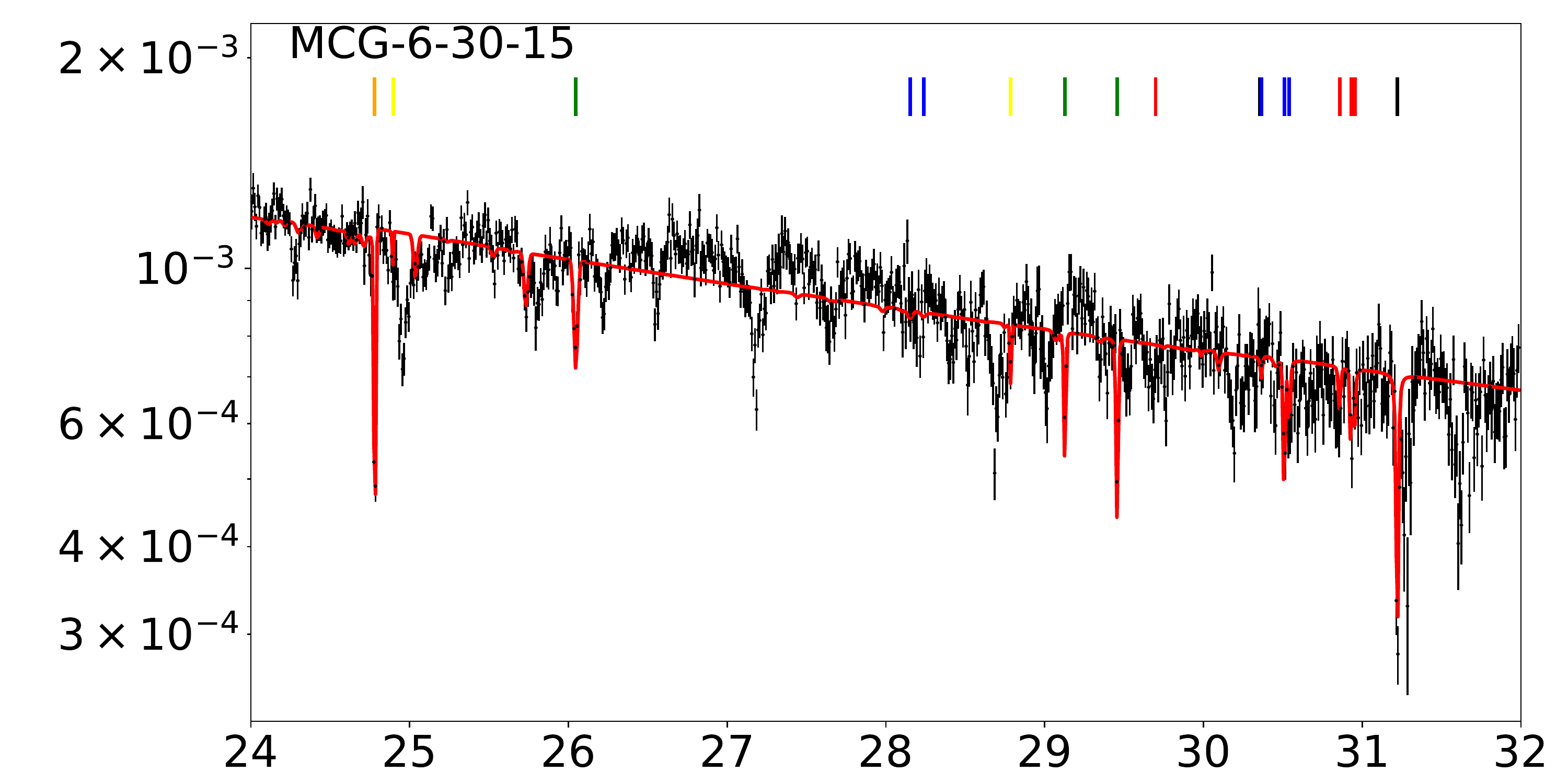}\\
\includegraphics[scale=0.195]{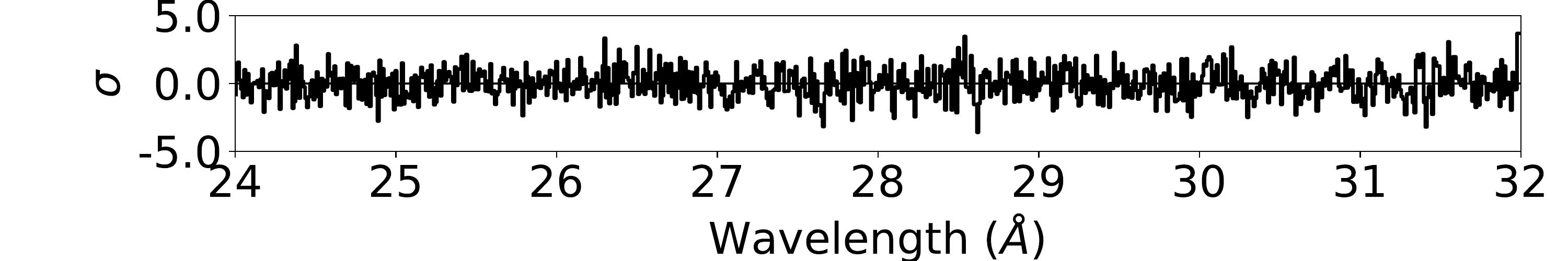}
\hspace{-2.5mm}
\includegraphics[scale=0.195]{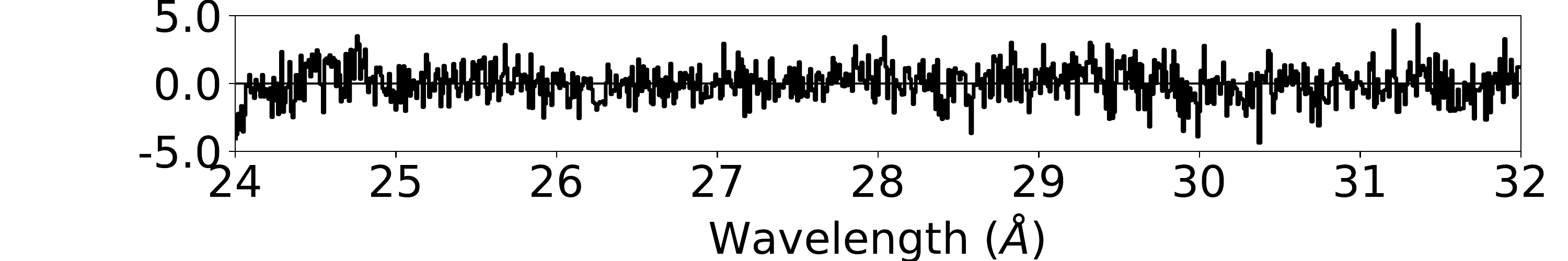} 
\hspace{-2.5mm}
\includegraphics[scale=0.195]{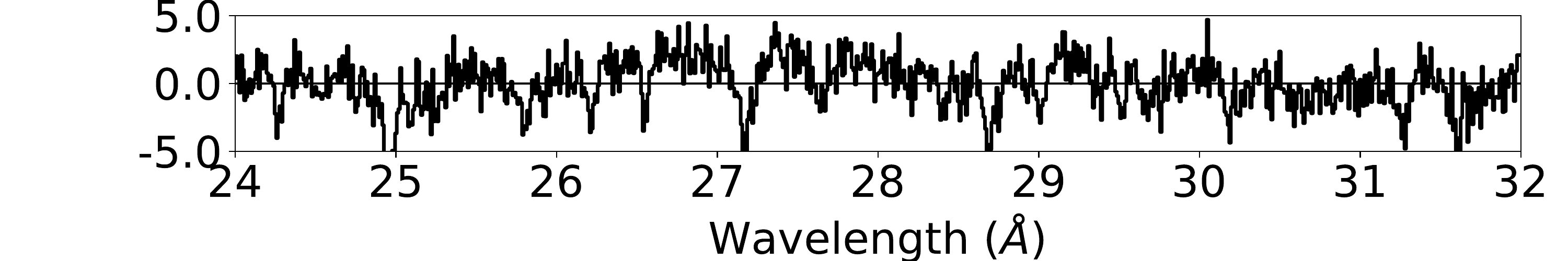}\\

\includegraphics[scale=0.195]{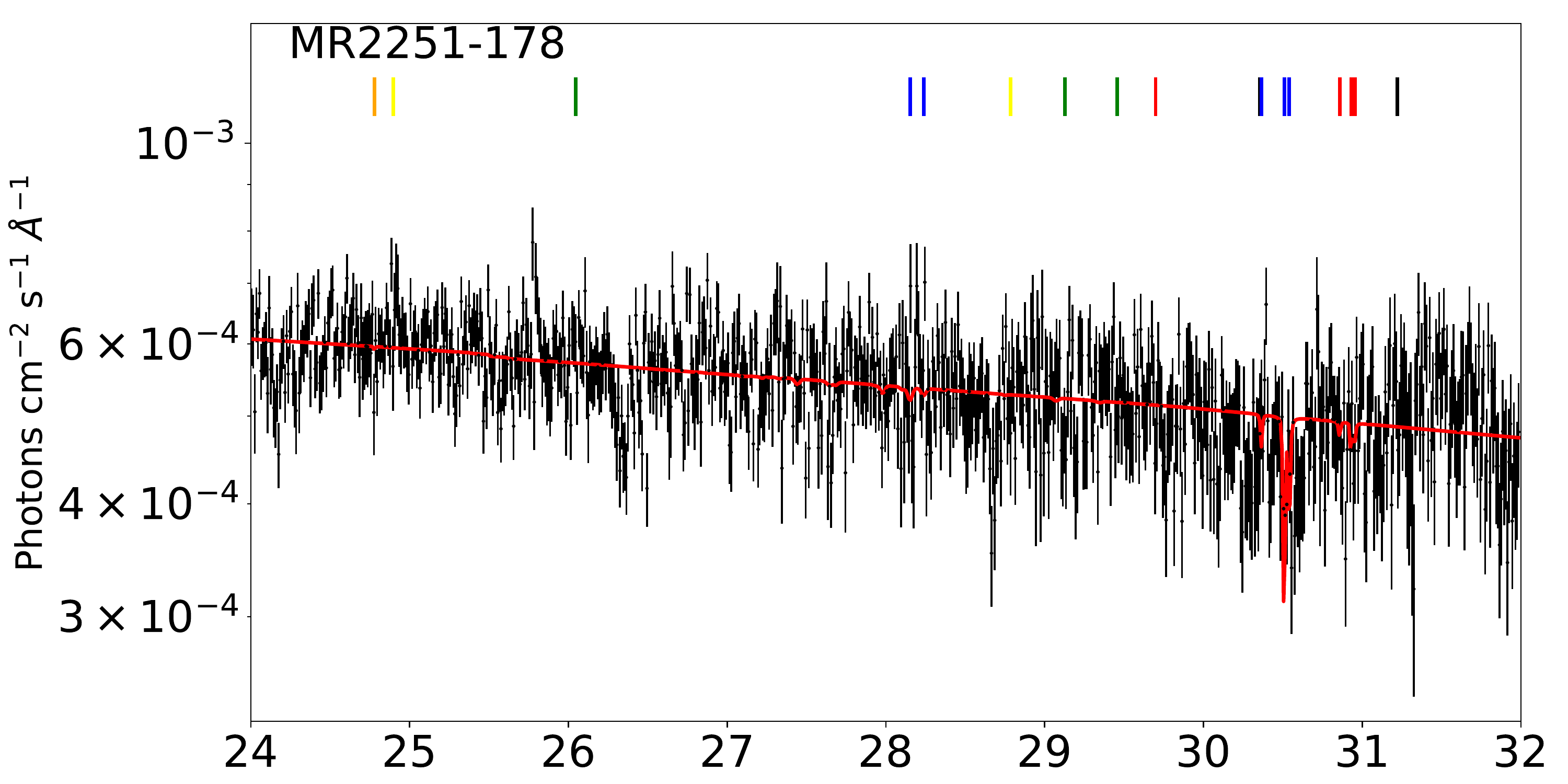}
\hspace{-2.5mm}
\includegraphics[scale=0.195]{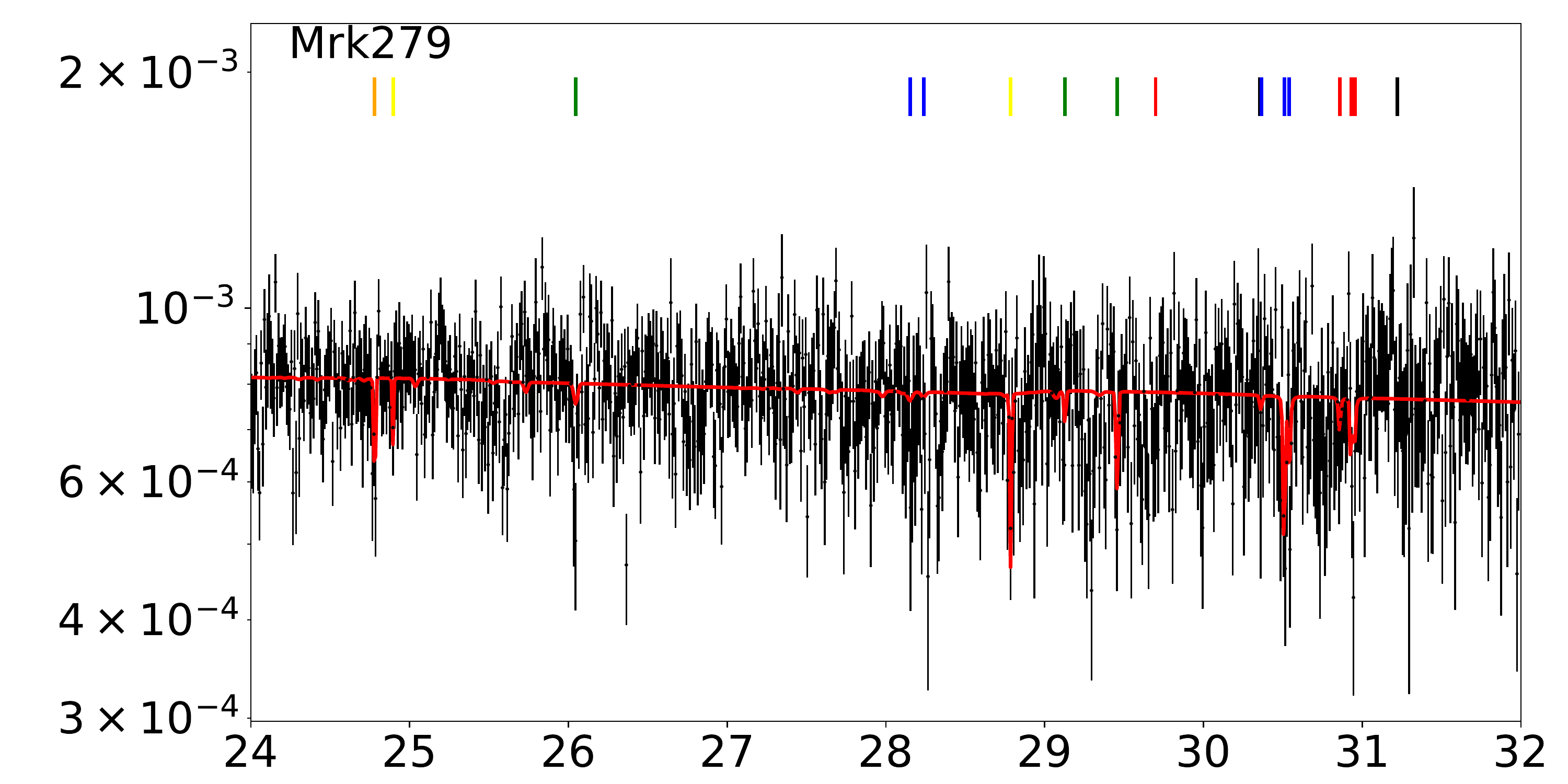} 
\hspace{-2.5mm}
\includegraphics[scale=0.195]{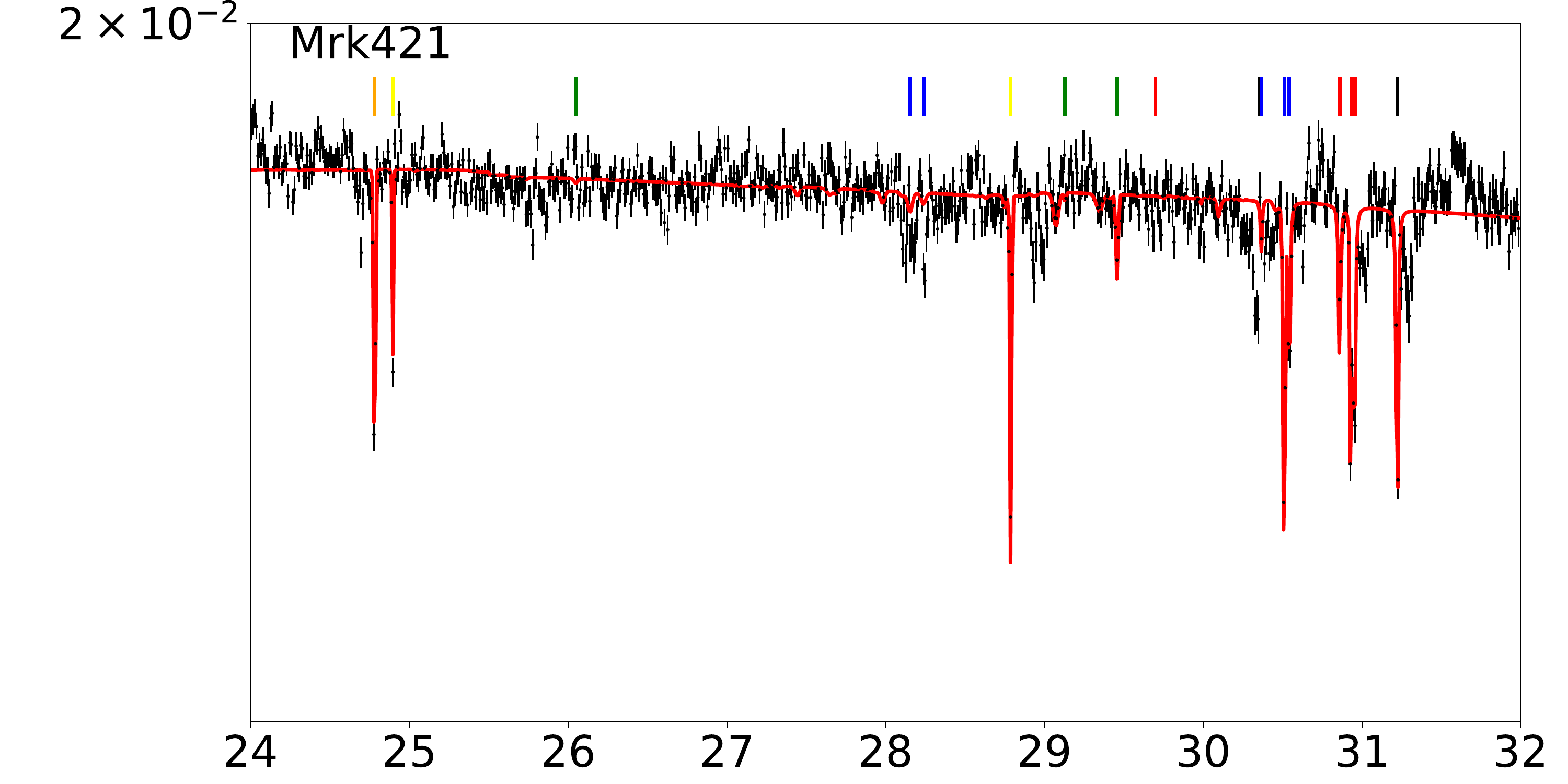}\\
\includegraphics[scale=0.195]{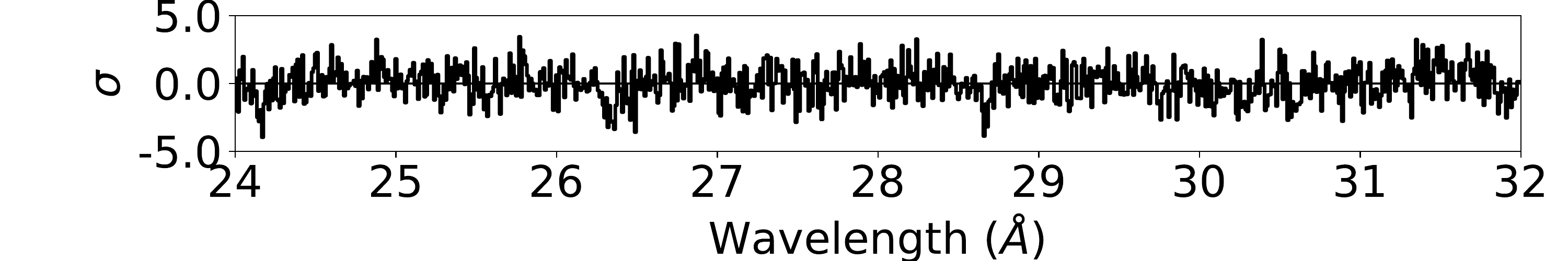}
\hspace{-2.5mm}
\includegraphics[scale=0.195]{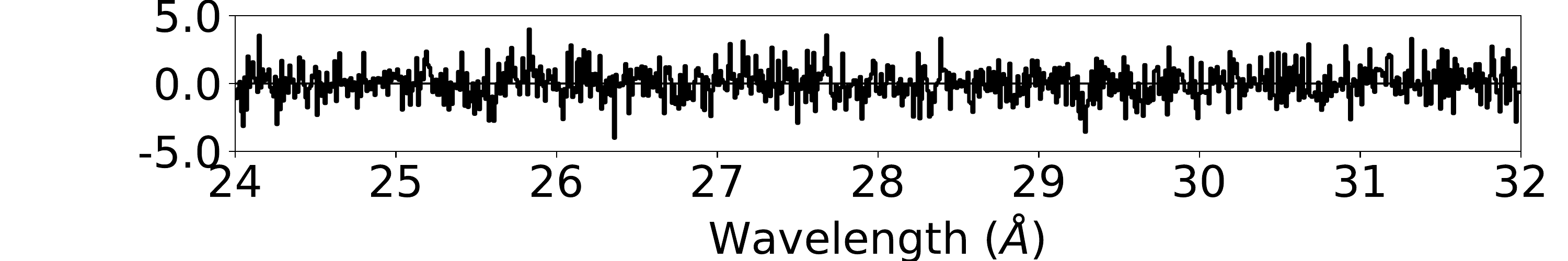} 
\hspace{-2.5mm}
\includegraphics[scale=0.195]{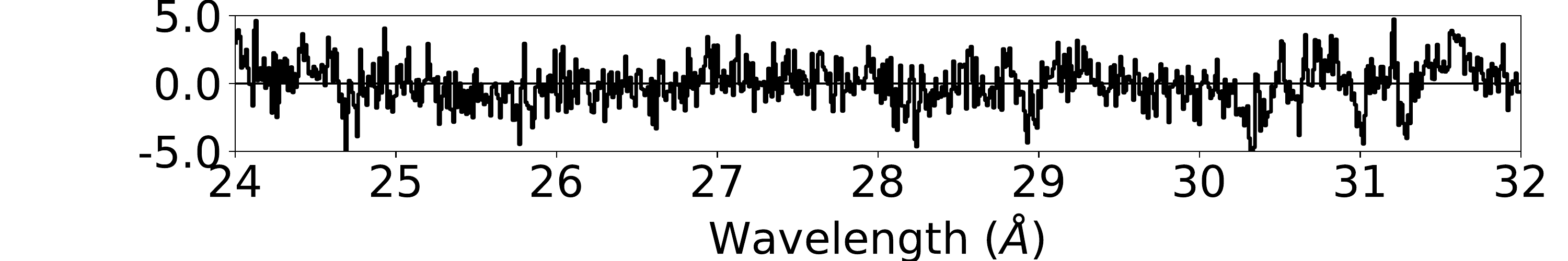}\\

\includegraphics[scale=0.195]{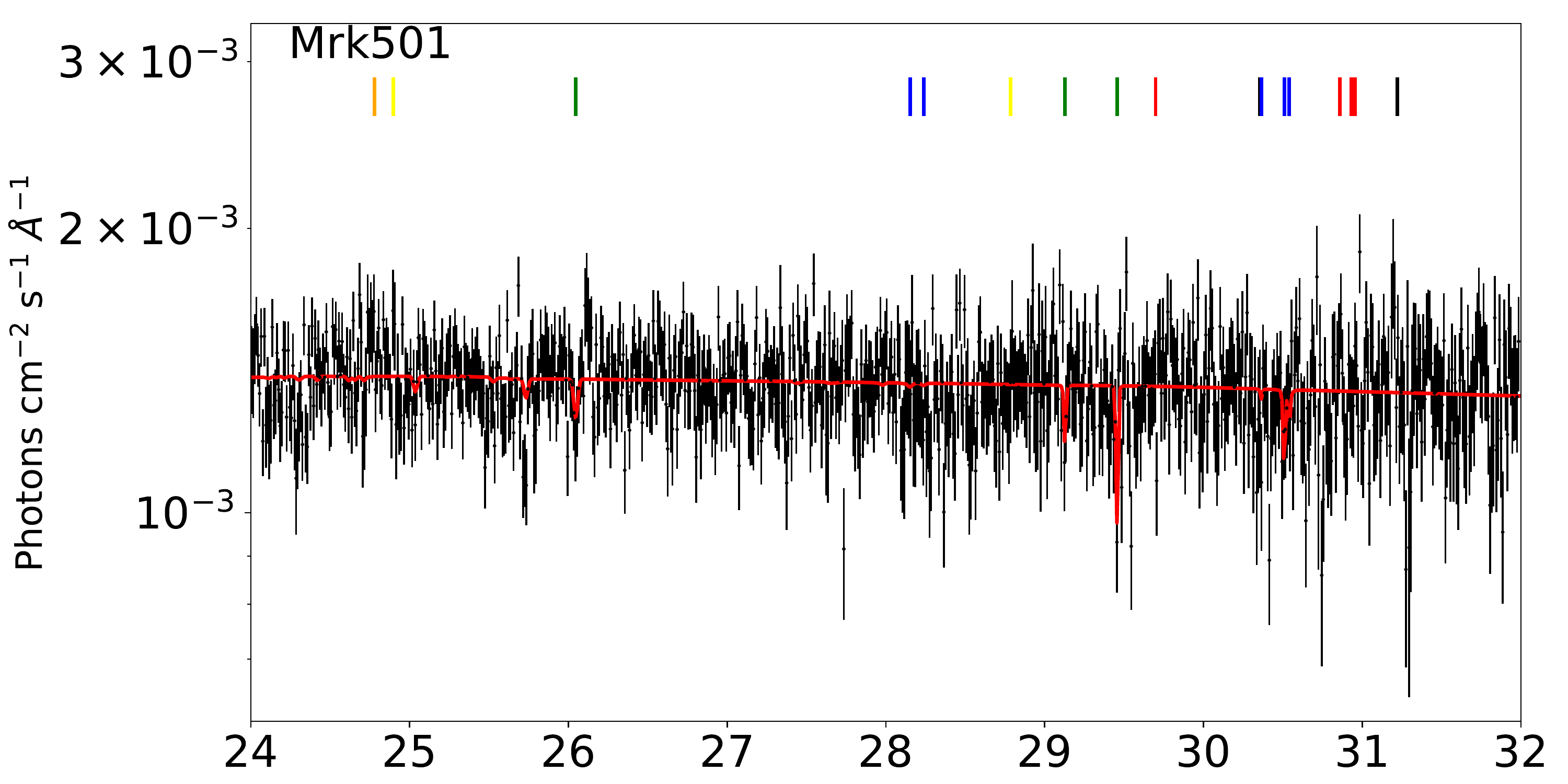}
\hspace{-2.5mm}
\includegraphics[scale=0.195]{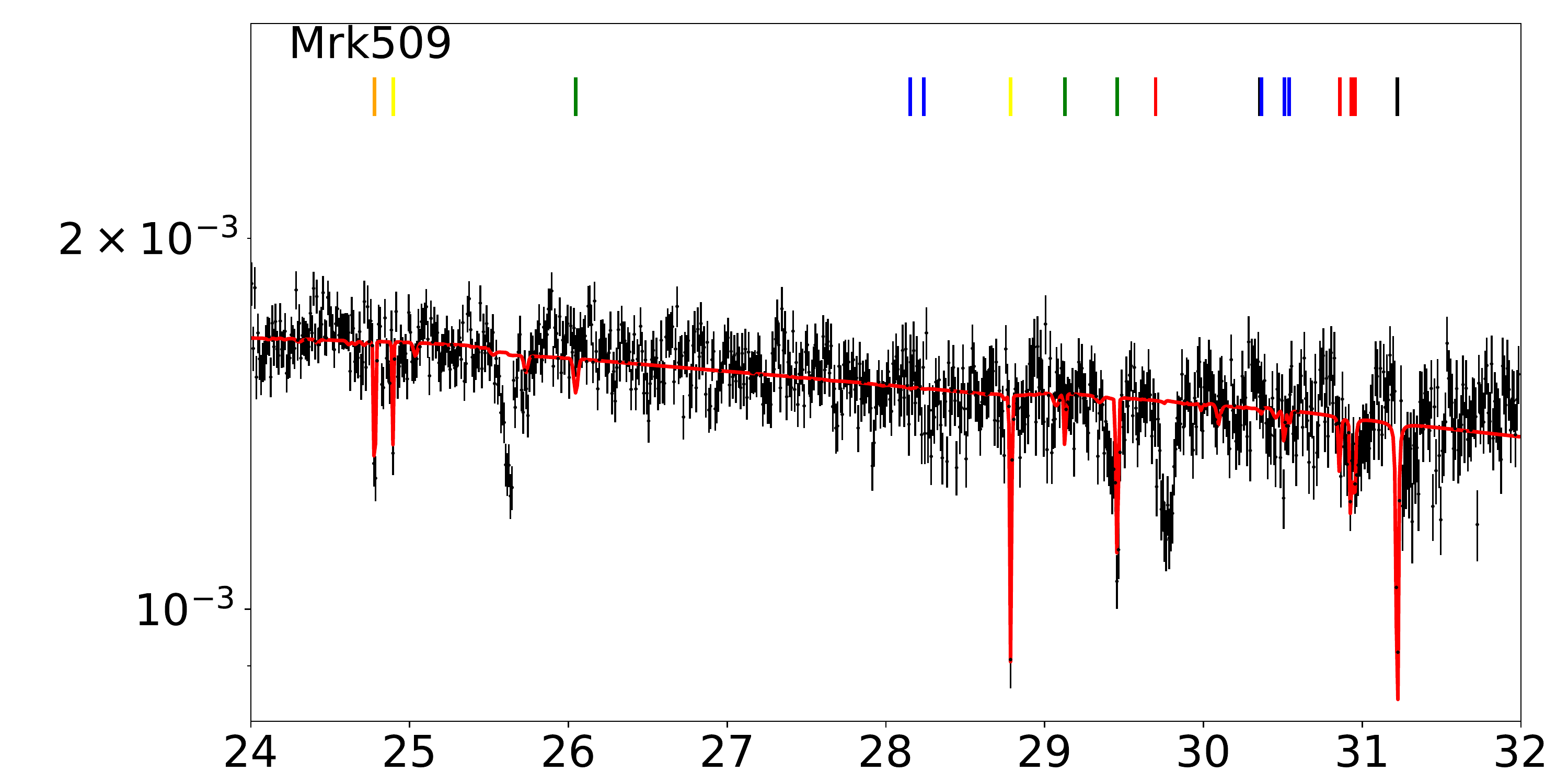} 
\hspace{-2.5mm}
\includegraphics[scale=0.195]{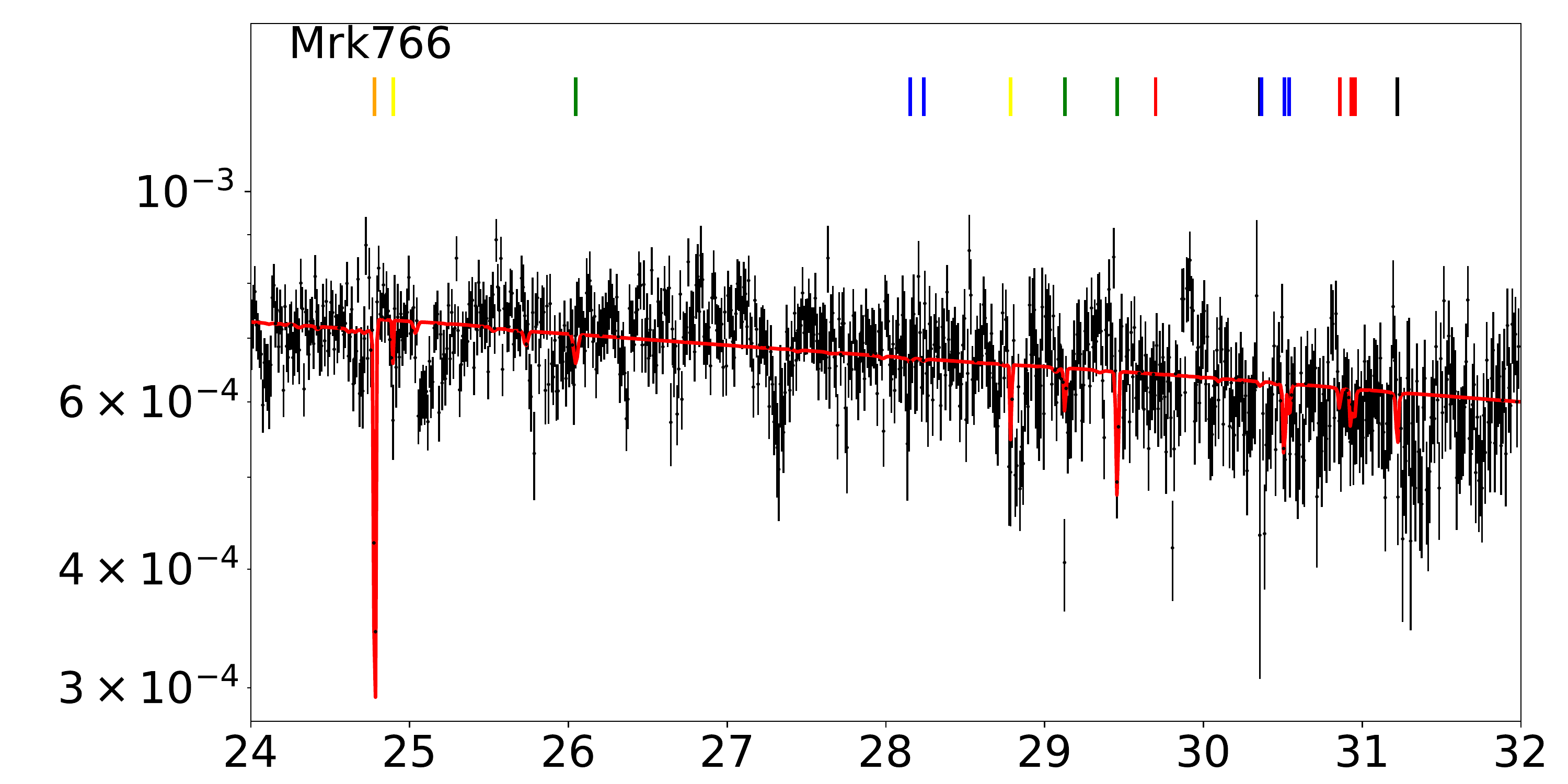}\\
\includegraphics[scale=0.195]{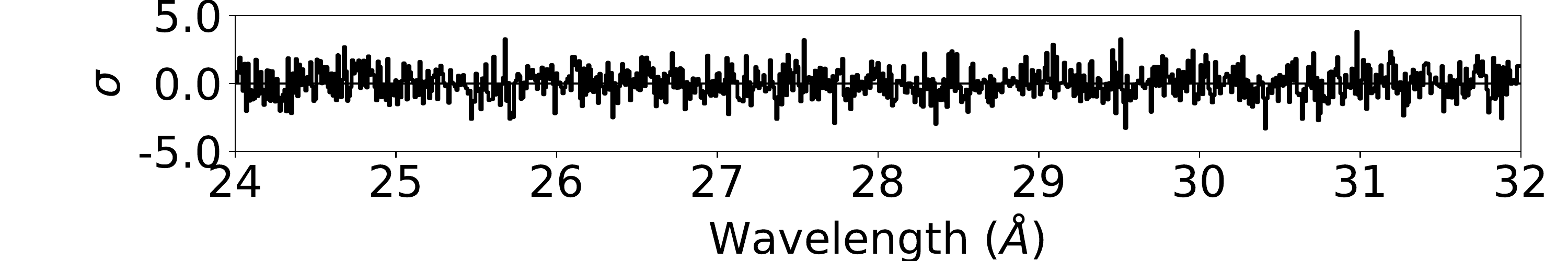}
\hspace{-2.5mm}
\includegraphics[scale=0.195]{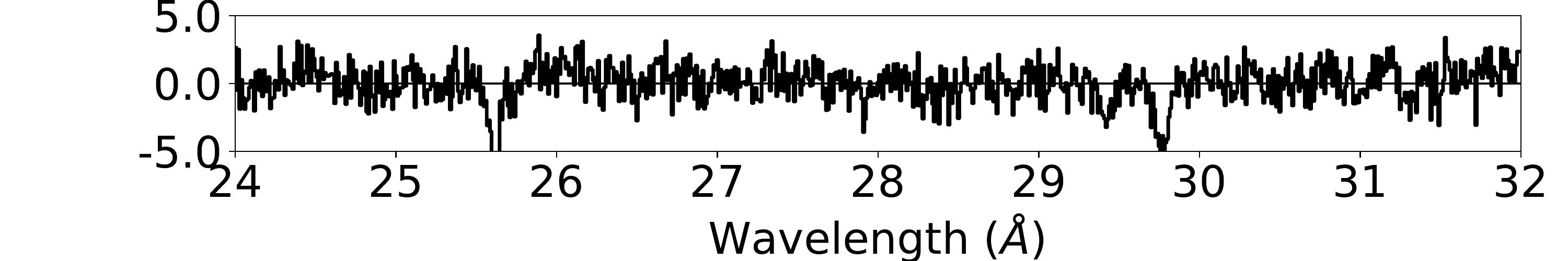} 
\hspace{-2.5mm}
\includegraphics[scale=0.195]{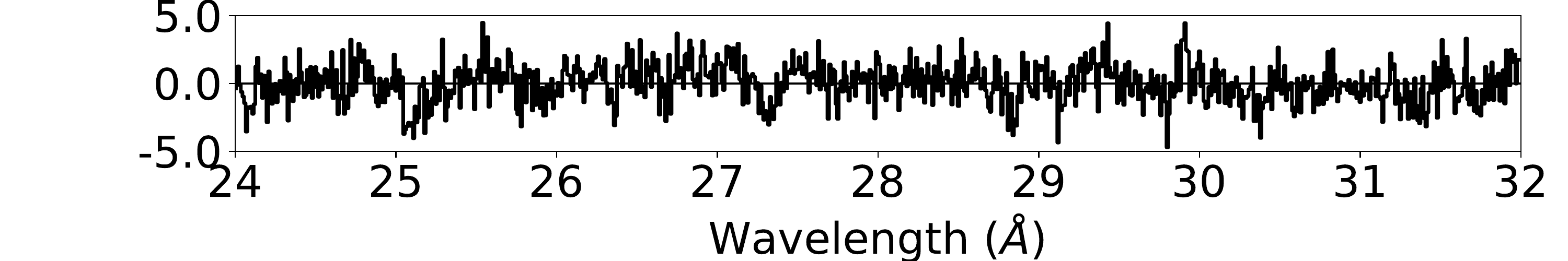}\\

\includegraphics[scale=0.195]{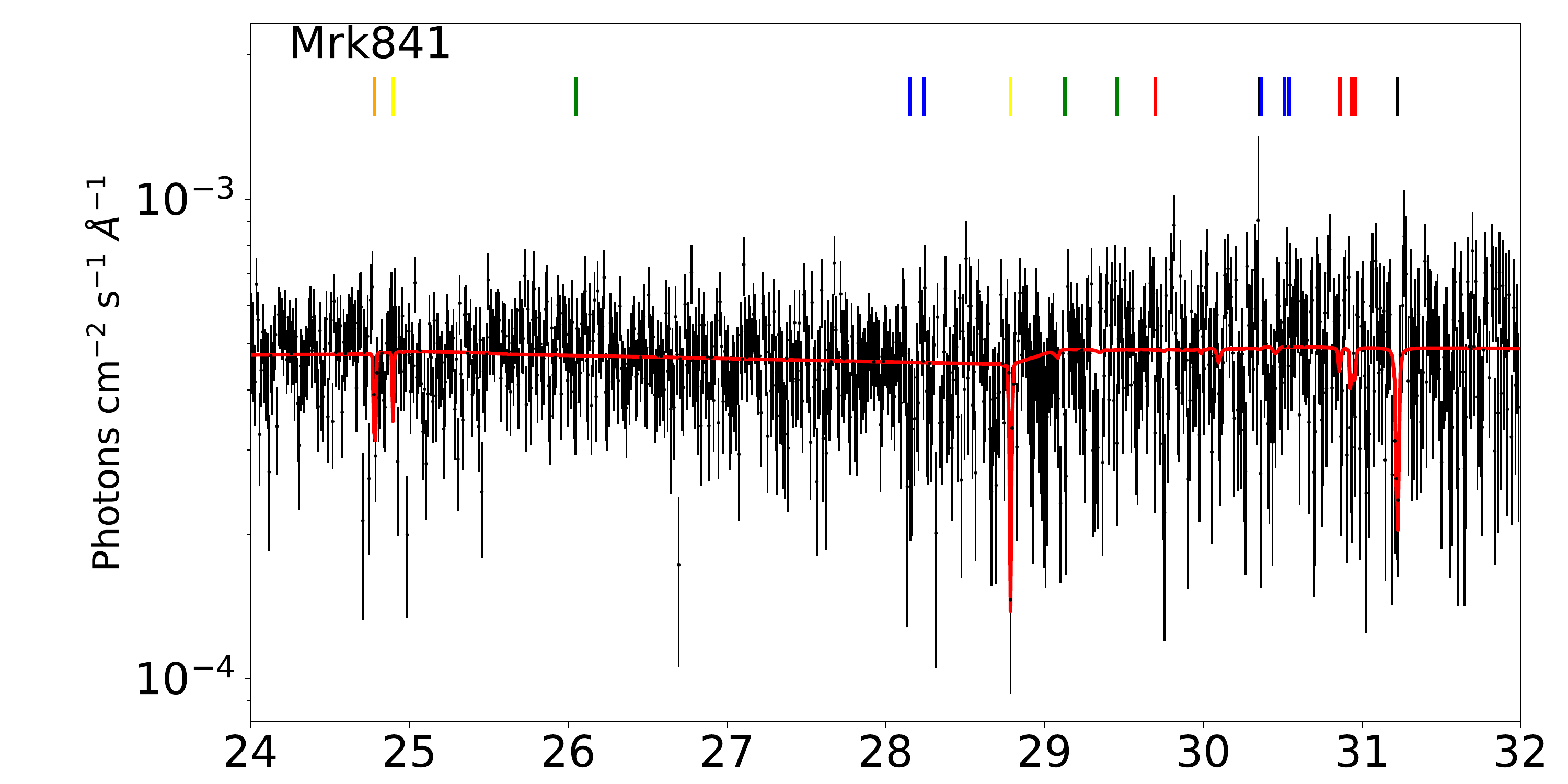}
\hspace{-2.5mm}
\includegraphics[scale=0.195]{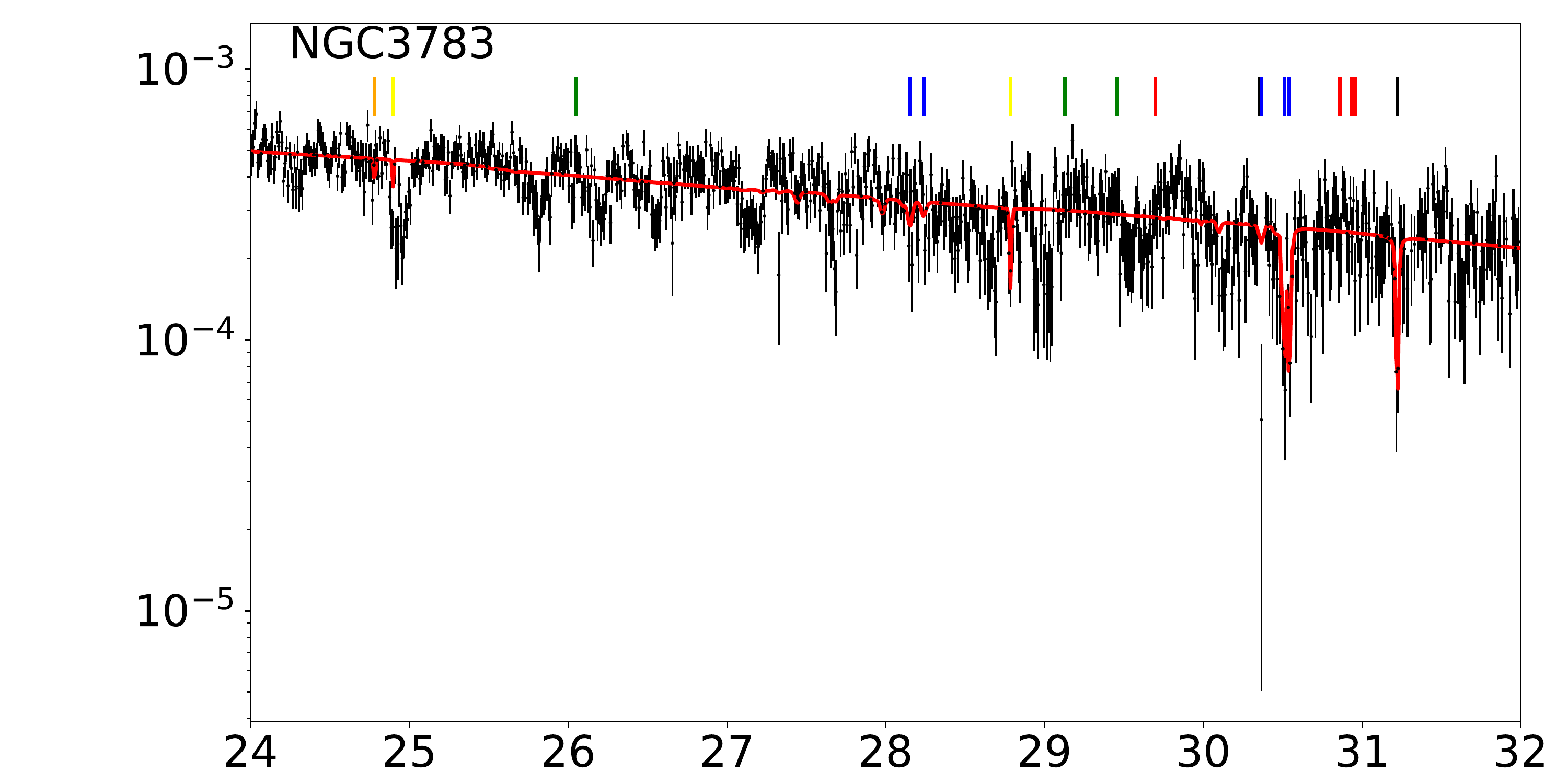}  
\hspace{-2.5mm}
\includegraphics[scale=0.195]{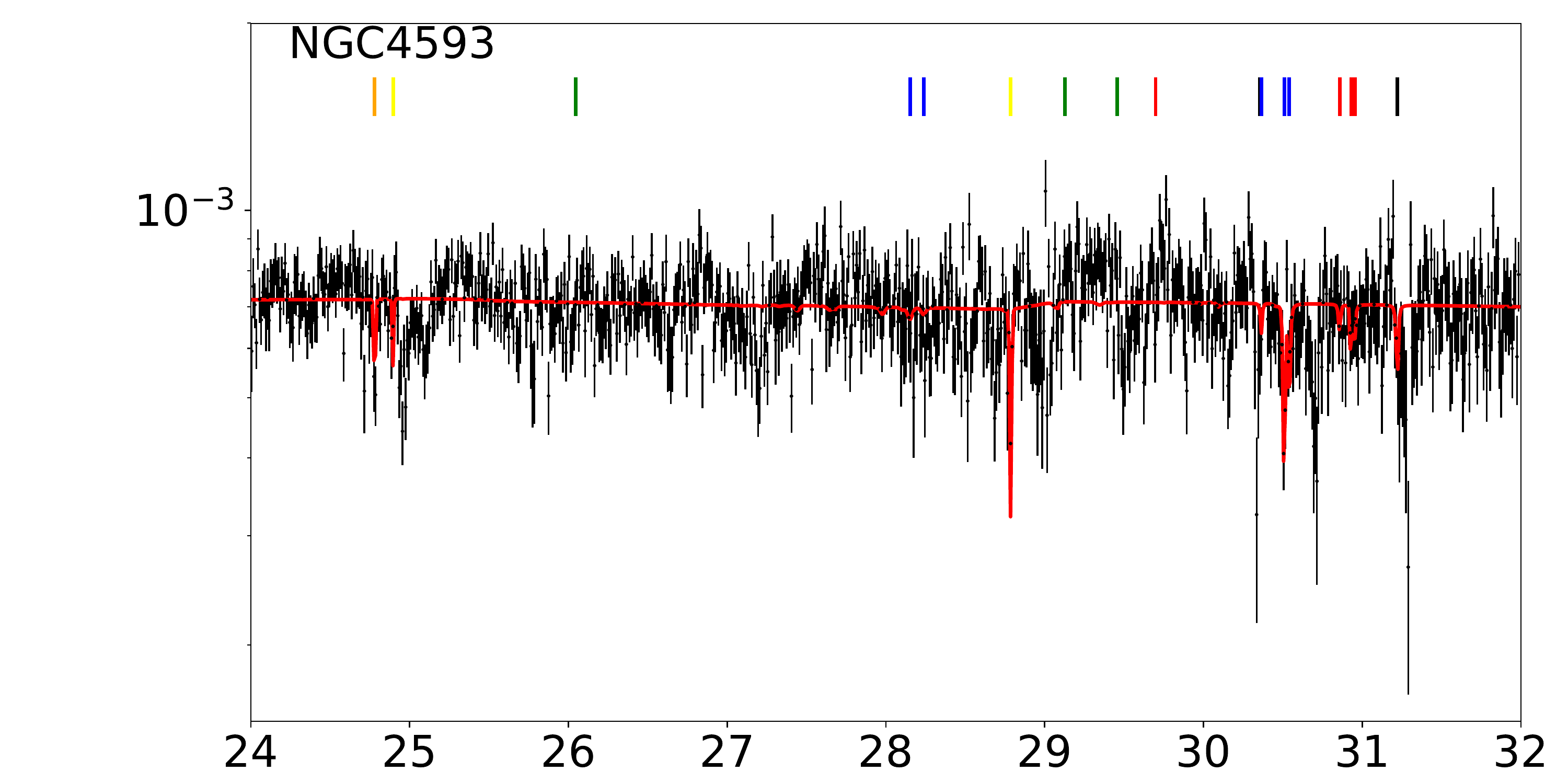}\\
\includegraphics[scale=0.195]{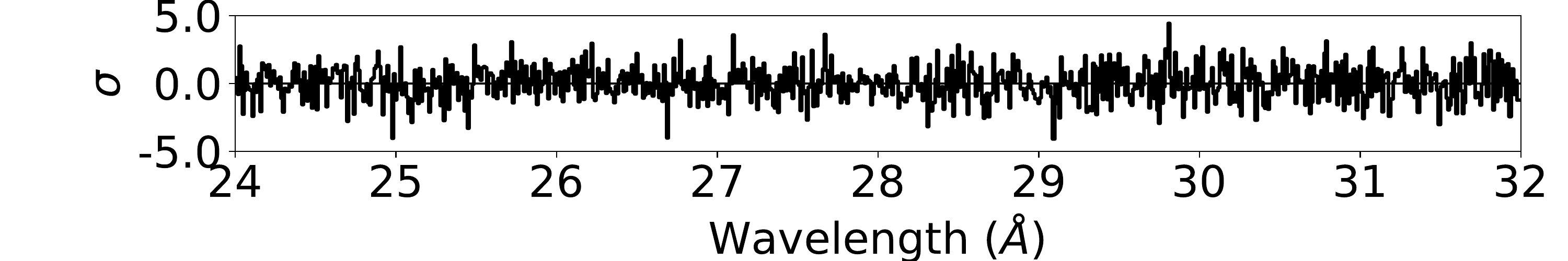}
\hspace{-2.5mm}
\includegraphics[scale=0.195]{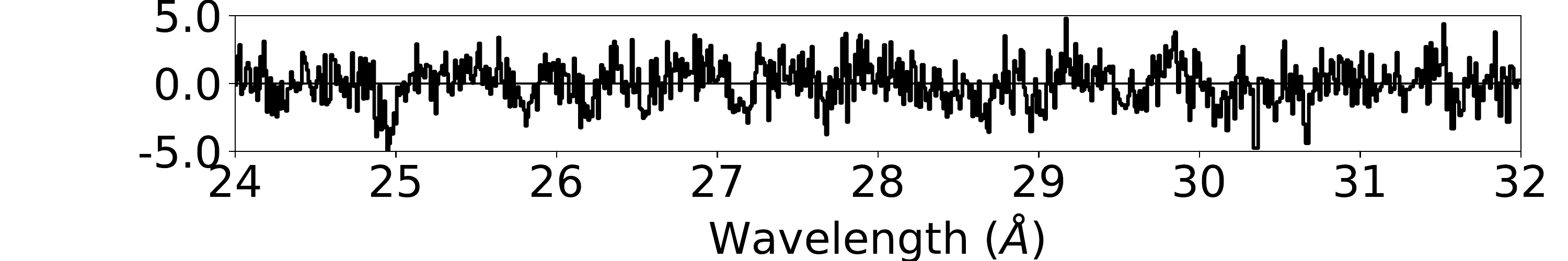} 
\hspace{-2.5mm}
\includegraphics[scale=0.195]{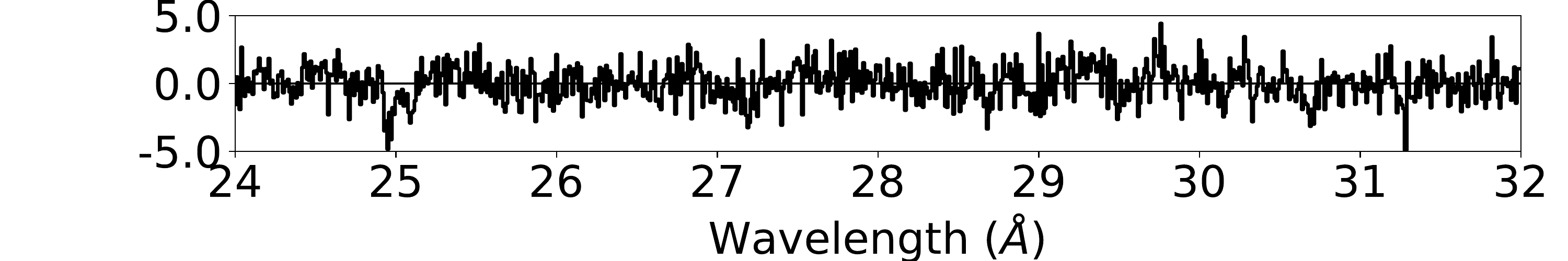}\\
      \contcaption{.}\label{fig_fits_ex2}
   \end{figure*}

        \begin{figure*}
          \centering

\includegraphics[scale=0.195]{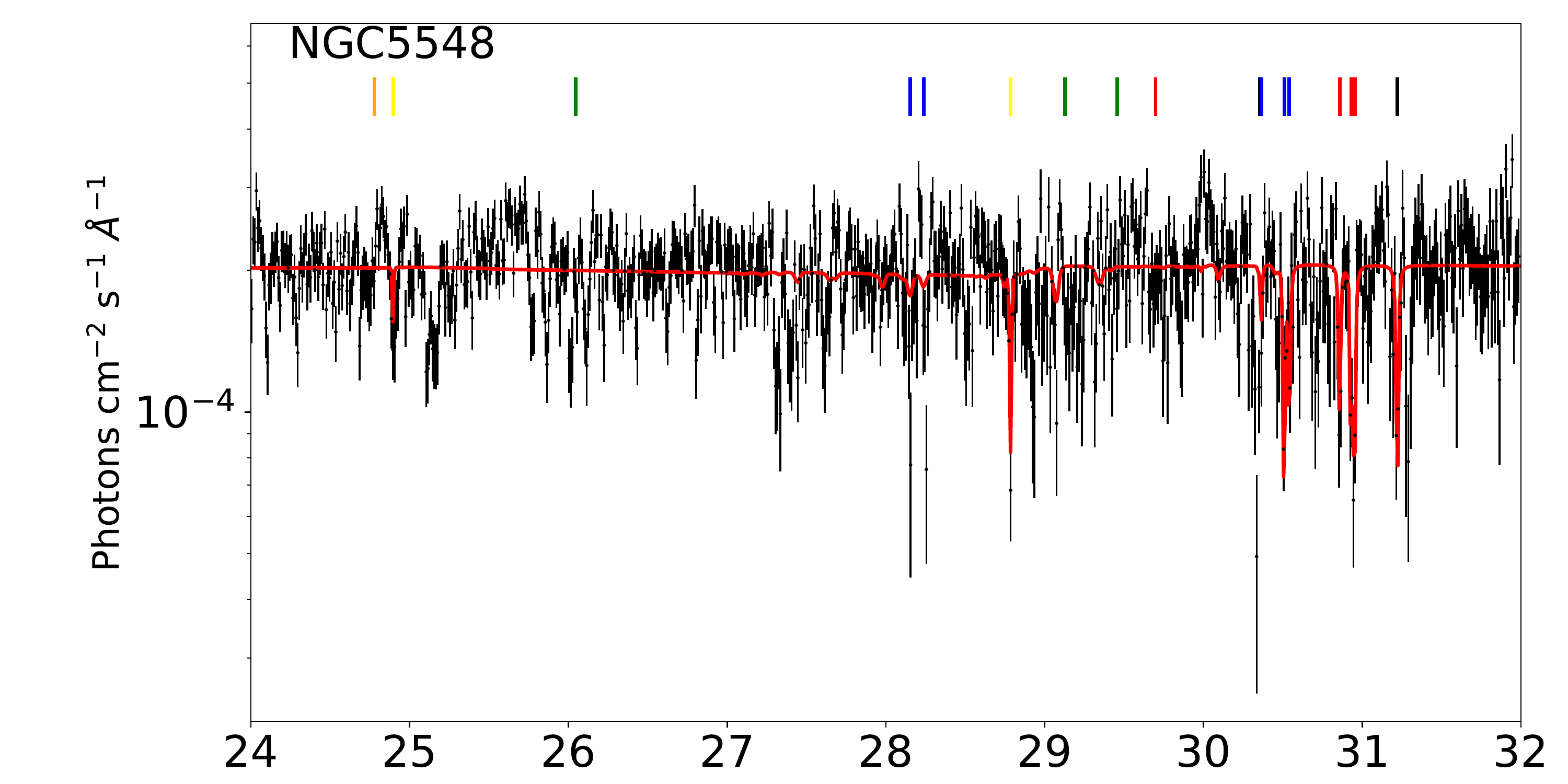}
\hspace{-2.5mm} 
\includegraphics[scale=0.195]{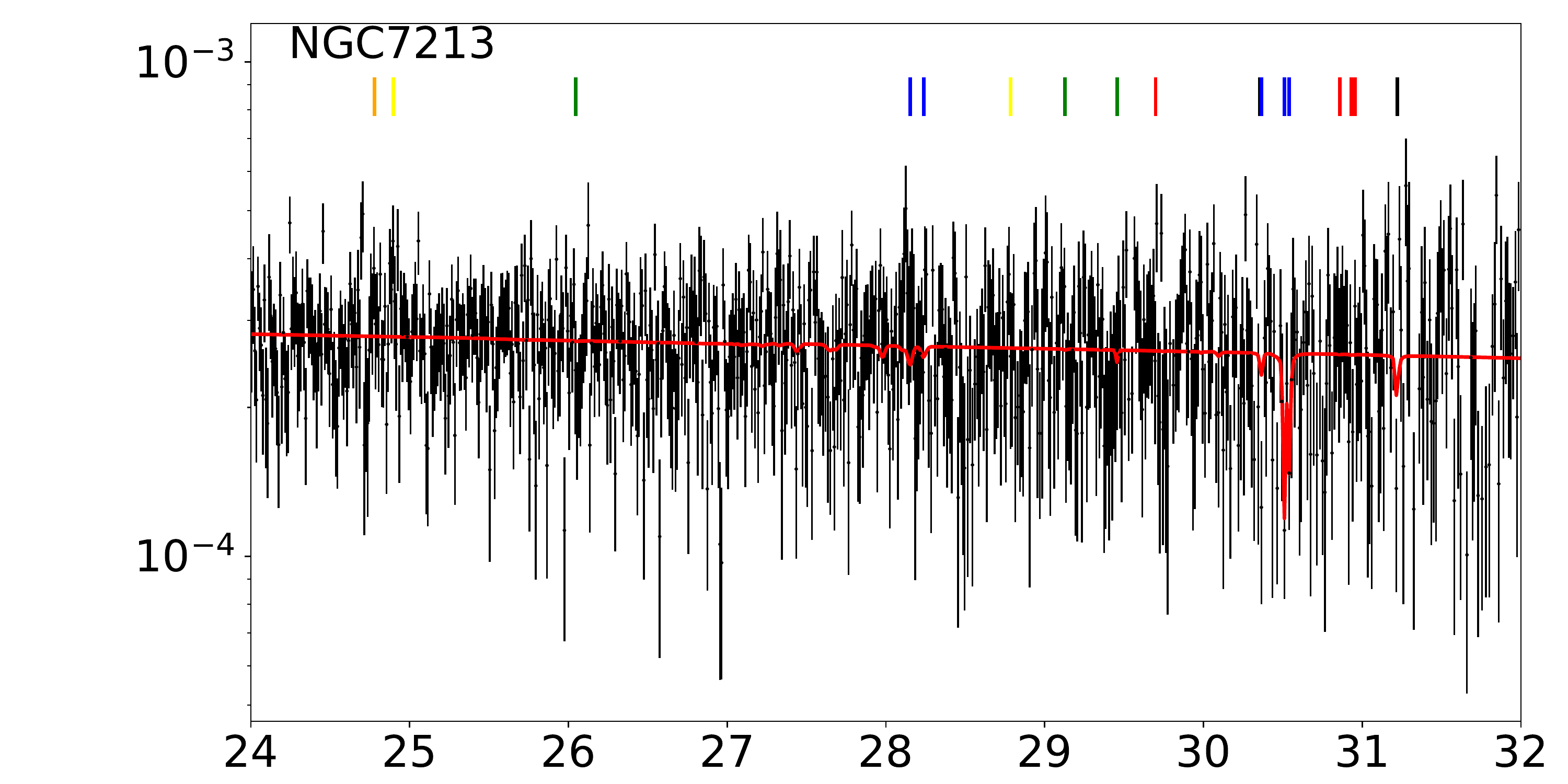}
\hspace{-2.5mm}
\includegraphics[scale=0.195]{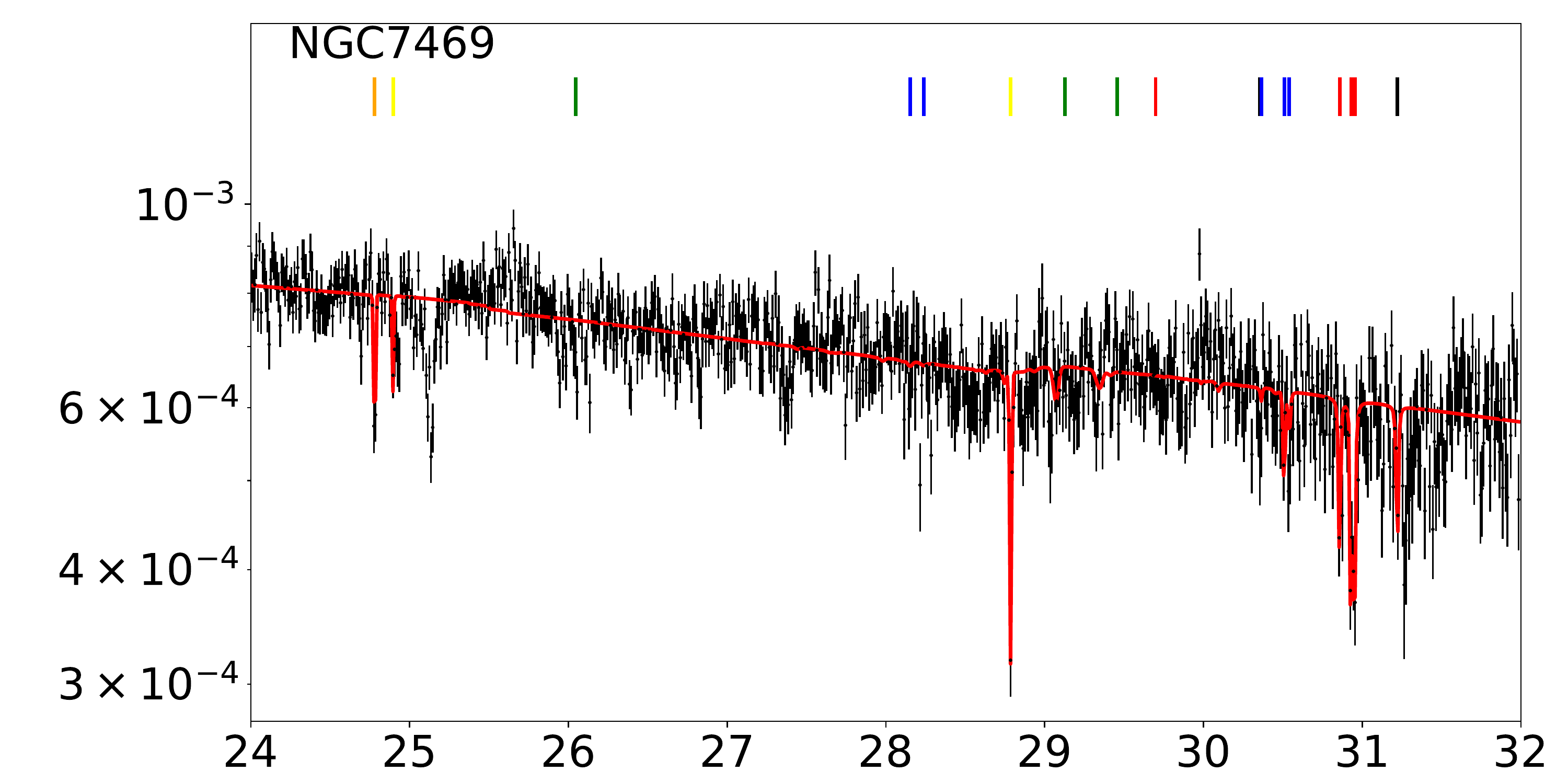}\\
\includegraphics[scale=0.195]{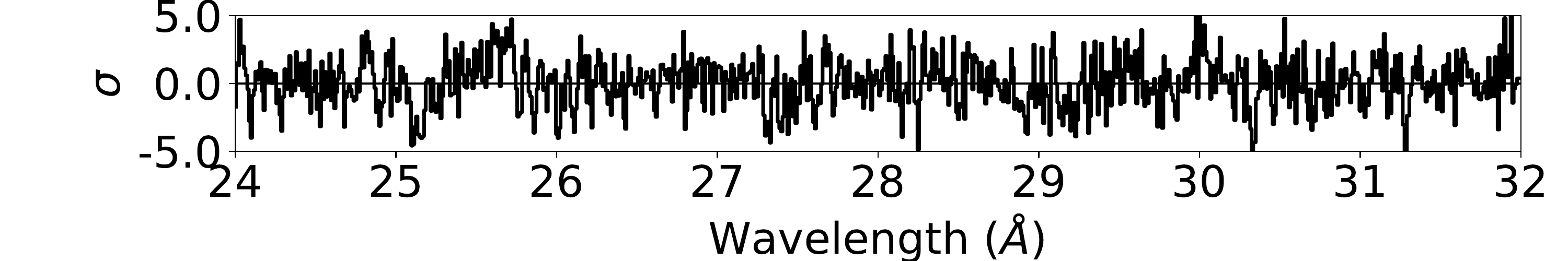} 
\hspace{-2.5mm}
\includegraphics[scale=0.195]{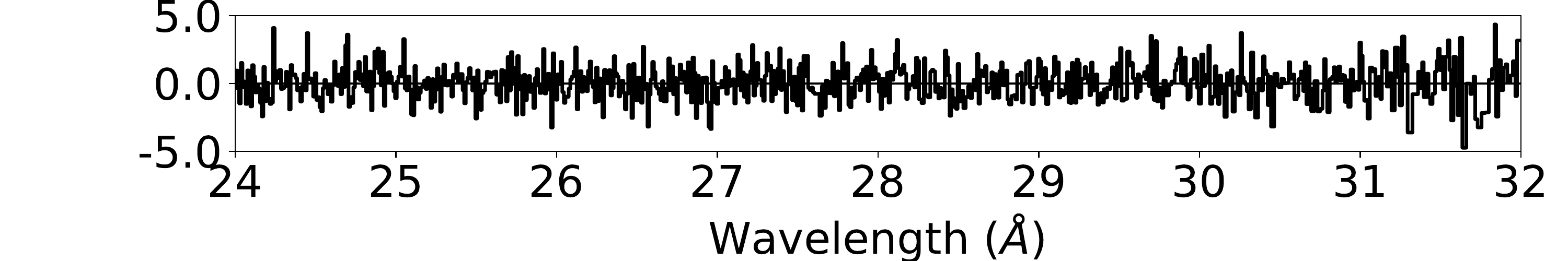} 
\hspace{-2.5mm}
\includegraphics[scale=0.195]{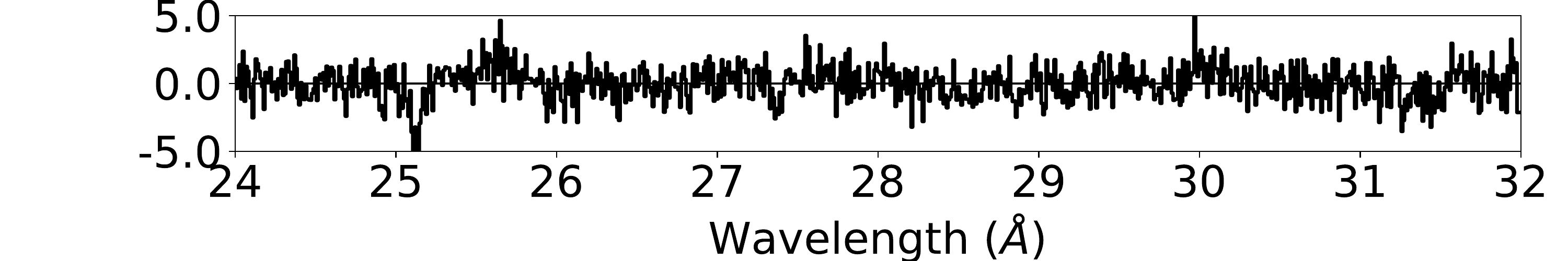}\\

\includegraphics[scale=0.195]{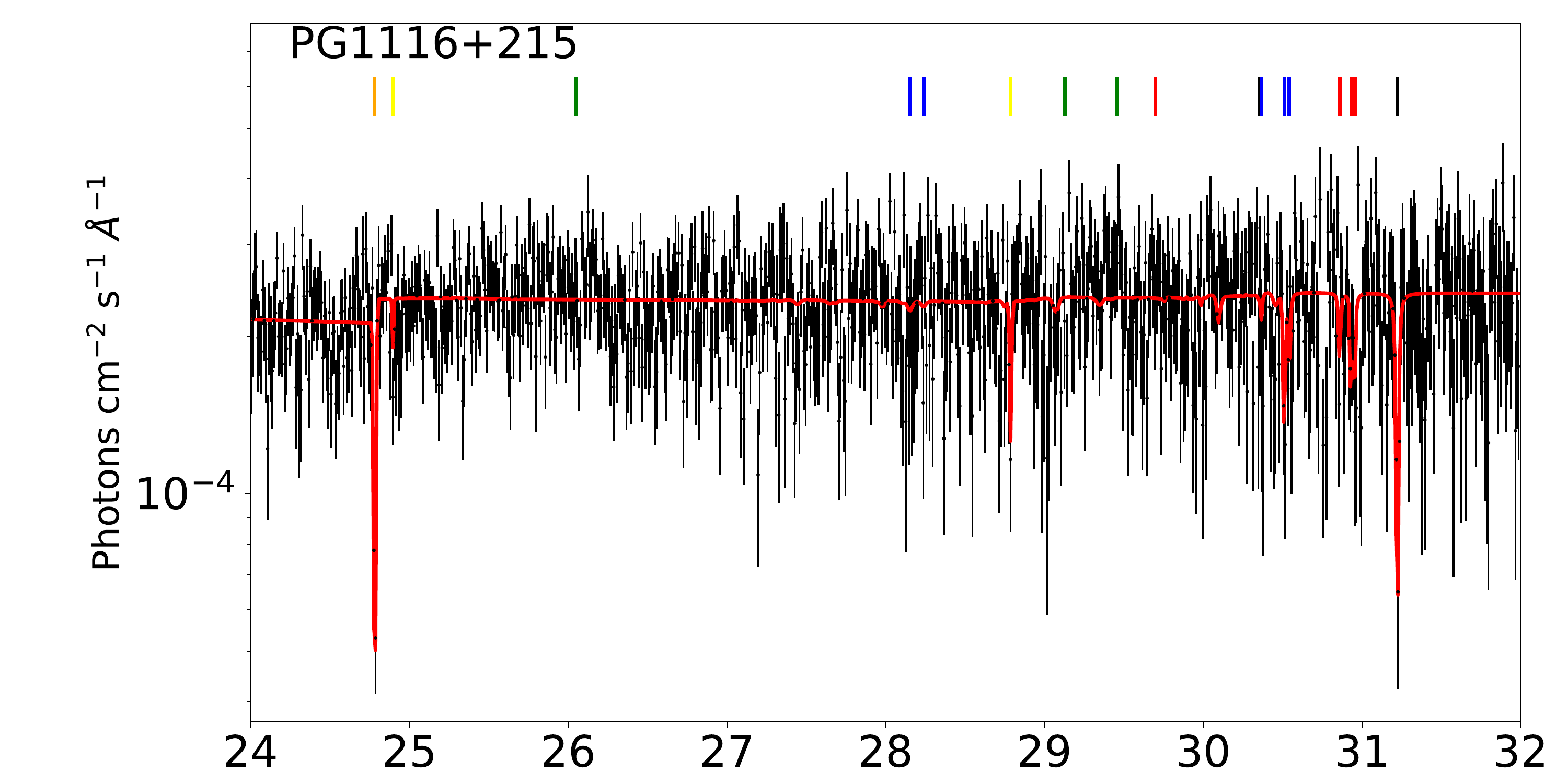} 
\hspace{-2.5mm}
\includegraphics[scale=0.195]{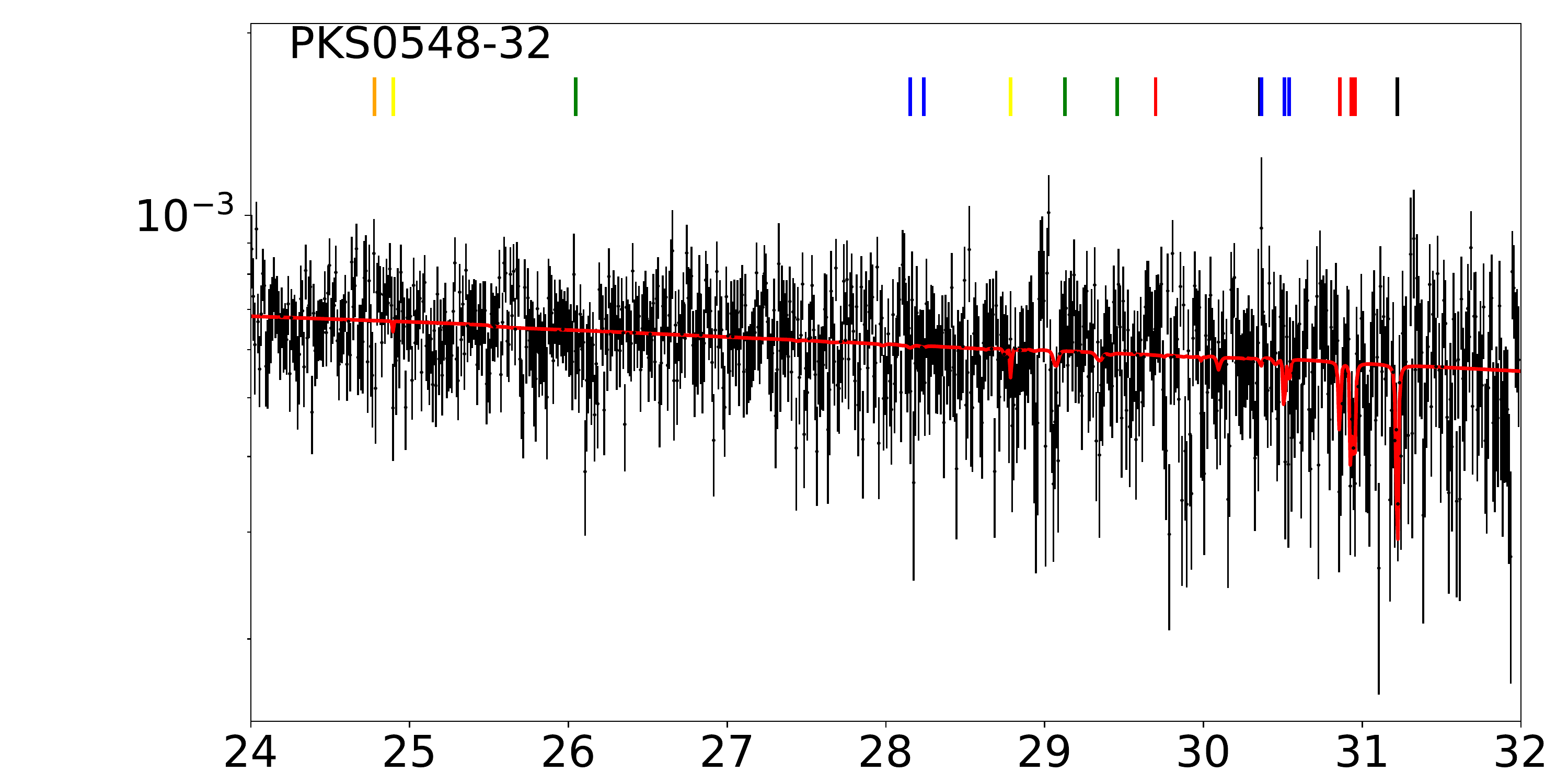}
\hspace{-2.5mm}
\includegraphics[scale=0.195]{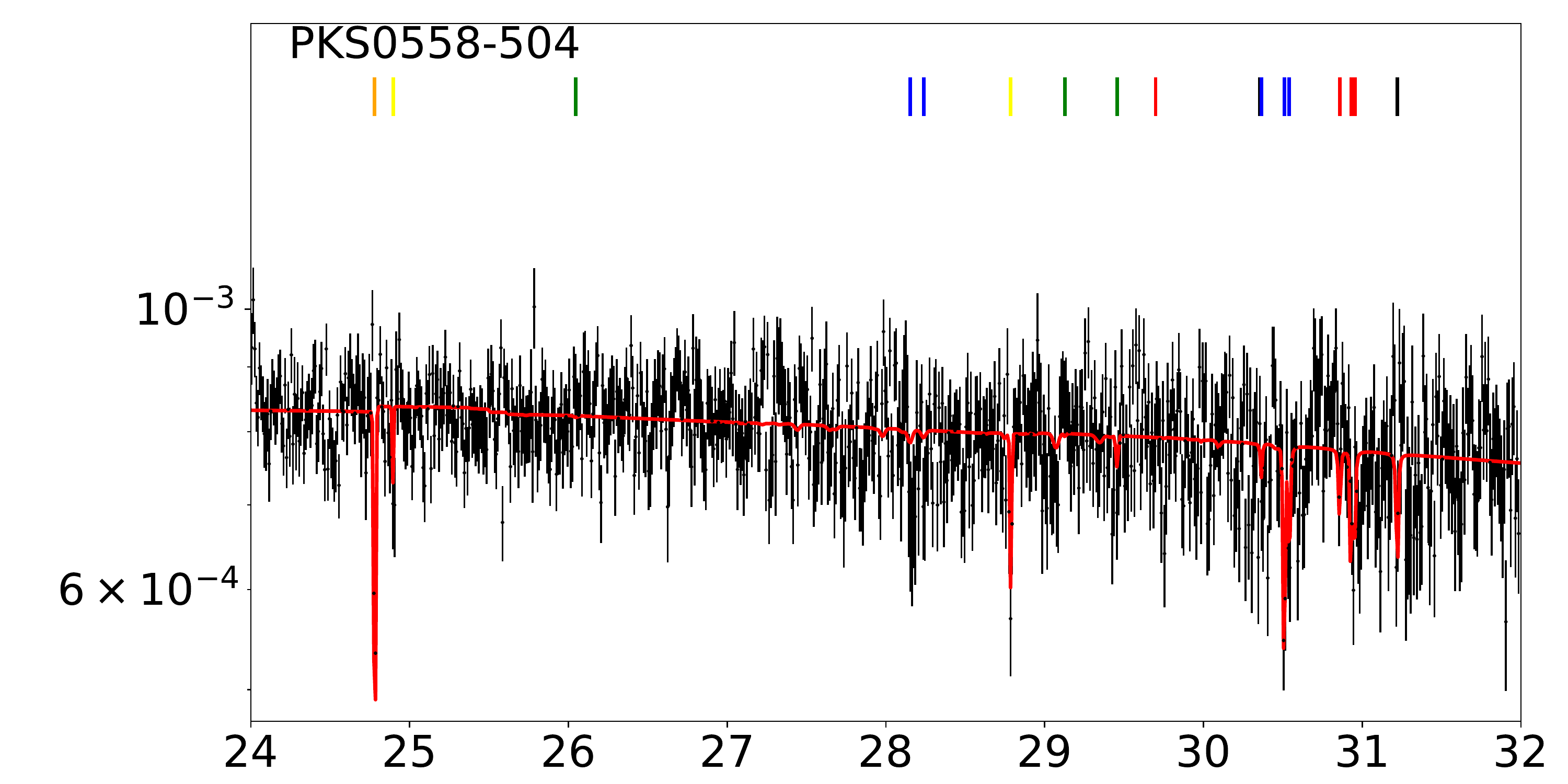}\\
\includegraphics[scale=0.195]{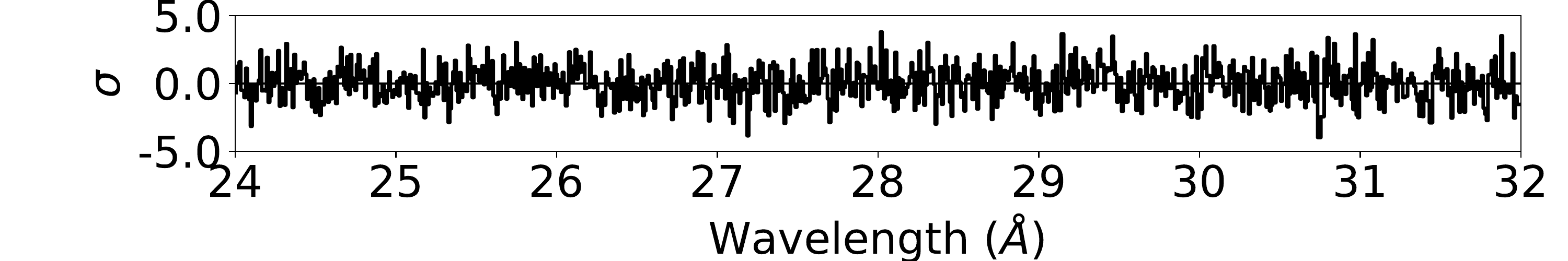} 
\hspace{-2.5mm}
\includegraphics[scale=0.195]{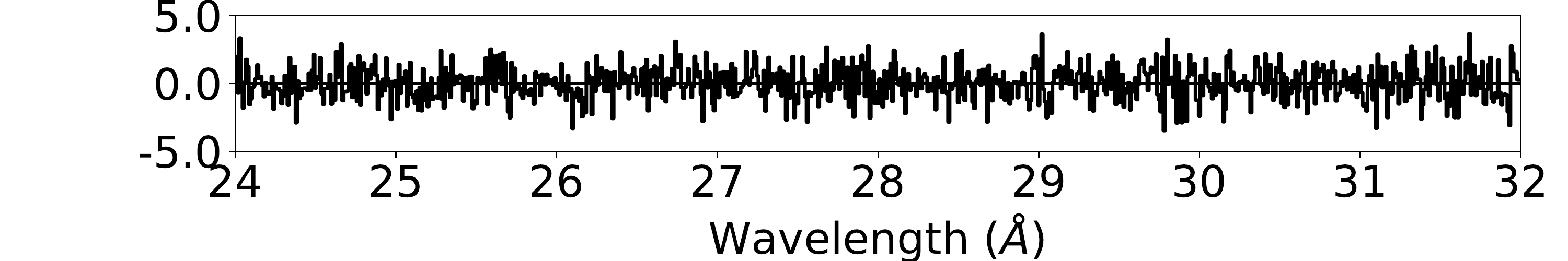}
\hspace{-2.5mm}
\includegraphics[scale=0.195]{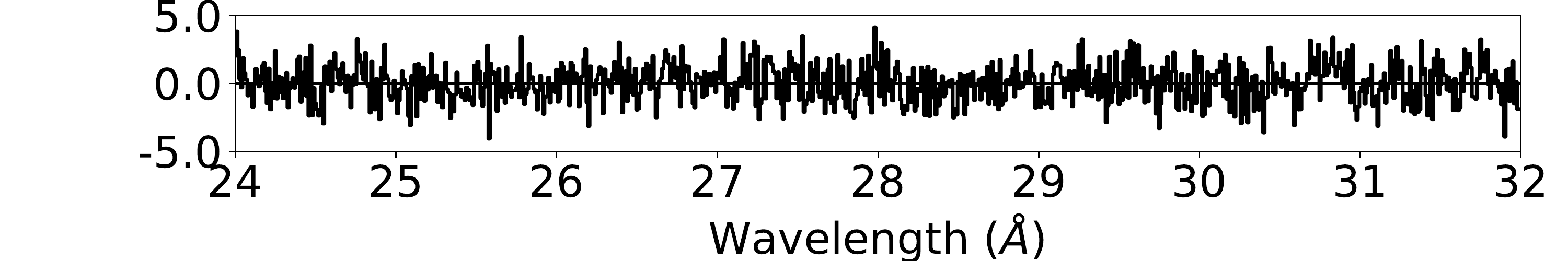}\\

\includegraphics[scale=0.195]{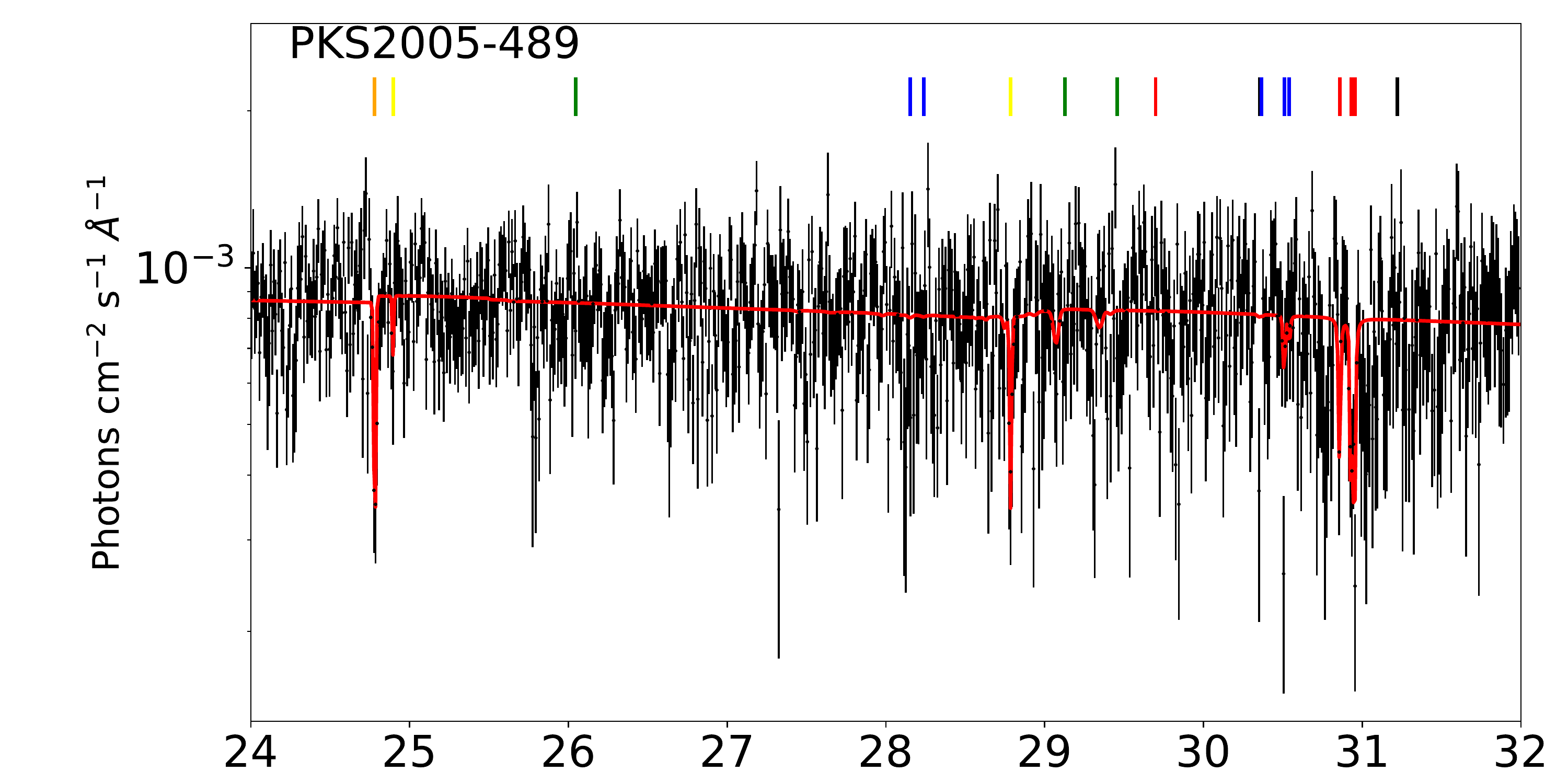} 
\hspace{-2.5mm}
\includegraphics[scale=0.195]{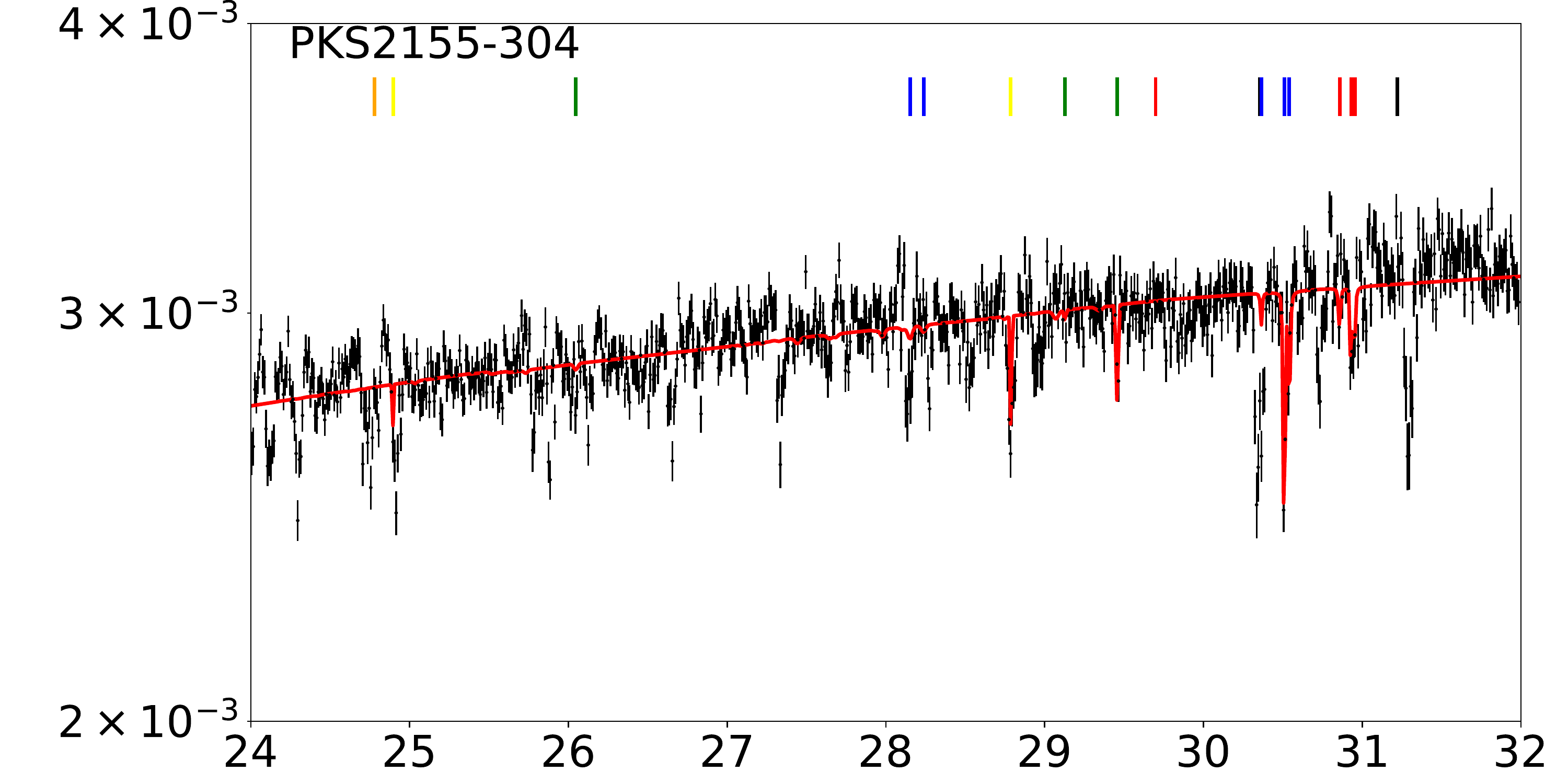}
\hspace{-2.5mm}
\includegraphics[scale=0.195]{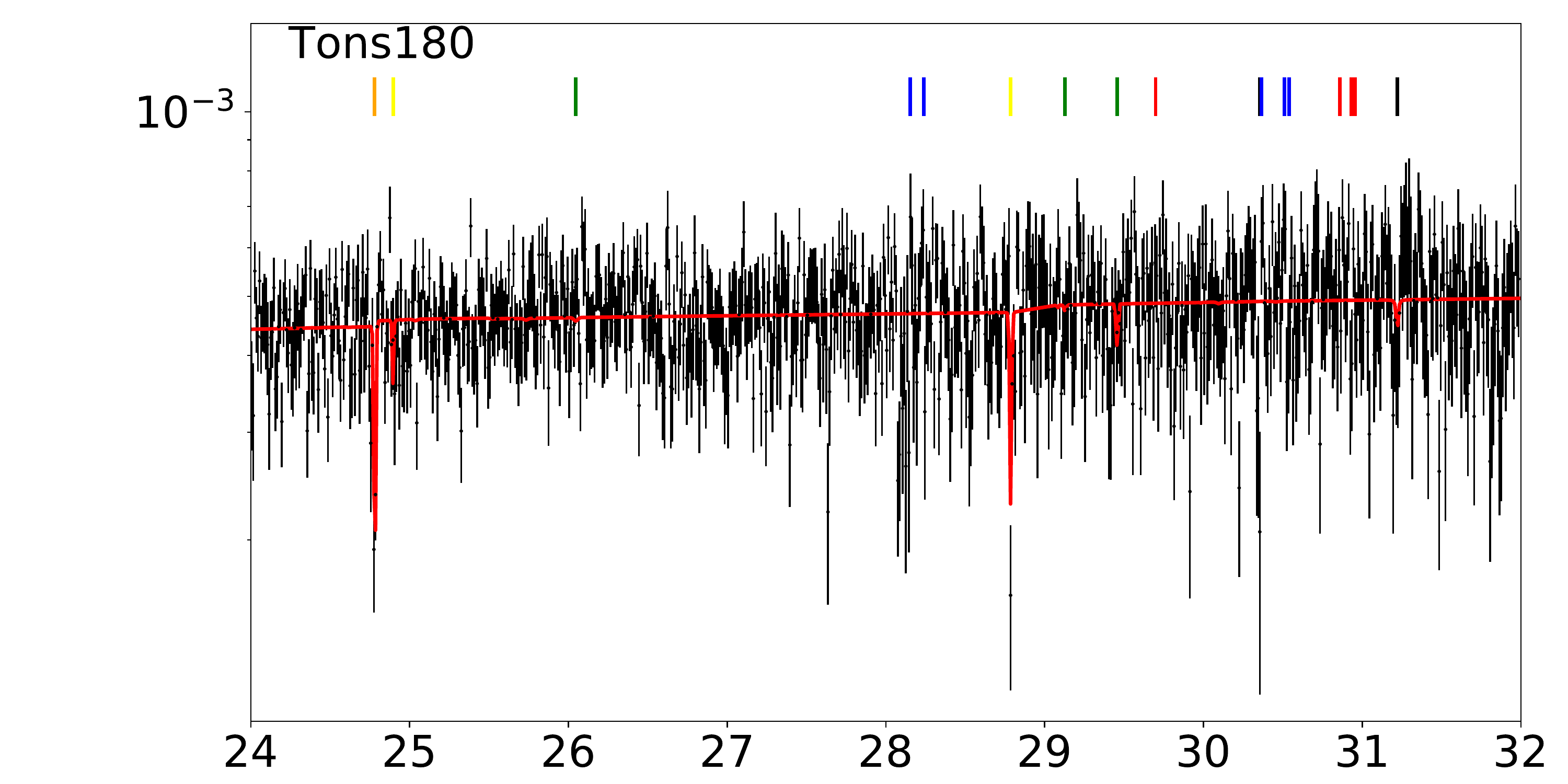}\\
\includegraphics[scale=0.195]{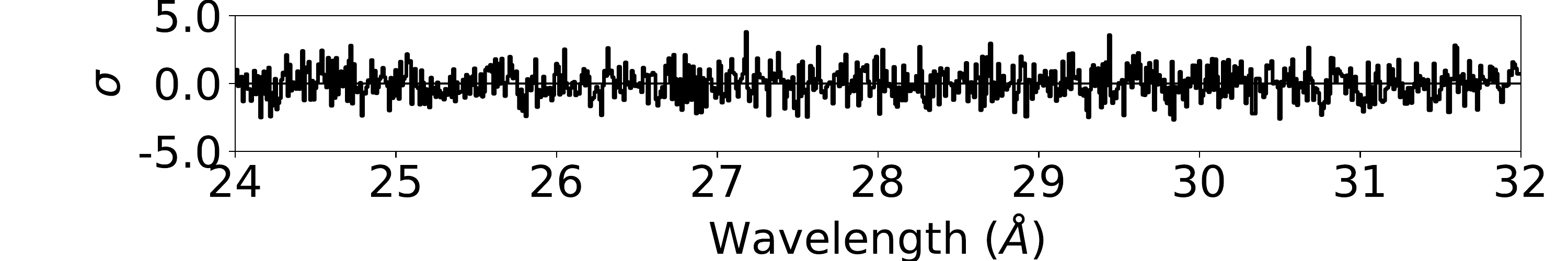}
\hspace{-2.5mm} 
\includegraphics[scale=0.195]{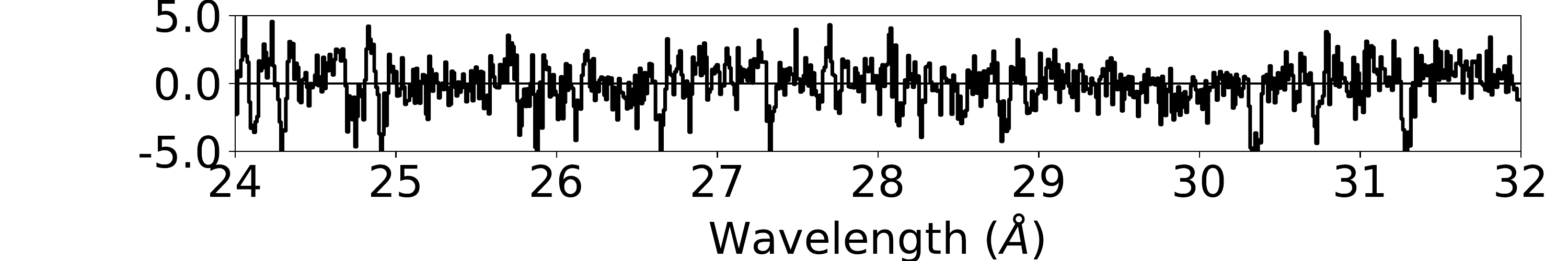}
\hspace{-2.5mm}
\includegraphics[scale=0.195]{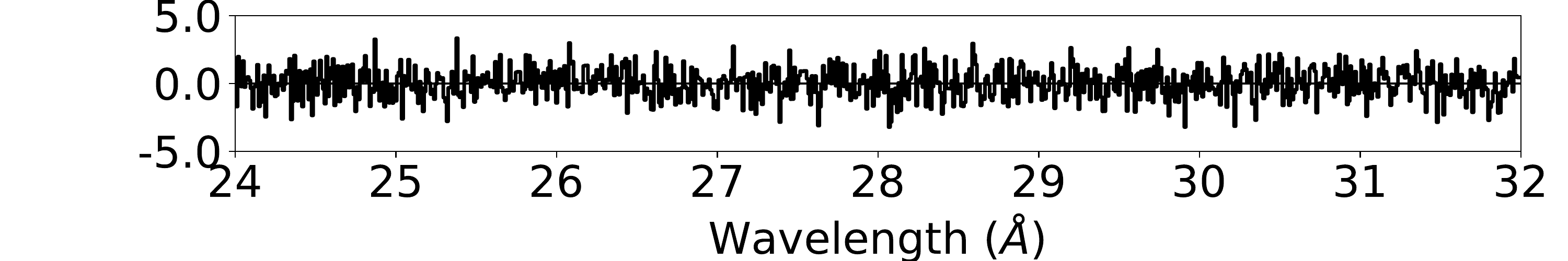}\\
 
      \contcaption{.}\label{fig_fits_ex3}
   \end{figure*}

\end{document}